# Electrically Conductive 2D Material Coatings for Flexible & Stretchable Electronics: A Comparative Review of Graphenes & MXenes

*Vicente Orts Mercadillo, Kai Chio Chan, Mario Caironi, Athanassia Athanassiou, Ian A. Kinloch, Mark Bissett *, Pietro Cataldi **

Vicente Orts Mercadillo, Kai Chio Chan, Prof. Ian Kinloch, Dr. Mark Bissett, Dr. Pietro Cataldi

Henry Royce Institute, National Graphene Institute, Department of Materials, The University of Manchester, Oxford Road, M13 9PL, U.K.

Email Address: mark.bissett@manchester.ac.uk; pietro.cataldi@iit.it

Dr. Mario Caironi, Dr. Pietro Cataldi

Center for Nano Science and Technology @PoliMi, Istituto Italiano di Tecnologia, Via Pascoli 70/3, Milan, 20133 Italy

Dr. Athanassia Athanassiou, Dr. Pietro Cataldi

Smart Materials, Istituto Italiano di Tecnologia, Via Morego 30, Genova, 16163 Italy

# Abstract

There is growing interest in transitioning electronic components and circuitry from stiff and rigid substrates to more flexible and stretchable platforms, such as thin plastics, textiles, and foams. In parallel, the push for more sustainable, biocompatible, and cost-efficient conductive inks to coat these substrates, has led to the development of formulations with novel nanomaterials. Among these, 2D materials, and particularly graphenes and MXenes, have received intense research interest due to their increasingly facile and scalable production, high electrical conductivity, and compatibility with existing manufacturing techniques. They enable a range of electronic devices, including strain and pressure sensors, supercapacitors, thermoelectric generators, and heaters. These new flexible and stretchable electronic devices developed with 2D material coatings are poised to unlock exciting applications in the wearable, healthcare and Internet of Things sectors. This review has surveyed key data from more than 200 articles published over the last 6 years, to provide a quantitative analysis of recent progress in the field and shade light on future directions and prospects of this technology. We find that despite the different chemical origins of graphenes and MXenes, their shared electrical properties and 2D morphology, guarantee intriguing performance in end applications, leaving plenty of space for shared progress and advancements in the future.

**Table 1**: Definition of acronyms used.

| Acronym | Definition |
|---|---|
| **AC** | Activated Carbon |
| **Ag NPs** | Silver Nanoparticles |
| **Ag NWs** | Silver Nanowires |
| **APP** | Ammonium Polyphosphate |
| **C60** | Buckminster Fullerene |
| **CB** | Carbon Black |
| **CCF** | Carbonised Cotton Fabric |
| **CMC** | Carboxymethyl Cellulose |
| **CNF** | Carbon Nanofibres |
| **CNTs** | Carbon Nanotubes |
| **Co NPs** | Cobalt Nanoparticles |
| **CTAB** | Cetrimonium bromide |
| **Cu NPs** | Copper Nanoparticles |
| **Cu NWs** | Copper Nanowires |
| **DMF** | Dimethylformamide |
| **DMSO** | Dimethyl sulfoxide |
| **ECG** | Electrochemically Exfoliated Graphene |
| **EG** | Ethylene glycol |
| **FLG** | Few Layer Graphene |
| **GNP** | Graphene Nanoplatelets |
| **GNR** | Graphene Nanoribbons |
| **GO** | Graphene Oxide |
| **Hf-$SiO_2$** | Superhydrophobic Fumed Silica |
| **HIPS** | High Impact Polystyrene |
| **HPMC** | Hypromellose |
| **IPA** | 2-Propanol |
| **MIT** | Methylisothiazolinone |
| **MOF** | Metal Organic Framework |
| **Ni-MOF** | Nickle Metal Organic Framework |
| **Ni/Fe LDH** | Nickle / Iron Layered Double Hydroxide |
| **NMP** | N-Methyl-2-pyrrolidone |
| **PAA** | Polyacrylic acid |
| **PAI** | Polyamide |
| **PANI** | Polyaniline |
| **PANI NWs** | Polyaniline Nanowires |
| **PDA** | Polydopamine |
| **PDAC** | Polydiallyldimethylammonium chloride |
| **PDMS** | Polydimethylsiloxane |
| **PEDOT** | Poly(3,4-ethylenedioxythiophene) |
| **PEDOT:PSS** | Poly(3,4-ethylenedioxythiophene) |
| **PEG** | Polyethylene Glycol |
| **PEI** | Polyethyleneimine |
| **PEN** | Polyethylene naphthalate |
| **PET** | Polyethylene terephthalate |
| **PI** | Polyimide |
| **PP** | Polypropylene |
| **PPy** | Polypyrrole |
| **PS** | Polystyrene |
| **PSS** | Poly(styrenesulfonate) |
| **PTFE** | Polytetrafluoroethylene |
| **PU** | Polyurethane |
| **PVA** | Polyvinyl alcohol |
| **PVDF** | Polyvinylidene fluoride |
| **PVP** | Polyvinylpyrrolidone |
| **rGO** | reduced Graphene Oxide |
| **SA** | Sodium Alginate |
| **SC** | Sodium Cholate |
| **SDBS** | Sodium Dodecylbenzene Sulfonate (SDBS) |
| **SDC** | Sodium Deoxycholate (SDC) |
| **SDS** | Sodium Dodecyl Sulfate (SDS) |
| **SrGO** | Sulphonated Reduced Graphene Oxide |
| **$Ti_3C_2Tx$** | Titanium Carbide |
| **TPU** | Thermoplastic polyurethane |
| **WPU** | Water-Borne Polyurethane |
| **ZnO NPs** | Zinc Oxide Nanoparticles |

# 1.0. Introduction

The growing demand for seamlessly integrated wearable devices and the Internet of Things has fuelled interest in transitioning electronic circuits away from traditional stiff and rigid substrates and onto flexible and stretchable materials.[1-2] Typically, printed circuit boards are fabricated using an epoxy glass-fibre panel onto which conductive copper tracks are patterned and etched to connect electronic components.[3] This approach is incompatible with flexible and stretchable platforms due to the difference in stiffness between the electronic connection and the substrates.[4-6] As such, the focus has turned to developing conductive inks to coat and pattern electronics circuits on flexible structures, that are capable of flexing with the substrate while remaining adhered to it, retaining their electrical properties, while also enabling synergistic effects through the combination of binders and complementing conductive fillers into hybrid systems.[7]

The closest substitute to etched copper tracks are metal-based conductive inks. Typically, these inks consist of metal (e.g., gold, silver, platinum, or copper) micro or nanoparticles, a binder to increase the inter-particle bonding and adhesion to the substrate, and a solvent to carry the dispersion.[8] The solvent and the binder often also stabilise the metal particle dispersion, restrict oxidation and agglomeration, and act as reducing agents in the post-printing sintering process, which also serves to form the conductive network and improves the electrical conductivity. These inks benefit from the low electrical resistance of the metal particles, which are usually only 2 – 3 times less conductive than their bulk counterparts, with sheet resistances reaching the $\mu\Omega$ /sq. range and resistivities below $10^{-7}$ $\Omega$ m.[9-10]

Currently, nonmetal inks are more electrically resistive than metal inks, generally reaching $10^{-1}$ $\Omega$ /sq. and $10^{-5}$ $\Omega$ m, as shown in section 2.3. However, they have other essential advantages.

The category of nonmetal conductive inks can encompass a wide range of organic conductive fillers, including conductive polymers (e.g., poly(3,4-ethylenedioxythiophene), polystyrene sulfonate (PEDOT: PSS) or polyaniline (PANI)) and different dimensionality nanocarbons such as 0D fullerenes, 1D nanotubes, or 2D platelets. [11] These nonmetal inks do not require a post-coating sintering step, which can reach damaging temperatures for common flexible polymer substrates such as polyethylene terephthalate (PET) and cellulose. They are generally easier to be dispersed and create a stable conductive ink with time, and in some instances, are biocompatible, enabling easier processing and a more comprehensive range of applications. They are less costly than inks that employ precious metal particles as, for example, gold and silver ones. Furthermore, generally they can also be more easily formulated using fewer and more environmental- and human-friendly solvents such as water or alcohols.

Conductive inks with 2D material-based formulations have drawn particular attention due to the features enabled by their flake-like shape and high aspect ratio.[12] In thin-film coatings on flat, transparent, and flexible polymer substrates (e.g., PET), they can display high optical transmissivity and good electrical conductivity, enabling their use in transparent, stretchable electronics. They adhere well to many textiles and foams, forming conductive networks consisting of nanoplatelet stacks that can slip and slide over one another, resulting in unique micromechanical properties that can be exploited for strain and pressure sensing. Advances in scalable manufacturing processes for 2D material nanoplatelets, particularly liquid-phase exfoliation, are enabling large production quantities and competitive price points[13]; graphene screen sprinting pastes can cost half the price of silver alternatives.[14] Furthermore, the large and growing family of solution-processable 2D materials encompasses materials with different electronic properties: from conductors, e.g., graphene, to semi-conductors, e.g., tungsten disulphide ($WS_2$) or molybdenum disulphide ($MoS_2$) and insulators, e.g., hexagonal boron nitride (hBN) or graphene oxide (GO). Additionally, MXenes, a family of conductive 2D transition metal carbides / nitrides, have been shown to combine the low resistivities of metal inks with the platelet morphology and easier processability of 2D inks. These factors are generating excitement, investments, and rapid technological advances.

Flexible, conductive coatings can be directly utilized (e.g., for EMI shielding capabilities) or controlled to perform a function (e.g., for sensing or heating). Furthermore, the coatings can be patterned to leverage conducting geometries (e.g., for interdigitated electrodes in supercapacitors or strain sensing). Many existing large area coating techniques can be used with compatible conductive inks such as screen printing, spray coating, or dip coating. The technique chosen will depend on the substrate being used, the desired coating properties and the necessity to scale production. Inkjet printing is among the most actively researched techniques for manufacturing flexible electronics. Progress in inkjet printing 2D nanomaterials-based inks has been detailed in multiple recent reviews [8, 15-21] and as such will not be considered here. The facile application of novel 2D nanomaterials-based ink formulations has enabled the quick prototyping of devices across a wide range of fast-growing fields, including motion/tactile sensing[22], supercapacitors[23-24], heat generation[25-27], strain/pressure sensors[28-29], thermoelectric generators[30-31], and EMI shielding[32].

The ground-breaking isolation of graphene at The University of Manchester in 2004 spurred a boom in 2D materials research. The discovery of MXenes at Drexel University in 2011, drove a subsequent explosion in growth in the last decade, which has particularly picked up pace over the last five years (see **Figure 1**).[33] As these two 2D material families bridge the gap from lab to industry, consolidating recent progress enables us to assess the state of the art. This review surveys more than 200 articles published in the last six years (2015 – 2021) that report coating of flexible or stretchable substrates with inks containing graphenes or MXenes for electronic applications. The sustainability of common ink formulations, considering solvent, binders, and conductive fillers, as well as the possible environmental implications of their adoption are explored. Reviewed articles are grouped by the principal application described: strain sensing, pressure sensing, supercapacitors, thermoelectric generation, joule heaters, or electromagnetic interference (EMI) shielding. Crucial values are summarised in tables for each application field. Where the data was available, essential information was gathered into composite plots to help establish aggregate trends and spot outliers and provide a reader-friendly and instant picture of the state-of-the-art.

# 2.0. Coatings

## 2.1 Techniques

**Table 2**: Categorisation of substrates into sheets, textiles, and foams, with a breakdown of materials and coating techniques used per substrate type.

| | ***Sheets*** | ***Textiles*** | ***Foams*** |
|---|---|---|---|
| | 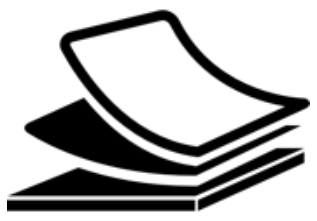 | 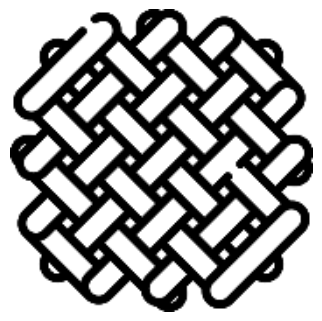 | 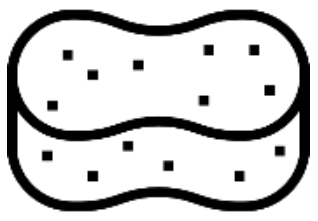 |
| *Articles Surveyed* | 103 | 121 | 16 |
| *Materials Used* | PET (35%), Paper (12%), PDMS (8%), PI (8%), PU (6%), Cellulose (5%), Latex (3%), Other (23%) | Cotton (36%), Carbon (15%), PET (7%), Nylon (7%), Glass (5%), Polyester (5%), Silk (4%), PU (3%), Other (18%) | PU (50%), Melamine (12%), Ecoflex (6%), Nickel (6%), PDMS (6%), PI (6%), Salt (6%), Sea Sponge (6%) |
| *Coating Technique* | Spray (39%), Dip (11%), Drop Casting (8%), Spin (8%), Blade (5%), Screen Printing (5%), Gravure Printing (4%), Other (20%) | Dip (70%), Spray (15%), Drop Casting (2%), Electrochemical Deposition (2%), Blade (2%), Filtration (2%), Spin (2%), Other (5%) | Dip (88%), Drop Casting (6%), Electrochemical Deposition (6%) |

There is a wide range of coating techniques available for coating flexible and stretchable substrates. The technique used are influenced by the substrate material and form, with 45 different materials being used across the surveyed literature. These have been assigned into three categories related to their form: sheets, textiles, and foams. Sheets consist of a flat surface, making for a smooth, laminar support on which to coat thin films. Textiles consist of threads arranged into a fabric, providing a permeable support network for inks to adhere onto. Foams consist of porous structures that readily soak up and adsorb inks. **Table 2** shows the material breakdown for these three forms. The same material may be engineered into multiple forms, e.g., PET can be woven into a fabric or made into flat transparent sheets. There are clear engineering preferences for certain materials across each category: PET sheets, cotton textiles, and polyurethane (PU) foams. Table 2 shows that the coating techniques can be used interchangeably between the categories. Specific techniques are adjusted to the properties of particular substrates. The data shows that spray coating was the most popular technique for coating sheets (~40%), while researchers mostly opted for dip coating when coating textiles (~70%) and foams (~88%). Other methods have key advantages for specific uses: screen printing enables complex patterns to be coated simply, repeatedly, and consistently, while blade coating allows for fine control of coating thickness over flat surfaces. These are summarised in **Table 3**.

**Table 3**: Main coating techniques used in surveyed literature.

| | ***Dip*** | ***Spray*** | ***Screen*** | ***Blade*** |
|---|---|---|---|---|
| *Application* | Holding the substrate to be coated in a bath of ink solution and allowing for the dispersed phase to adhere onto the substrate surface.[34] | Loading a spray gun with ink solution and using compressed air (or similar) to direct a cloud of airborne material onto the substrate surface.[34] | Squeezing an ink paste through a mesh placed above a substrate by applying a shear pressure with a squeegee.[35] | Dragging a pool of ink across the flat substrate surface using a mechanical blade.[34] |
| *Patterning* | Cutting substrate into shape. | Mask on the substrate. | Mask applied to screen. | Cutting substrate into shape. |
| *Ink Properties* | Low surface tension. | Fast evaporating and low viscosity of 1 – 10 mPa s.[36] | High viscosity ~1 Pa s, to resemble a paste | High viscosity 1 – 10 Pa s.[37] |
| *Articles Surveyed* | 111 | 57 | 7 | 7 |
| *Solvents Used* | Water (89%), Ethanol (2%), IPA (2%), NMP (2%), Other (5%) | Water (52%), DMF (13%), Ethanol (11%), Chloroform (10%), IPA (8%), Acetone (3%), NMP (3%) | Water (71%), Dibasic Esters (14%), NMP (14%) | Water (50%), NMP (38%), Diethylene Glycol (12%) |

## 2.2 Formulations, Challenges and Sustainability

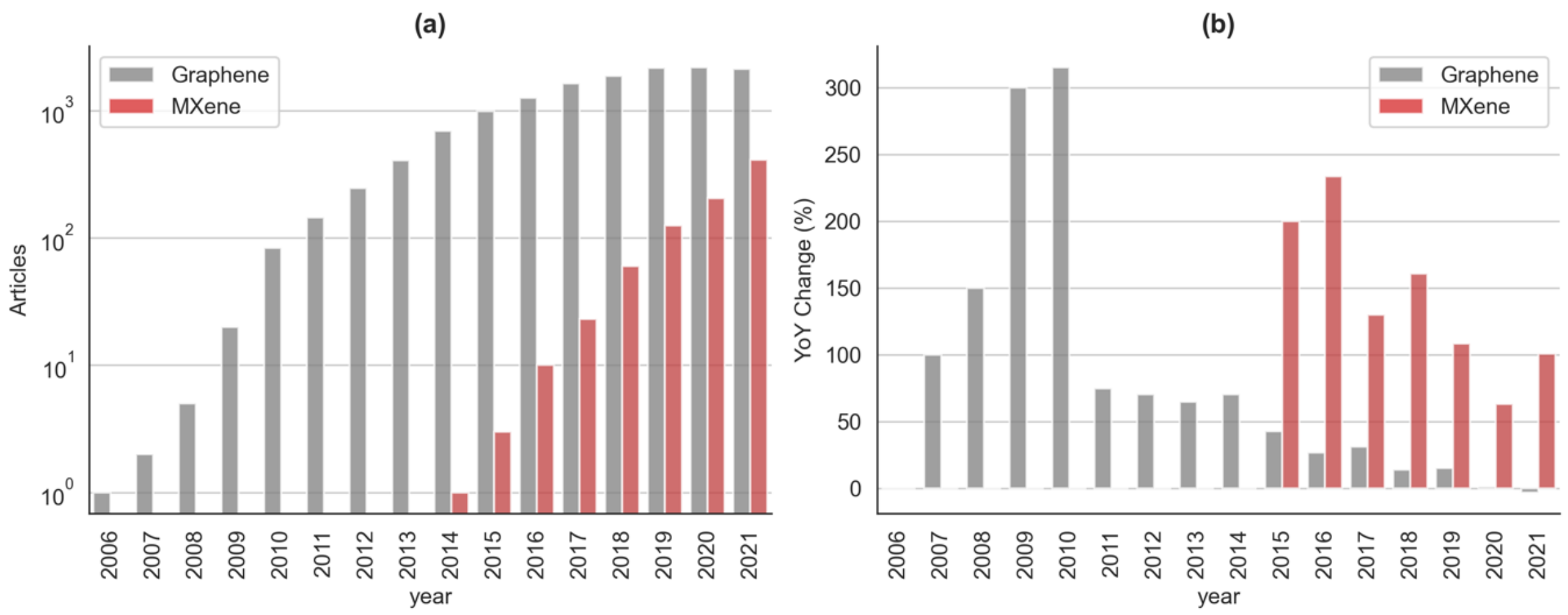

**Figure 1**: (a) The absolute number and (b) year-on-year percentage change of published articles with graphene, MXene (or $Ti_3C_2T_x$) and flexible or stretchable in the title, keywords or abstract over the last 15 years from the Web of Science (data collected on 27/05/2022).

All 2D materials share an atomically thin flake-like shape, which at the platelet level unlocks a high aspect ratio (i.e., the ratio of lateral size to thickness) and accessible surface area (~2600 $m^2$ $g^{-1}$ in completely exfoliated graphene[38]). There are billions of individual platelets at the powder macroscale, each with its lateral size, atomic thickness, and surface chemistry. Understanding and controlling the unique interactions brought about by this two-dimensionality, both between platelets and between their surroundings, is crucial to successfully leverage their properties. Conductive inks benefit from stable, homogeneous formulations that allow for consistent coating behaviour and performance. Achieving a good dispersion of 2D platelets in a solvent, while controlling stability, electrical conductivity and viscosity with stabilisers, binders, and additives can make for a challenging formulation. Furthermore, different 2D material families have different atomic compositions which affect these properties, and thus their use in practice as summarised in **Table 4**. **Figure 1** shows the absolute number (a) and year-on-year percentage change (b) of articles referencing graphene and MXenes, alongside keywords ‘flexible’ or ‘stretchable’ over the last 15 years to show how academic interest in these materials has evolved. MXenes, first isolated in 2011, are experiencing an almost identical rate of uptake in the field of stretchable and flexible electronics to graphene, isolated 7 years earlier. In absolute terms graphene articles outpace MXene articles by an order of magnitude, although growth has plateaued. It will be interesting to observe whether MXenes will continue to follow graphene’s trend as the field develops.

**Table 4**: Breakdown of 2D materials in surveyed literature. GNP and GO chemical structures adapted with permission.[39] Copyright (2021) MDPI. Ti3C2Tx chemical structure adapted with permission.[40] Copyright (2021), Wiley–VCH.

| | ***Graphene*** | ***Graphene Oxide*** | ***$Ti_3C_2T_x$*** |
|---|---|---|---|
| *Chemical Structure* | 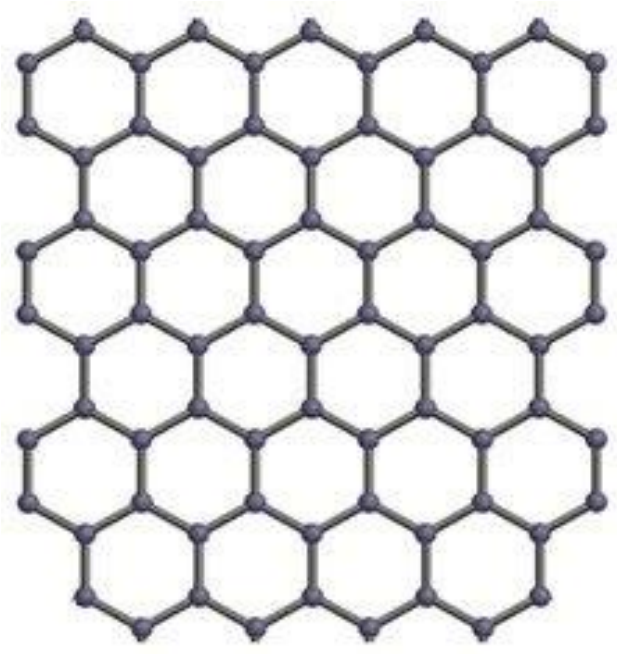 | 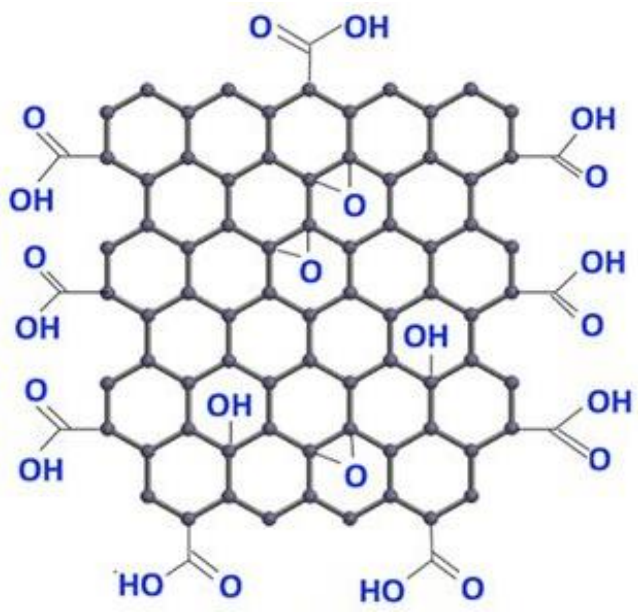 | 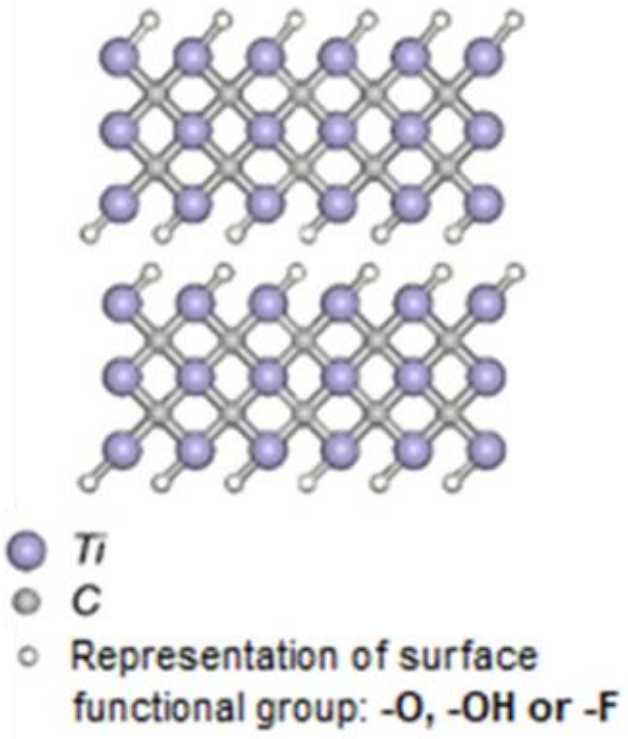 |
| *Articles Sampled* | 68 | 96 | 50 |
| *Solvents Used* | Water (55%), Chloroform (11%), IPA (7%), Ethanol (5%), DMF (4%), NMP (4%), Acetone (3%), Other (11%) | Water (78%), DMF (9%), Ethanol (7%), Other (6%) | Water (76%), NMP (12%), Ethanol (4%), IPA (4%), DMF (2%), DMSO (2%) |
| *Polymer Additives Used* | None (48%), PEDOT:PSS (10%), WPU (5%), Acrylic Binder (4%), PU (4%), PVP (4%), TPU (4%), Other (12%) | None (87%), PVDF (3%), PDA (2%), PSS (2%), Other (6%) | None (88%), PVDF (8%), PPy (2%), PTFE (2%) |

Graphene is a 2D allotrope of graphite consisting of a single atomic layer of carbon – carbon bonds arranged in a honeycomb lattice.[41] This $sp^2$ hybridisation imparts strong mechanical properties (namely a 1 TPa Young's modulus) from the σ bonding, and low electrical resistivity (~ $10^{-8}$ Ω m) from the π bonding.[42] It is possible to isolate close to ideal single-layer graphene (SLG) sheets through both top-down, e.g., micromechanical exfoliation such as the scotch-tape method), and bottom-up techniques, e.g., chemical vapour deposition (CVD). The graphene produced via mass production techniques (i.e., kilotons/year)[43], e.g., liquid phase, shear or electrochemical exfoliated, mainly consists of thicker flakes categorised as few-layer (4 to 10 layers) or multi-layer (> 10 layers), with lateral sizes ranging from 1 to 50 μm[44], and is generally commercially distributed as graphene nanoplatelets (GNPs)[45]. Obtaining SLG from these mass-production techniques requires additional ultracentrifugation steps to single out only the thinner flakes.[46] While GNPs do not match the exceptional properties of SLG, they can reach a high surface area (up to 1000 $m^2$ $g^{-1}$[44]) and low resistivity (~ $10^{-4}$ Ω m[47]). GNPs are usually sold in batches of similar lateral size, e.g., First Graphene Ltd. sells their PureGraph GNP in 5, 10, or 20 μm sizes, or surface area, e.g., XGSciences sells their grade C GNP in 300, 500, or 750 $m^2$ $g^{-1}$ areas.[44] The lateral size, thickness, and surface chemistry of these flakes will influence how well they disperse. For example, GNPs that are thinner, smaller, or decorated with hydrophilic surface chemistry will best disperse in water. However, 86% of surveyed articles used organic solvents, binders, or surfactants to assist with GNP dispersibility in water. Whereas this dropped to 27% when using GO or its reduced form rGO.

GO is a chemical derivative of graphene where the basal plane and edges are decorated with defects and oxygen-containing functional groups. This surface chemistry allows GO to disperse homogeneously in water, and thus is often used in inks without stabilising additives. It can also assist adherence to the substrate surface through improved chemical compatibility. The rise in $sp^3$ character affects the material properties, notably drastically increasing the electrical resistivity (~ $10^8$ Ω m) through the loss of conducting π bonds. As such, GO cannot be used as a conductive coating until it has been reduced to recover more $sp^2$ hybridisation, resulting in resistivity values up to five orders of magnitude higher, ~$10^3$ **Ω**.[48] GO can be manufactured at scale (i.e., hundred tonnes / year)[49] via a modified Hummer's method, involving the use of heavy acids to surface treat exfoliated graphitic material and selectively filtering thin flakes.[50] It can also be produced via simultaneous electrochemical exfoliation and oxidation.[51]

Reduction to rGO can be carried out using thermal (16%), chemical (76%), electrochemical (5%), or plasma/microwave (3%) techniques (the values in brackets referencing surveyed articles), applied to either the GO dispersion or coated substrate. It is usually preferred to first coat with GO and subsequently reduce the coated platelets to rGO, making the most of GO's improved dispersibility. Achieving sustainable chemical production and reduction of GO requires manufactures to displace dangerous ingredients and chemicals.[52] Green reducing agents actively being researched as alternatives include microorganisms, amino acids, and ascorbic acid, to name a few.[53] These green substitutes often use also low temperatures (below 50°C), further improving their sustainability. However, this comes at the cost of speed and reduction extent. [54] A hybrid approach that uses green reducing agents in combination with fewer quantities of traditional reducing supporting agents, could be a positive compromise.

MXenes are 2D metal carbides or nitrides. They can reach lower resistivities than any other solution-processed 2D material, down to ~ $10^{-6}$ Ω m,[55] due to their metallic nature, have strong mechanical properties (330 GPa Young's Modulus), and feature an oxide-like surface that imparts hydrophilic properties for good dispersibility in water. Most surveyed papers (60%) dispersed MXenes in water alone. There are many MXene materials, typically following the formula $M_{n+1}X_nT_x$, where M is an early transition metal (e.g., titanium, vanadium or molybdenum), X is carbon and / or nitrogen, T is a functional group (e.g., -OH, -F, -Cl, or =O), and n can vary from 1 to 4.[56] The first MXene to be isolated was $Ti_3C_2T_x$[57], and despite 29 other forms having been experimentally synthesised[33], it was used in all surveyed MXene articles. MXenes are mainly manufactured via top-down etching of $M_{n+1}AX_n$, i.e., MAX phases, where $Ti_3AlC_2$ is used for $Ti_3C_2T_x$. The element represented by A (usually found in groups 13 or 14) is selectively etched away, leaving the metal carbide or nitride layers, which can subsequently be delaminated in solutions of cation intercalants, down to single-layer sheets, reaching lateral sizes of up to 10 μm.[58] Acidic solutions containing fluorides tend to be used as a chemical etchant, primarily hydrofluoric acid (HF)[56]. Alternative etching techniques include molten salt[59], hydrothermal[60], halogen[61] and electrochemical etching[62]. MXenes have yet to reach the tons / year scale of GNPs, with current scale reaching 50g batch reactors.[63] The surface terminations that impart hydrophilicity also cause MXenes to oxidise, limiting their environmental stability. An aqueous $Ti_3C_2T_x$ dispersion exposed to air will completely oxidise to $TiO_2$ within days, and this can be observed by a colour change from brown to white due to the precipitation of titania from the flakes.[64] To minimise this issue, dispersions can be stored at low temperatures extending stability to over a month, and under argon for over 6 months.[65]

The research effort of the research community towards more sustainable production of MXenes is still at an embryonal stage, which is not surprising considering the more recent development of MXenes. This effort includes the production of MXene with more safe chemicals, ideally water-based, compared to hydrogen fluoride, which is toxic.[66] Researchers recently also proposed using the weight percent of residues consisting of unetched MAX and unexfoliated multi-layered MXene (that can be up to 80–90% of the total starting weight) instead of trashing this sediment away.[67]

There is a clear preference among surveyed articles (70%) to use water as a solvent where possible, since it is easily accessible, not hazardous, and environmentally friendly. The next most popular solvents were ethanol (8%), dimethylformamide (DMF) (6%), isopropyl alcohol (IPA) (6%) and n-methyl-2-pyrrolidone (NMP) (5%). The latter three are polar aprotic solvents that have been shown to aid graphene exfoliation and make the dispersed flakes less prone to agglomerate. On the other hand, water and ethanol are polar protic solvents, thus dispersing hydrophilic material (e.g., GO and $Ti_3C_2T_x$) well. Additives were generally used to either tune rheological properties and improve binding between flakes and substrate (e.g., polyvinylidene fluoride (PVDF), polyurethane (PU), or polyvinylpyrrolidone (PVP)) or to stabilise dispersions (e.g., sodium dodecyl sulphate (SDS), sodium dodecylbenzene sulfonate (SDBS), sodium deoxycholate (SDC) or sodium cholate (SC)). Their use has to be moderate, as they can also degrade electrical properties and reduce the flake active surface area and thus functionality.[68] Degradable and bio-based dispersing agents and binders should be preferred compared to others.[69] An example is the valorization of wool wastes from the agricultural sector to extract keratin that is employed as a dispersant and binder for making environmentally friendly water-based conductive inks based on GNPs.[70]

Hybrid formulations combine 2D materials with other conductive fillers to enhance electrical properties or create synergies between conduction mechanisms. Around 40% of surveyed coatings used a hybrid ink formulation, these included conductive polymers, e.g., PEDOT:PSS, PANI, nanometals, e.g., silver nanowires (AgNWs) and nanoparticles (AgNPs), and other nanocarbons, e.g., carbon black (CB) and carbon nanotubes (CNTs). Very rarely graphene and MXenes were used in the same coating [71-74].

Understanding the biodegradability of the 2D nanomaterials and thus their end-of-life toxicity to humans and wildlife will be a crucial hurdle in enabling their mass adoption.[75] It is known that low concentrations of horseradish peroxidase (HRP) could degrade GO, whereas rGO was not affected.[76] Kurapati et al.[77] exhibited enhanced degradation of GO if functionalized with catechol and coumarin, which naturally bond with HRP. The same authors also displayed that myeloperoxidase, a human enzyme, can degrade GO. The size (lateral dimension and thickness) was critical, with aggregated samples more challenging to be degraded.[78] Microbials can also degrade graphenes, and this does not demand specific conditions such as a control on the temperature and pH.[75] For example, Lalwani et al.[79] documented that GO and rGO nanoribbons were degraded by lignin peroxidase, an enzyme released from fungi. Regarding MXene, the toxicity and effects of MXenes on human beings and the environment have not been thoroughly investigated, and the toxicity mechanism is not clear up till now.[80]

### 2.3 Electrical Properties

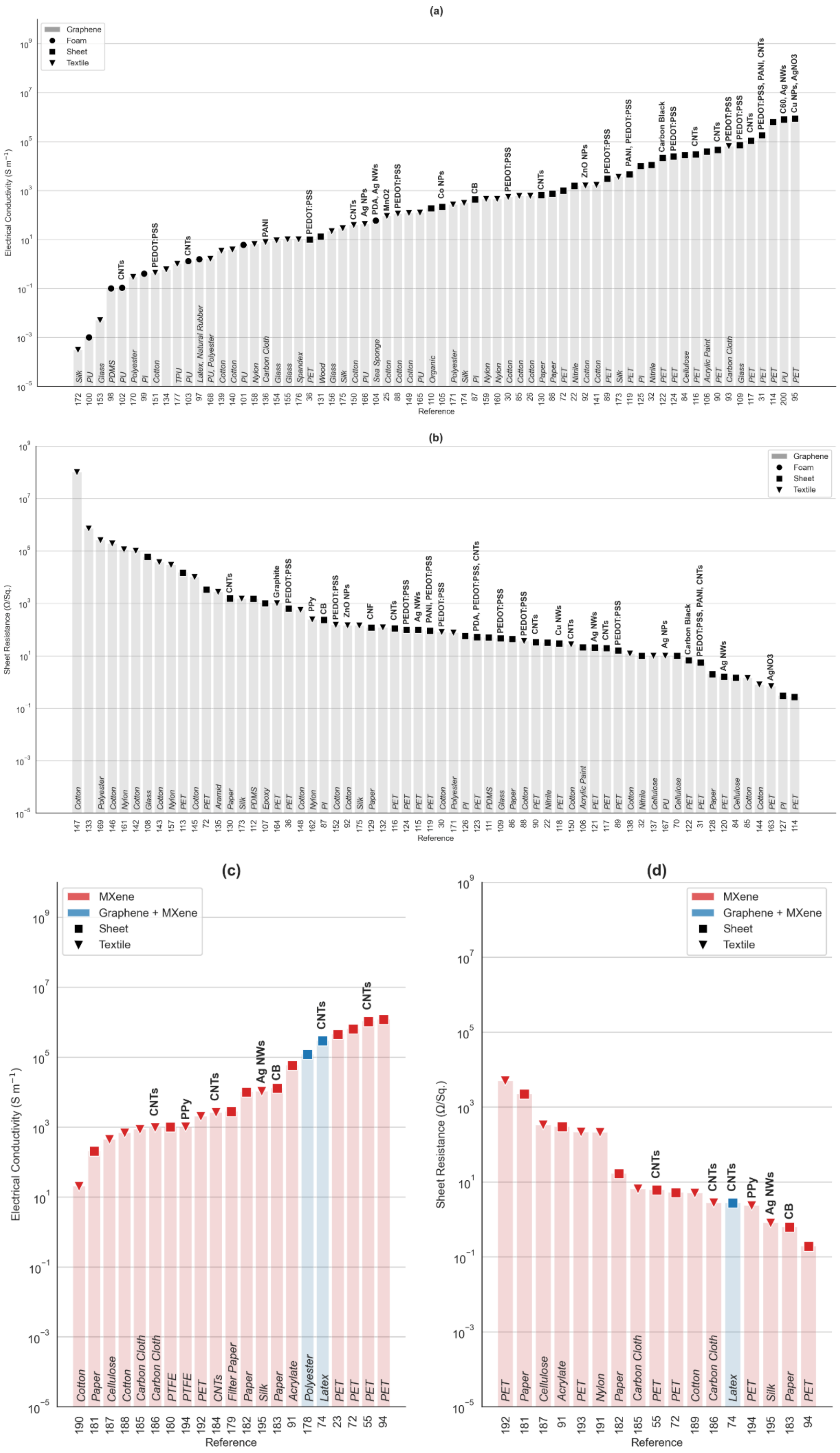

**Figure 2**: Plots of (a,b) electrical conductivity and (c,d) sheet resistance for coatings surveyed in the literature (reference number labelled in the x-axes). The marker shape corresponds to the substrate type used: square (■) for sheets, triangle (▼) for textiles, and circle (●) for foams. The values in bold are conductive materials in addition to graphenes or MXenes used in the coating. The values in italics are the substrate, followed by any polymer binders in brackets.

The electrical properties of thin coatings can be measured accurately using the four-point probe technique, whereby equidistant, linearly separated, electrical connections are made with the material surface. Current ($I$) flows across the outer pair of electrodes, and the voltage drop ($V$) is measured at high impedance (i.e., at $I$ = 0 A) across the inner pair. The determination of a value for electrical conductivity then requires considering the sample physical dimensions relative to the probe set-up. Both the thickness ($t$) of the sample and its length ($l$), must be considered. If the thickness of the sample is a factor of ten lower than the inter-probe spacing ($t/l < 0.1$), for the sake of simplicity the sample can be considered 'infinitesimally thin', and the sheet resistance ($R_s$) can be approximated by using[81]:

$$R_s = \frac{\pi}{\ln 2} \cdot \frac{V}{I} \qquad (1)$$

Any thicker sample ($t/l > 0.1$) requires the bulk conductivity ($\sigma$) to be determined, using the appropriate equations[82] and correction factors, which relate to the sample geometry.[83] The sheet resistance can then be found from the measured bulk conductivity, assuming the thickness is also known:

$$R_s = \frac{1}{\sigma t} \qquad (2)$$

In sheet substrates, the coating is usually deposited on top as a thinner layer and the sheet resistance is typically reported. In textiles, the coating can infiltrate deeper into the network of fibres, so either sheet or bulk values are given. Whereas the 3D nature of foams means the electrical properties can only be reported as bulk conductivity.

The bulk conductivities and sheet resistances of the surveyed articles are reported in **Figure 2** and **Table 5**. Conductive inks formulated with binders have tuneable electrical properties depending on the proportion of conductive fillers used. Some articles report conductivity as a function of filler loading[22, 31, 55, 84-92], and the most conductive formulations are shown for these cases. Sheet based coatings show the highest electrical conductivities, reaching $10^6$ S $m^{-1}$, with the closest textile displaying two orders of magnitude lower values[93] than the best performing sheet[94] in terms of bulk conductivity. This is due to the coating not having to form a percolative conductive network around the substrate, as is the case in textiles or foams. A sheet-based coating can thus also require less or no binder, since the 2D nature of the nanosheets can lead to good adherence onto compatible surfaces.

The most electrically conductive coating surveyed ($1.2 \times 10^6$ S $m^{-1}$) consisted of a 4.3 µm thick $Ti_3C_2T_x$ film spray coated onto a PET sheet.[94] Coatings enhanced with additional conductive fillers tend to reach higher conductivities: generally, graphene based coatings are more often used alongside additional conductive fillers (39%), over MXene based coatings (12%). The most electrically conductive graphene based coating surveyed ($8.7 \times 10^5$ S $m^{-1}$) was enhanced with metal based conductive fillers, copper nanoparticles (Cu NPs) and silver nitrate ($AgNO_3$), blade coated onto a PET sheet and sintered.[95] Similarly, the most electrically conductive based textile coating surveyed ($6.4 \times 10^4$ S $m^{-1}$) used rGO enhanced with PEDOT:PSS.[93] While the most electrically conductive foam (59 S $m^{-1}$) used rGO enhanced with silver nanowires (AgNWs) and polydopamine (PDA).

The tendency of MXenes to oxidise can be problematic not just for dispersion stability, but also conductivity performance within a matrix. Habib et al. (2019)[96] demonstrated that while $Ti_3C_2T_x$ oxidises faster in liquid media, a notable decrease in performance was observed in a PVA matrix at 10 and 50 wt.% loadings with conductivity values dropping to 7 and 40% of the initial conductivity respectively. Unfortunately, most studies do not consider the long-term stability of these MXene nanocomposites. The issue is further complicated as the humidity, UV exposure, voltages used, and frequency of use may all affect rate and extent of oxidation. Future work on conductive MXene coatings should account for long-term electrical stability, enabling a wider understanding of the extent of the problem across different matrix systems and with different flake surface chemistries.

**Table 5**. Electrical conductivity values from surveyed literature

| 2D Material | Substrate | Other Conductive Materials | Sheet Resistance (Ohm/Sq.) | Electrical Conductivity (S m-1) | Ref. |
|---|---|---|---|---|---|
| **Graphene** | Foam (Latex, Natural Rubber) | | | 1.56E+00 | [97] |
| | Foam (PDMS) | | | 1.00E-01 | [98] |
| | Foam (PI) | | | 4.00E-01 | [99] |
| | Foam (PU) | | | 1.00E-03 | [100-101] |
| | Foam (PU) | CNTs | | 1.06E-01 | [102] |
| | Foam (PU) | CNTs | | 1.30E+00 | [103] |
| | Foam (Sea Sponge) | PDA, Ag NWs | | 5.90E+01 | [104] |
| | Sheet () | Co NPs | | 2.17E+02 | [105] |
| | Sheet (Acrylic Paint) | | 2.10E+01 | 3.97E+04 | [106] |
| | Sheet (Cellulose) | | 1.48e+00 to 1.00e+01 | | [70, 84] |
| | Sheet (Epoxy) | | 1.00E+03 | | [107] |
| | Sheet (Glass) | | 6.00E+04 | | [108] |
| | Sheet (Glass) | PEDOT:PSS | 4.63E+01 | 7.20E+04 | [109] |
| | Sheet (Nitrile) | | 1.00e+01 to 3.22e+01 | 1.55E+03 | [22, 32] |
| | Sheet (Organic) | | | 1.86E+02 | [110] |
| | Sheet (PDMS) | | 5.00e+01 to 1.50e+03 | | [111-112] |
| | Sheet (PET) | | 2.67e-01 to 1.50e+04 | | [72, 113-114] |
| | Sheet (PET) | Ag NWs | 1.00E+02 | | [115] |
| | Sheet (PET) | CNTs | 1.98e+01 to 1.11e+02 | 3.00E+04 | [90, 116-117] |
| | Sheet (PET) | Cu NWs | 3.00E+01 | | [118] |

| | | | | |
|---|---|---|---|---|
| Sheet (PET) | PANI, PEDOT:PSS | 9.00E+01 | 4.57E+03 | [119] |
| Sheet (PET) | Ag NWs | 1.60e+00 to 2.06e+01 | | [120-121] |
| Sheet (PET) | Carbon Black | 6.64E+00 | 2.15E+04 | [122] |
| Sheet (PET) | Cu NPs, AgNO3 | | 8.70E+05 | [95] |
| Sheet (PET) | PDA, PEDOT:PSS, CNTs | 5.22E+01 | | [123] |
| Sheet (PET) | PEDOT:PSS | 1.60e+01 to 6.40e+02 | 1.00E+01 | [36, 89, 124] |
| Sheet (PET) | PEDOT:PSS, PANI, CNTs | 5.60E+00 | 1.79E+05 | [31] |
| Sheet (PI) | | 5.76E+01 | | [125-127] |
| Sheet (PI) | CB | 2.35E+02 | 4.37E+02 | [87] |
| Sheet (Paper) | | 2.00e+00 to 4.41e+01 | | [86, 128] |
| Sheet (Paper) | CNF | 1.20E+02 | | [129] |
| Sheet (Paper) | CNTs | 1.54E+03 | 6.50E+02 | [130] |
| Sheet (Wood) | | | 1.30E+01 | [131] |
| Textile () | | 7.10E+05 | | [132-134] |
| Textile (Aramid) | | 2.70E+03 | | [135] |
| Textile (Carbon Cloth) | PANI | | 7.70E+00 | [136] |
| Textile (Carbon Cloth) | PEDOT:PSS | | 6.40E+04 | [93] |
| Textile (Cellulose) | | 1.00E+01 | | [137] |
| Textile (Cotton) | | 1.00E+08 | | [138-140] |
| | | | | [26, 85, 141] |
| | | | | [142-144] |
| | | | | [145-147] |
| | | | | [148-149] |
| Textile (Cotton) | Ni | 8.00E-01 | | [144] |
| Textile (Cotton) | ZnO NPs | 1.44E+02 | 1.58E+03 | [92] |
| Textile (Cotton) | CNTs | 2.70E+01 | 3.86E+01 | [150] |
| Textile (Cotton) | MnO2 | | 9.00E+01 | [25] |
| Textile (Cotton) | PEDOT:PSS | 1.50E+02 | | [30, 88, 151] |
| | | | | [152] |
| Textile (Glass) | | | 5.00E-03 | [153-155] |
| | | | | [156] |
| Textile (Nylon) | | 1.12E+05 | | [157-159] |
| | | | | [160-162] |

| | | | | | |
|---|---|---|---|---|---|
| | Textile (Nylon) | PPy | 2.40E+02 | | [162] |
| | Textile (PET) | AgNO3 | 6.78E-01 | | [163] |
| | Textile (PET) | Graphite | 1.00E+03 | | [164] |
| | Textile (PU) | | | 1.24E+02 | [165] |
| | Textile (PU) | Ag NPs | 1.00E+01 | | [166-167] |
| | Textile (PU, Polyester) | | | 1.60E+00 | [168] |
| | Textile (Polyester) | | 2.53E+05 | | [169-171] |
| | Textile (Silk) | | 1.50E+03 | 3.06E-04 | [172-174] |
| | | | | | [175] |
| | Textile (Spandex) | | | 1.00E+01 | [176] |
| | Textile (TPU) | | | 1.00E+00 | [177] |
| **Graphene + MXene** | Sheet (Polyester) | | | 1.20E+05 | [178] |
| **MXene** | Sheet (Acrylate) | | 2.99E+02 | 5.77E+04 | [91] |
| | Sheet (Filter Paper) | | | 2.76E+03 | [179] |
| | Sheet (PET) | | 5.21E+00 | 4.50E+05 | [23, 72, 94] |
| | Sheet (PET) | CNTs | 6.15E+00 | 1.05E+06 | [55] |
| | Sheet (PTFE) | | | 1.00E+03 | [180] |
| | Sheet (Paper) | | 1.67e+01 to 2.23e+03 | 2.04E+02 | [181-182] |
| | Sheet (Paper) | CB | 6.25E-01 | 1.28E+04 | [183] |
| | Textile (CNTs) | CNTs | | 2.60E+03 | [184] |
| | Textile (Carbon Cloth) | | 6.50E+00 | 8.50E+02 | [185] |
| | Textile (Carbon Cloth) | CNTs | 2.77E+00 | 9.55E+02 | [186] |
| | Textile (Cellulose) | | 3.31E+02 | 4.40E+02 | [187] |
| | Textile (Cotton) | | 5.00E+00 | | [188-190] |
| | Textile (Nylon) | | 2.10E+02 | | [191] |
| | Textile (PET) | | 2.15e+02 to 5.00e+03 | | [192-193] |
| | Textile (PET) | PPy | 2.33E+00 | 1.00E+03 | [194] |
| | Textile (Silk) | Ag NWs | 8.00E-01 | 1.04E+04 | [195] |

# 3 Applications

## 3.1 Sensors

### *3.1.1 Strain Sensors*

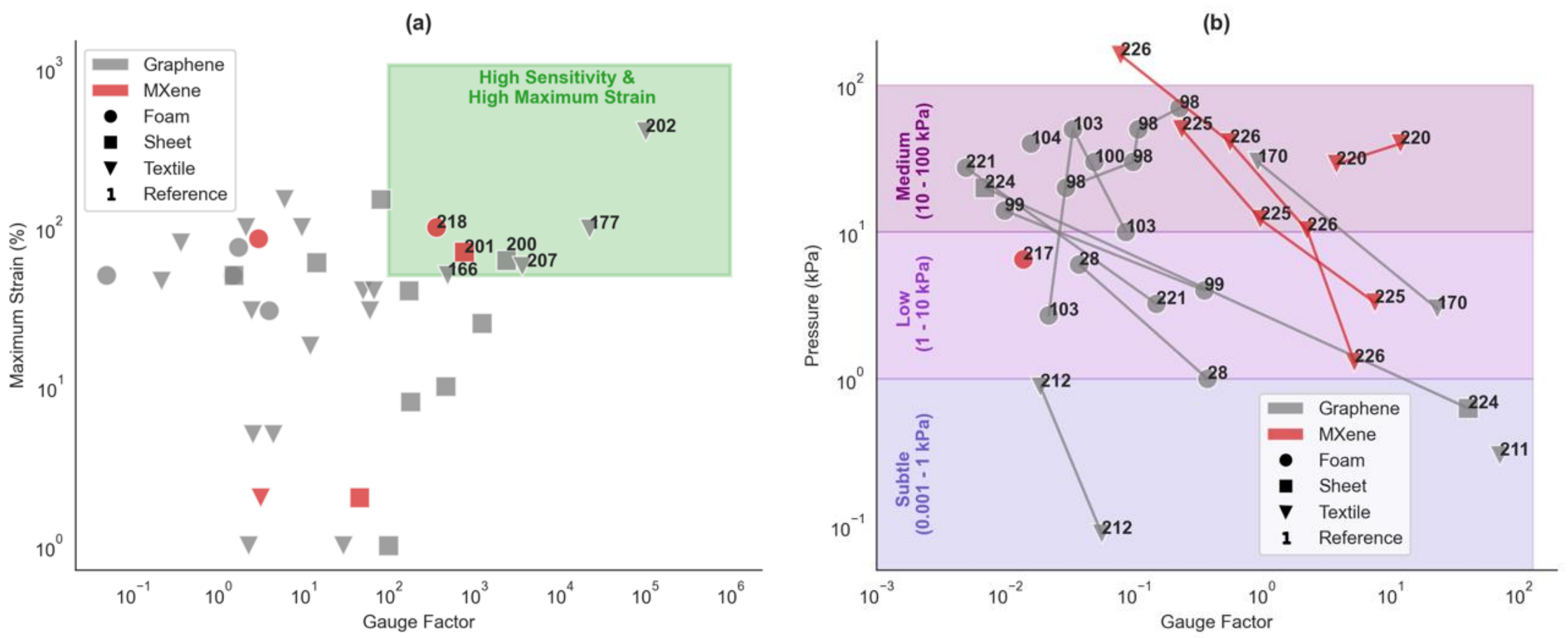

**Figure 3**: (a) Maximum strain vs. gauge factor (GF) for strain gauges surveyed in literature. (b) Pressure vs. sensitivity for pressure sensors surveyed in literature. The marker shapes correspond to the substrate type used: square (■) for sheets, triangle (▼) for textiles, and circle (●) for foams. The highlighted area in (a) marks where the GF > 100 and maximum strain > 50%. . In (b) the lines link values from the same device.

Materials experience strain due to applied stress. Measuring these strains can allow devices to monitor, respond, and react to the external forces that interact with them. Strain gauges are used to detect changes in strain, and since strain often accompanies movement, there are countless applications for these devices. The advent of highly stretchable nanomaterial-based strain sensors has led to exciting developments in wearable sensors for healthcare and electronic skin for soft robotics, detailed in multiple recent reviews. [196-199]

Most commonly, strain gauges measure the change in electrical resistance over a design of conductive tracks induced by a tensile or compressive force. These sensors are called piezoresistive since they use a change of resistance ($R$) to quantify the strain, and they differ from, e.g., capacitive strain sensors, which use a change of electrical capacity as a feedback mechanism to quantify strain. Thus, in piezoresistive strain sensors, the modification in resistance is measured against a known applied strain, and the relationship is modelled using a fitting function. Typically, a linear trend with high linearity is sought after as it allows for facile calculations. The gradient of this linear fit is a measure of the sensor sensitivity, known as the gauge factor (GF), and expressed as:

$$GF = \frac{\Delta R / R_0}{\epsilon} \qquad (3)$$

$\Delta R$ is the modification in electrical resistance after the strain $\epsilon$ was applied, and $R_0$ is the resistance at zero strain.[198] The strain range over which the sensor can operate (i.e., working range) is a measure of its stretchability. A sensor might operate across multiple strain ranges, with a different fit for each one. Ideally, a strain sensor should have a high sensitivity, over an extended linear working range, experience minimal hysteresis, good cyclability, be biocompatible and robust enough to withstand its environment, all while being small and cheap. In practice, achieving all these properties simultaneously presents a significant challenge.[198]

Conventional metal- and semiconductor-based piezoresistive strain sensors are limited by their intrinsically low stretchability ($< 0.6\%$), defined as the maximum strain at which the sensor can reliably and repeatedly measure strain.[198] Coating flexible and stretchable platforms (e.g., elastomers or textiles) with conductive nanomaterials enables strain sensors with large elastic regions achieving stretchability values between 10% to 800%.[198] As with conventional sensors, the working principle behind these conductive coatings is a function of the intrinsic piezoresistive effect and sensor geometry, i.e., the change in resistance due to the strain disrupting the substrate's mechanical integrity, most evident in the case of woven textiles where adjacent fibres might be pulled apart, creating gaps in the electrical network. However, there are also contributions from the conductive filler interactions with itself and surroundings. There are three main mechanisms of note; crack propagation, slippage disconnection, and tunnelling.[198] In the first, the conductive particles are pulled apart, causing cracks to appear and propagate through the network as the applied strain intensifies, increasing the electrical resistance. Once the sensor is relaxed, it pulls the network back together again. In the second, the applied strain causes stacks of overlapping conductive material to move out of contact in a slipping motion, reducing the inter-particle connectivity of the network and increasing resistance. In the third, electrons can travel through a non-conductive matrix by quantum tunnelling between densely packed conductive fillers. An applied strain can change the packing density and thus change the probability of tunnelling. An issue with resistive nanomaterial-based strain sensors is that strain cycling can lead to permanent and significant changes in the network that affect operation (hysteresis) over time. Moreover, it is difficult for these micro-mechanical deformations to affect the sensor area homogeneously and maintain sensor linearity.[197]

Inherent to this micromechanical deformation is the dimensionality of the materials involved. The same application of strain will affect a network of 0D fullerenes differently from 1D nanotubes and 2D nanoplatelets. Indeed, each distinct particle changes shape and re-arrange in the bulk network differently. Conductive 2D materials, and in particular graphenes and MXenes, have gathered particular interest as their large aspect ratio, resulting from their paper-like nanoplatelet shape, is particularly efficient in strain sensing application because of the possibility to benefit on the three mechanism of strain detection, i.e., crack, slippage, and tunnelling.

The performance of strain sensors from 36 different articles has been summarised in **Figure 3**(a). The bulk of these surveyed literature values fall outside the green box that delineates a high sensitivity (GF $>$ 100) and high maximum strain ($>$ 50%). Generally, they either achieve a higher maximum strain at lower sensitivities or a high sensitivity at a lower maximum strain. This behaviour has been reported for thin-film sensors[200] and is a consequence of the three mechanisms discussed previously. When crack propagation dominates, the sensor is susceptible to minimal changes in strain (< 20%).[200] However, the propagation of cracks can quickly disrupt the network to the point where the increase in resistance cannot be managed. When slipping dominates, a considerable strain is required to induce significant enough resistance changes to measure, resulting in low sensitivities (GF < 100). [200] Therefore, achieving high sensitivity in conjunction with high stretchability is a significant challenge.

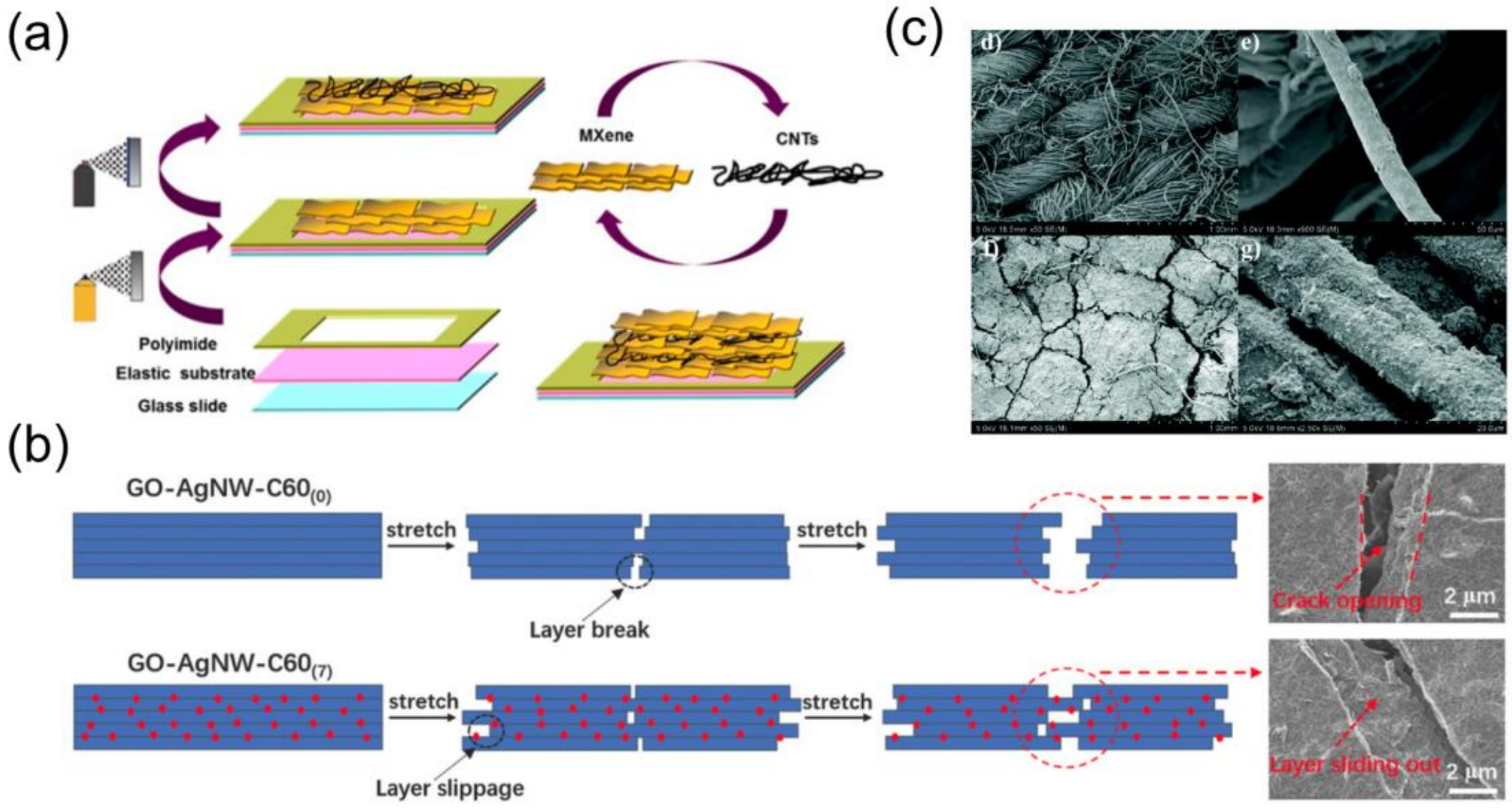

**Figure 4**: (a) Device fabrication process. Adapted with permission.[201] Copyright 2018, American Chemical Society. (b) Sensing mechanism schematics, with accompanying SEM images, showcasing the layer slippage induced due to addition of C60. Adapted with permission.[200] Copyright 2018, Wiley. (c) SEM images of the uncoated fabric and fibres (top), and graphene + CB coated fabrics and fibres (bottom). Adapted with permission.[202] Copyright 2018, Royal Society of Chemistry.

One of the strain sensors surveyed with the most extensive working range is Cai et al.'s [201] MXene + CNT hybrid on latex. In order to retain the advantages of both crack formation and slipping disconnection mechanisms, the authors spray-coated bilayers of $Ti_3C_2T_x$ and CNTs onto a 20 x 5 $mm^2$ latex sheet with a polyimide mask, as shown in **Figure 4**(a). The MXene nanoplatelets are sensitive to crack propagation, ensuring good sensitivity at low strains (GF = 4.35 at 0.1 – 0.6 % strain). While the CNTs act as conductive bridges that block layers of MXene from re-stacking and agglomerating, allowing the sensor to retain both structural and electrical integrity over higher strains (GF = 772.6 at 30 – 70 % strain) where slipping dominates. The authors have demonstrated the sensor range by measuring physiological signals of different magnitudes, monitoring the stretching of skin at the neck to distinguish different spoken words, and flexing the knee joint while exercising.

Shi et al. [200] also leveraged the different nanomaterials dimensionality, going further and incorporating a 0D material, C60 fullerenes, alongside 1D AgNWs and 2D GO. In a one-step screen printing process, all three nanomaterials were combined in a hybrid water-based ink with fluorosurfactant and coated onto a 200 x 3.5 mm polyurethane sheet. The AgNWs form a conductive network supported by the layered GO structure. As expected, this combination allows for high sensitivity to low strains due to crack propagation (GF = 466.2 at 3 – 35 % strain). However, the addition of 0D fullerenes suppresses the density and magnitude of crack propagation by acting as a lubricant to promote interlayer slipping instead, as shown in Figure 4(b). This combination of mechanisms increases the working strain range of the sensor while maintaining the high sensitivity (GF = 2392.9 at 52 – 62 % strain).

The most sensitive strain sensor at higher loadings surveyed is Souri et al.'s [202] graphene + CB hybrid on cotton fabric. The sensor was manufactured out of a 100 x 7 $mm^2$ dog bone-shaped cotton fabric cut-out, with a notch at the centre, dip-coated in graphene + CB water-based ink, with sodium dodecyl benzene sulfonate (SDBS) surfactant. The coated cotton was subsequently encapsulated in ecoflex silicone polymer. Before undergoing characterisation, the sensor underwent rupture training whereby the cotton was cracked and even completely fractured at the notch following 5 cycles at 400% strain. Consequently, these rupture trained sensors operate via a substrate-driven mechanism, whereby breaking these weakened yarns at high strain levels disrupts the electrical network and increases resistance, enabling high sensitivity at high strain (GF = 102351 at 342 – 400 % strain). This mechanism also yields good sensitivities at low strain (GF = 11.82 at 0.5 – 1 % strain) due to the interaction of the fractured fibres, arranged in a tentacle-like manner at the notch, that are distanced from each other with strain.

**Table 6**: Performance of Surveyed Strain Sensors.

| 2D Materials | Substrate | Other Conductive Materials | Maximum Strain (%) | Gauge Factor | Ref. |
|---|---|---|---|---|---|
| **Graphene** | Foam (PU) | | 3.00E-01 | 4.00E+00 | [100] |
| | Foam (PU) | CNTs | 7.50e-01, 5.00e-01, 2.50e-01 | 1.75e+00, 8.60e-01, 9.60e-01 | [102] |
| | Foam (PU) | CNTs | 1.00e+00, 8.60e-01, 5.00e-01 | -2.30e+00, -2.13e+00, 5.00e-02 | [103] |
| | Foam (Sea Sponge) | PDA, Ag NWs | 5.00E-01 | 1.50E+00 | [104] |
| | Sheet (Ecoflex) | | 1.50e+00, 1.00e+00 | 8.16e+01, 2.97e+01 | [203] |
| | Sheet (Epoxy) | | 1.00E-02 | 1.00E+02 | [107] |
| | Sheet (PDMS) | | 5.00E-01 | 1.55E+00 | [204] |
| | Sheet (PDMS) | | 1.00E-01 | 4.66E+02 | [205] |
| | Sheet (PDMS) | | 4.00e-01, 3.00e-01, 2.00e-01, 1.00e-01 | 1.73e+02, 3.70e+01, 1.10e+01, 7.00e+00 | [112] |
| | Sheet (PI) | | 2.50e-01, 1.70e-01 | 1.24e+03, 2.36e+02 | [126] |
| | Sheet (PU) | CNTs | 8.00E-02 | 1.81E+02 | [206] |
| | Sheet (PU) | C60, Ag NWs | 6.20e-01, 5.20e-01, 3.50e-01, 3.00e-02 | 2.39e+03, 1.00e+03, 4.66e+02, 2.50e+01 | [200] |
| | Sheet (TPU) | | 6.00E-01 | 1.44E+01 | [29] |
| | Textile (Cotton) | | 5.70e-01, 4.00e-01 | 3.67e+03, 4.16e+02 | [207] |
| MXene | Textile (Cotton) | | 3.00E-01 | 2.49E+00 | [140] |
| | Textile (Cotton) | CB | 1.50e+00, 1.20e+00 | 6.05e+00, 1.67e+00 | [208] |
| | Textile (Cotton) | CB | 1.00e-02, 2.00e-02, 2.31e+00, 3.42e+00, 4.00e+00 | 1.18e+01, 1.73e+01, 9.56e+01, 2.18e+03, 1.02e+05 | [202] |
| | Textile (Glass) | | 0.00e+00, 1.00e-02 | 1.13e+02, 2.95e+01 | [156] |
| | Textile (Hair) | | 5.00E-02 | 4.46E+00 | [209] |
| | Textile (Lycra) | PANI | 4.00E-01 | 6.73E+01 | [210] |
| | Textile (Nylon) | | 1.00e-01, 1.80e-01 | 1.85e+01, 1.21e+01 | [161] |
| | Textile (Nylon) | | 4.60e-01, 3.30e-01, 1.20e-01 | 2.20e-01, 7.00e-02, 8.00e-02 | [158] |
| | Textile (PAI) | | 1.00e-01, 8.00e-01 | 1.63e+00, 3.70e-01 | [211] |
| | Textile (PET) | | 1.00E-02 | -7.10E+00 | [212] |
| | Textile (PU) | Ag NPs | 5.00E-01 | 4.90E+02 | [166] |
| | Textile (PU) | CNTs, CB | 1.00E+00 | 2.14E+00 | [213] |
| | Textile (PU, Polyester) | | 1.00e+00, 1.00e-01, 5.00e-02 | 9.65e+00, 2.50e+00, 2.60e+00 | [168] |
| | Textile (PVA) | PDA | 1.00E-02 | 2.30E+00 | [214] |
| | Textile (Polyester) | | 1.00e-02, 5.00e-02 | 6.42e+00, 2.58e+00 | [215] |
| | Textile (Spandex) | | 4.00E-01 | 5.01E+01 | [176] |
| | Textile (Spandex) | PANI | 3.00e-01, 1.50e-01 | 6.03e+01, 6.80e+00 | [216] |
| | Textile (TPU) | | 9.80e-01, 1.50e-01 | 2.25e+04, 1.80e+02 | [177] |
| | Foam (PU) | | 8.50e-01, 4.50e-01, 3.10e-01 | 3.00e+00, 2.30e-01, 1.70e-01 | [217] |
| | Foam (Salt) | CNTs | 1.00E+00 | 3.63E+02 | [218] |
| | Sheet (Latex) | CNTs | 7.00e-01, 3.00e-01, 1.00e-02 | 7.73e+02, 6.46e+01, 4.35e+00 | [201] |
| | Sheet (PI) | | 1.00e-02, 2.00e-02 | 9.48e+01, 4.59e+01 | [219] |
| | Textile (Cellulose) | | -2.00E-01 | 6.20E+00 | [187] |
| | Textile (Cotton) | | 2.00e-02, 1.00e-02 | 3.18e+00, 1.16e+00 | [189] |

### *3.1.2 Pressure Sensors*

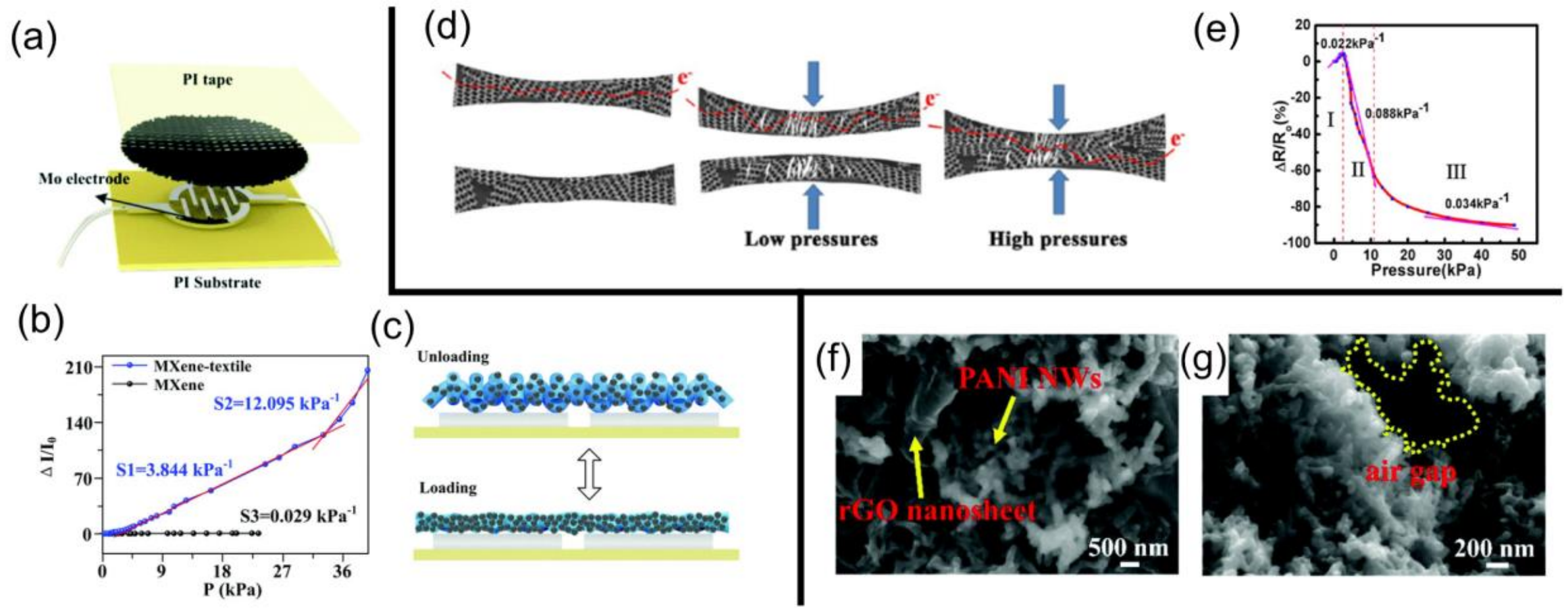

**Figure 5**: (a) Schematic of MXene textile device. (b) Change in relative current against applied pressure for the device. (c) Operating principle of device under pressure. Adapted with permission.[220] Copyright 2019, Royal Society of Chemistry. (d) Deformation schematic of the rGO + MWNT coated sponge, under no, low and high pressure. The red lines show the conductive network, the white lines indicate crack formation in the rGO. (e) Change in relative resistance against applied pressure for the device. Adapted with permission.[103] Copyright 2018, American Chemical Society. (f,g) SEM images of the rGO + PANI NW coated sponge. Adapted with permission.[221] Copyright 2018, Royal Society of Chemistry.

Pressure sensing enables the measurement of interfacial pressure between two surfaces. The growing interest in robotics and wearable healthcare increased demand for low-power, lightweight and flexible pressure sensors.[222] These applications demand sensitivity within different pressure ranges, categorised by Zang et al. [223] into the following regimes: Ultra-low, below 0.001 kPa, e.g., sound pressure, for applications in hearing aids; subtle, from 0.001 to 1 kPa; low, from 1 kPa to 10 kPa, e.g., light to gentle touch pressures, for applications in e-skin (electronic skin); medium, from 10 kPa to 100 kPa, e.g., blood pulse and body-weight pressures, for heart rate monitoring and body pressure mapping applications. Several technologies are available to sense pressure, such as capacitive, electromagnetic, and piezoelectric, but piezoresistive is the most diffuse.

Piezoresistive pressure sensing works analogously to strain sensing, the key difference being that a compressive force ($F$) is applied perpendicular to the sensor surface area ($A$) and reported as a unit of pressure ($P$), where $P = F/A$. Consequently, rather than the dimensionless GF, the sensitivity is given in units of $\mathrm{kPa}^{-1}$ as strain is displaced by pressure in Equation 3. Note that the same mechanism can be reported under the category of strain sensing by recording a negative strain.[103] The chosen approach generally depends on the application being targeted and the measurement tools available. As with strain sensing, the use of conductive nanomaterial coatings on flexible and stretchable platforms enables high sensitivities through micro-mechanical changes in the filler network. Although typically applying compressive stress causes the resistance to decrease as inter-filler distance also decreases, rather than vice-versa.

The performances of 17 pressure sensors from surveyed literature are shown in Figure 3(b). Most papers reported limits of detection as high as the medium regime and sensitivities below $0.5\ \mathrm{kPa}^{-1}$. Generally, their working range is split over at least two linear regions, with many devices operating across both medium and low regimes. Most of the pressure sensing devices surveyed are based on foams, and this is due to their ability to modulate the operating range of the device through their stiffness, recover their initial shape once the load is released, and readily adsorb the conductive coating into their porous network. The latter is particularly important as it increases the surface area over which the filler network can form and break connections, increasing sensitivity. The higher surface area of 2D materials over 0D or 1D materials, further promotes this effect as it enables better adhesion to the pores' surface.[220]

Li et al.'s [220] was the best performing MXene device among the pressure sensors surveyed. The piezoresistive sensor was manufactured by dip-coating cotton fabric in $Ti_3C_2T_x$ solution and subsequently placing it over an interdigitated molybdenum electrode, on top of a polyimide (PI) substrate, secured with PI tape as shown in **Figure 5**(a). An array design is also presented, combining 16 pixels for pressure mapping. The sensors had a working range up to 40 kPa, split into two regions with high sensitivities: 3.84 kPa$^{-1}$ and 12.10 kPa$^{-1}$, per region, respectively, shown in Figure 5(b). The first of these is attributed to a reduction in contact resistance between the coated cotton and electrode ($R_{contact}$), and the second builds on this with a reduction in inter-filler distance between $Ti_3C_2T_x$ flakes in the cotton fabric ($R_{film}$), summarised as: $R_{total} = R_{contact} + R_{film}$.

Tewari et al. [103] combined different dimensionality nanomaterials, rGO and multi-walled CNTs (MWNTs), into a low-cost, polyurethane (PU) sponge through dip-coating a hybrid aqueous (10:1 rGO to MWNTs by volume) dispersion, connected via indium tin oxide coated PET electrodes attached to either end. Silver paste was applied to the connections in order to eliminate $R_{contact}$. MWNTs were used to increase the device conductivity (from $2 \times 10^{-3}$ to 1.3 S m$^{-1}$). The sensor response stretches the medium and low regimes, and is broken down into three sections, shown in Figure 5(e). Increasing pressure to 2.7 kPa caused resistance to increase slightly, and this is attributed to the expansion of microcracks in the rGO flakes breaking conductive pathways. Beyond 2.7 kPa, the effect was the opposite: as the coated PU foam skeleton is compressed tightly against itself, the resistance drops significantly. The proposed mechanism is shown in Figure 5(d).

Ge et al. [221] also leverage different dimensionality nanomaterials, in this case growing PANI nanowires in-situ on an rGO coated melamine sponge, wired up using copper tape and silver paste. This manufacturing technique resulted in a rough coating morphology, with porous cavities among the forests of nanowires observed under SEM, shown in Figure 5(f, g), further increasing the accessible contact area. The compression of these small airgaps was attributed to the second pressure regime observed, with its broad working range (from 12.32 kPa to 27.39 kPa) and low sensitivity (0.0049 $kPa^{-1}$). Furthermore, mechanical analysis was carried out to demonstrate that coating the sponges lowers their elastic modulus and increases their compressibility.

As with the surveyed strain sensors, these devices were used to demonstrate applications in wearable human-machine interfaces, recording finger movements and speech. **Table 7** summarises all the pressure sensing applications surveyed, displaying the variety of substrates and conductive material combinations used. Textile substrates were most used for strain-sensors, as their component fibres provide a network for deformation, enabling higher sensitivities. Foam substrates were most used for pressure-sensors, as they can modulate the applied force through their porous structure. Dip-coating with a water-based dispersion was most common across both sensor types, as textiles and foams readily adsorb graphenes and Mxenes into their structures, making it a facile approach to large-area coating.

While there is unlikely to be a one-size-fits-all solution to flexible strain and pressure sensing, the use of hybrid conductive fillers, combining different dimensionality nanomaterials, is unlocking increased sensitivities over more extensive working ranges, expanding the possible potential applications. The ability to tune conductive filler loadings and binders, enables devices to be optimised for certain regimes and applications. A key challenge will be in developing a better understanding of how these micro-mechanical electrical networks make and break conductive pathways both under extension and compression, when hybrid dimensionality fillers are used. Industrial uptake will require reliability and reproducibility; as such, more focus should be given to cyclic stretch-release tests, establishing the effects of hysteresis and changes to sensor response and recovery through repeated wear. The use of appropriate encapsulation materials will enable better mechanical durability, washability, and UV and humidity resistance.

**Table 7**: Performance of Surveyed Pressure Sensors

| 2D Materials | Substrate | Other Conductive Materials | Maximum Pressure (kPa) | Maximum Gauge Factor | Ref. |
|---|---|---|---|---|---|
| **Graphene** | Foam (Melamine) | PANI NWs | 3.24e+00, 2.74e+01 | 1.52e-01, 5.00e-03 | [221] |
| | Foam (PDMS) | | 7.00e+01, 5.00e+01, 3.00e+01, 2.00e+01 | 2.30e-01, 1.10e-01, 1.00e-01, 3.00e-02 | [98] |
| | Foam (PI) | | 4.00e+00, 1.40e+01 | 3.60e-01, 1.00e-02 | [99] |
| | Foam (PU) | | 3.00E+01 | 5.00E-02 | [100] |
| | Foam (PU) | | 1.00E+00 | 3.80E-01 | [28] |
| | Foam (PU) | CB | 6.00E+00 | 3.80E-02 | [28] |
| | Foam (PU) | CNTs | 5.00e+01, 1.00e+01, 2.70e+00 | 3.40e-02, 8.80e-02, 2.20e-02 | [103] |
| | Foam (Sea Sponge) | PDA, Ag NWs | 4.00E+01 | 1.60E-02 | [104] |
| | Sheet (PEN) | | 6.30e-01, 2.00e+01 | 4.08e+01, 7.00e-03 | [224] |
| | Textile (PAI) | | 3.00E-01 | 7.20E+01 | [211] |
| | Textile (PET) | | 8.80e-01, 8.80e-02 | 1.90e-02, 5.70e-02 | [212] |
| | Textile (Polyester) | | 3.00e+01, 3.00e+00 | 9.37e-01, 2.34e+01 | [170] |
| **MXene** | Foam (PU) | | 6.50e+00, 8.51e+01, 2.46e+02 | 1.40e-02, -1.50e-02, -1.00e-03 | [217] |
| | Textile (Airlaid Paper) | | 3.30e+00, 1.22e+01, 5.00e+01 | 7.65e+00, 9.80e-01, 2.40e-01 | [225] |
| | Textile (Cotton) | | 4.00e+01, 2.90e+01 | 1.21e+01, 3.84e+00 | [220] |
| | Textile (Cotton) | | 1.60e+02, 4.07e+01, 1.02e+01, 1.30e+00 | 8.00e-02, 5.70e-01, 2.27e+00, 5.30e+00 | [226] |
| | Textile (Nylon) | | 1.40e+02, 1.10e+02, 4.50e+01 | 2.88e+03, 3.63e+04, 6.41e+03 | [191] |

## 3.2 Energy Generation & Storage

### *3.2.1 Supercapacitors*

Supercapacitors (SCs) or *electrochemical capacitors* are an important energy storage component capable of storing 10,000 times the energy of regular capacitors, while still operating at higher power densities than batteries. Typically, SCs are made up of two electrodes one charged negative and the other one positive. There are two principal SC operating principles: nonfaradaic electrochemical double layer capacitance (EDLC) and faradaic in origin pseudocapacitance. In nonfaradaic (capacitive), processes charge is progressively stored at the interphase, with no electron exchange between bulk phases.[227] Whereas, in faradaic processes, a continuous current will flow, as long as a supply of ions at the interphase exchange electrons with the conducting bulk phase.

In EDLC, electrolyte ions diffuse towards and adsorb onto the surface of the oppositely charged electrode via electrostatic attraction. This creates an electric double layer at the interphase, consisting of electrolyte anions and cations, and an excess or deficit of conduction band electrons on the electrode surface.[228] The speed at which charge is released by the SC when discharged (i.e. the power), is limited by the movement of these ions between the electrode surfaces. This process is both faster than the diffusion limited faradaic redox reactions that power traditional lithium-ion batteries, and takes place in a lighter package, enabling higher power densities in SCs.[229] The electrodes are generally made up of a highly conductive, porous and high surface area material, as the charge stored is proportional to the specific surface area, i.e. to the capacitance.[230] For such reason, extensive use has been made of high surface area nanocarbons, including activated carbon (AC), CNTs and CB. Despite this, their energy density is typically lower than lithium-ion batteries, due to the limited accessible surface space relative to intercalated ions. The high surface area and electrical conductivity of graphene, alongside its many favourable structural thin-film properties, e.g., mechanical strength, flexibility and transparency, make them desirable EDLC active materials.[231] The theoretical gravimetric capacitance of a fully utilised graphene layer is 550 F $g^{-1}$.[232] The main processing challenge involves keeping its high surface area accessible, as it is prone to stack and re-agglomerate.

In pseudocapacitors, electrosorbed or intercalated ions undergo reversible chemical redox reactions with the electrode interphase surface. Despite the faradic nature of this process, it is termed a *pseudo* capacitor and distinguished from battery charging or discharging, due to the finite active surface, which leads to complex electrochemical behaviour with capacitive properties.[228, 233] This is observed experimentally with quasi-rectangular cyclic voltammetry (CV) curves and linear galvanostatic charge-discharge plots.[56] There is ongoing debate as to the appropriate scope of pseudocapacitance.[234] Pseudocapacitive materials include transition metal oxides, e.g., ruthenium oxide ($RuO_2$) and manganese oxide ($MnO_2$), as well as conductive polymers, e.g., PANI, poly(3,4-ethylenedioxythiophene) (PEDOT) and polypyrrole (PPy). More recently, the transition metal oxide surface of MXenes was found to be redox active, unlocking pseudocapacitive behaviour in acidic electrolytes, e.g., sulphuric acid ($H_2SO_4$).[230, 235-236] This mechanism is enhanced through the intercalation of the electrolyte between MXene layers, providing a source of protons to the redox active surface, and enabling for example $Ti_3C_2T_x$ to reach a theoretical gravimetric pseudocapacitance of 1116 F $g^{-1}$.[237] Achieving this capacitance in practice is difficult due the low oxidation potential of MXenes, which restricts their operating voltage range, and thus the energy density of the SC. [230, 233]

Generally, EDLCs have higher power densities but lower energy densities than pseudocapacitors since pseudocapacitors are similar to batteries enabled by kinetically slow redox reactions.[238] In practice these two operating principles occur simultaneously to different extents[228], e.g., functional moieties on the surface of graphene can undergo redox reactions, and $Ti_3C_2T_x$ sheets can serve as an EDLC surface. Engineering hybrid systems that combine both mechanisms can result in significantly improved SC performance. The chemical compatibility, mechanical stability, high capacitance, surface area and conductivity of 2D materials[239] such as graphenes and MXenes make them ideal electrode materials for SC applications, in particular when incorporating them onto flexible platforms.

There are various ways of characterising these devices: the principal tests include cyclic voltammetry (CV) and galvanostatic charge discharge (GCD). The shape of both plots can be used to identify the charge storage mechanisms used by the device. EDLCs show a linear voltage-time response and rectangular CV profile. The redox reactions occurring in pseudocapacitors are several, resulting in a quasi-rectangular CV profile which is the interpolation of the single peak pairs. On the other hand, the CV profile in batteries shows a single pair of peaks. [240] Energy storage devices can also be characterised by the power and energy they store per unit size, visualised on a Ragone plane. Batteries have high specific energy densities and low specific power densities, capacitors the inverse. The characteristic time, i.e. energy-power ratio, is representative of the charge / discharge period: high for batteries (minutes – hours) and low for capacitors (milliseconds – seconds).[241]

Most devices within the surveyed literature fall under the sub-category of *micro*-supercapacitors (MSCs) since the application of flexible conductive coatings is well suited to low-power electronics, as such they will be the focus of the subsequent analysis. However, there were also examples of MSCs with solution-based electrolytes[127, 149-150, 162, 242-246], parallel plate capacitors[22, 70], and fibre-supercapacitors[184, 187, 245, 247-251]. The use of small mass loadings (< 1 mg $cm^{-2}$) and thin electrodes (< 100 µm) in MSCs, necessitates the capacitance value to be normalised over its dimensions, e.g., area ($C_A$, mF $cm^{-2}$) or volume ($C_V$, F $cm^{-3}$) if the thickness can be accurately measured.[240]

The areal capacitance values for the surveyed devices are summarised in **Figure 6** and **Table 8**. The test conditions have been quoted where possible to discern between the use of CV or GCD methods. Most surveyed articles use either MXene or graphene-based electrodes, around a quarter of surveyed devices used asymmetric configurations, and the setup was evenly divided between cofacial and coplanar configurations (in the former electrodes are vertically stacked and in the latter electrodes are parallel). Volumetric capacitance results are also summarised in **Table 9**.

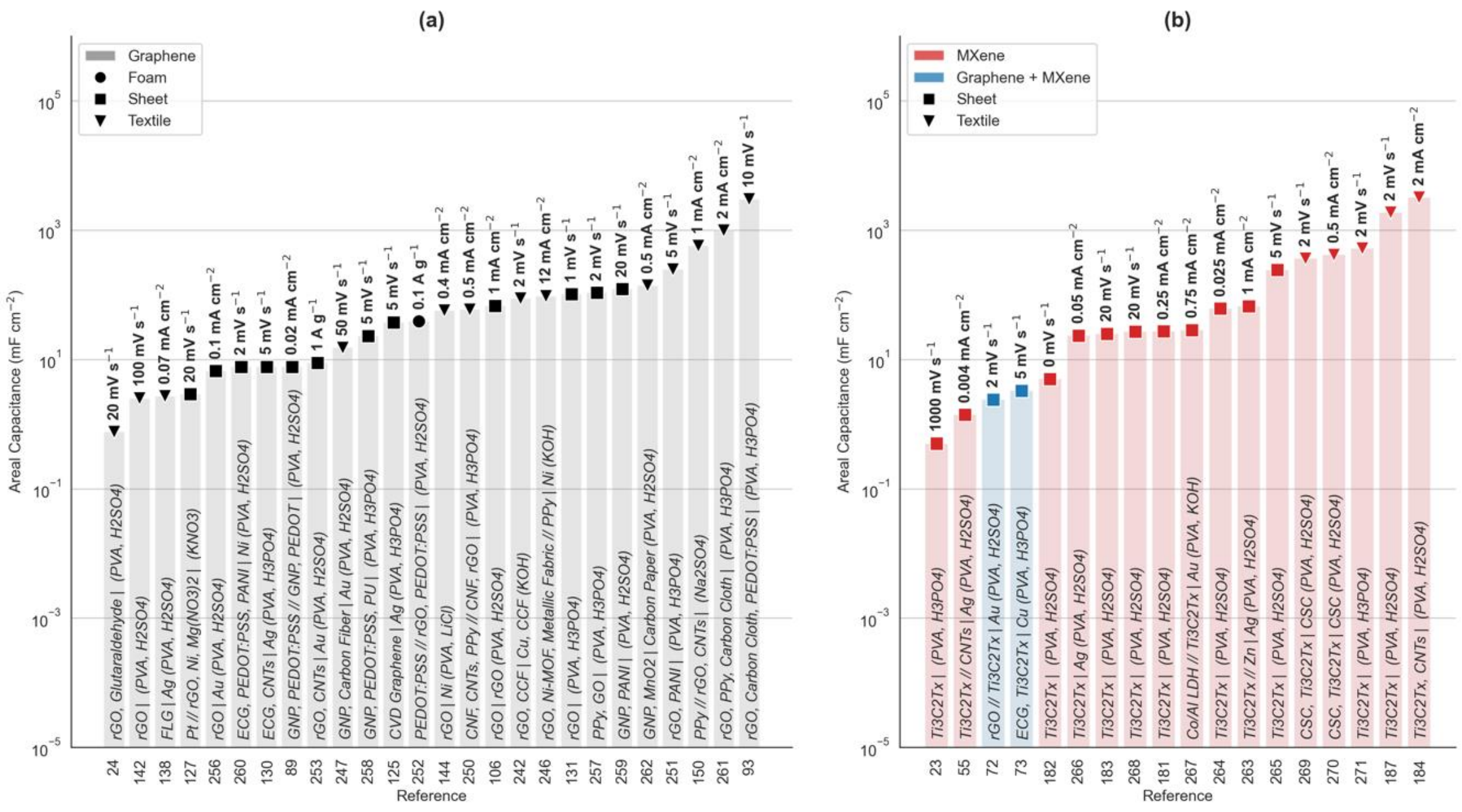

**Figure 6**: Areal capacitance values for surveyed supercapacitors. The marker shape corresponds to the substrate type used; square (▫) for sheets, triangle (▿) for textiles, and circle (∘) for foams. The bar annotation details the device set-up, beginning with the conductive electrode materials, asymmetric electrodes are separated by '//'. After the '|' any current collectors are detailed, as well as the electrolyte composition in brackets. The test conditions are quoted above the bars.

**Table 8**: Areal Capacitance ($C_A$) Values for Surveyed Articles

| 2D Material | Substrate | Other Conductive Materials | CA mF cm-2 | Ref. |
|---|---|---|---|---|
| **Graphene** | Foam (Ecoflex) | | 3.87E+01 | [252] |
| | Sheet (Acrylic Paint) | | 6.72E+01 | [106] |
| | Sheet (PET) | CNTs | 8.80E+00 | [253] |
| | Sheet (PET) | Carbon Aerogel, MnO2 | 8.70E+00 | [254] |
| | Sheet (PET) | PEDOT:PSS | 7.70E+00 | [89] |
| | Sheet (PI) | | 1.50e+00 to 3.75e+01 | [127, 255-256] [125] |
| | Sheet (PP) | | 1.08E+02 | [257] |
| | Sheet (PU) | PEDOT:PSS | 2.30E+01 | [258] |
| | Sheet (Paper) | PANI | 1.23E+02 | [259] |
| | Sheet (Paper) | CNTs | 7.70E+00 | [130] |
| | Sheet (Paper) | MnO2, PEDOT:PSS | 3.60E+00 | [260] |
| | Sheet (Paper) | PANI, PEDOT:PSS | 7.63E+00 | [260] |
| | Sheet (Paper) | PEDOT:PSS | 4.89E+00 | [260] |
| | Sheet (Wood) | | 1.02E+02 | [131] |
| | Textile (CNF) | | 5.88E+01 | [250] |
| | Textile (Carbon Cloth) | | 1.51E+01 | [247] |
| | Textile (Carbon Cloth) | PPy | 9.85E+02 | [261] |
| | Textile (Carbon Cloth) | PEDOT:PSS | 3.00E+03 | [93] |
| | Textile (Cellulose) | MnO2 | 1.39E+02 | [262] |
| | Textile (Cotton) | | 2.50e+00 to 8.75e+01 | [138, 142, 242] |
| | Textile (Cotton) | Ni | 5.70E+01 | [144] |
| | Textile (Cotton) | PANI | 2.46E+02 | [251] |
| | Textile (Cotton) | CNTs | 5.70E+02 | [150] |
| | Textile (PET) | | 7.56E-01 | [24] |
| | Textile (Polyester) | Ni-MOF | 9.50E+01 | [246] |
| **Graphene + MXene** | Sheet (PET) | | 2.40e+00 to 3.26e+00 | [72-73] |
| **MXene** | Sheet (Ni) | | 6.65E+01 | [263] |
| | Sheet (Ni) | CB | 5.20E+01 | [263] |
| | Sheet (PET) | | 5.00e-01 to 2.41e+02 | [23, 264] [265] |
| | Sheet (PET) | CNTs | 1.40E+00 | [55] |
| | Sheet (Paper) | | 5.00e+00 to 2.73e+01 | [181-182, 266] |
| | Sheet (Paper) | CB | 2.50e+01 to 2.85e+01 | [183, 267] |
| | Sheet (Scotch Tape) | | 2.70E+01 | [268] |
| | Textile (CNTs) | CNTs | 5.54e+02 to 3.19e+03 | [184] |
| | Textile (Carbon Cloth) | | 3.62e+02 to 4.13e+02 | [269-270] |
| | Textile (Cellulose) | | 1.87E+03 | [187] |
| | Textile (Cotton) | | 5.19E+02 | [271] |

**Table 9**: Volumetric Capacitance Values (Cv) for Surveyed Articles.

| 2D Material | Substrate | Other Conductive Materials | CV F cm-3 | Ref. |
|---|---|---|---|---|
| **Graphene** | Sheet (PET) | PANI, PEDOT:PSS | 6.10E+01 | [119] |
| | Sheet (PET) | Carbon Aerogel, MnO2 | 4.37E+01 | [254] |
| | Sheet (Paper) | CNTs | 7.73E+01 | [130] |
| | Sheet (Paper) | MnO2, PEDOT:PSS | 1.46E+01 | [260] |
| | Sheet (Paper) | PANI, PEDOT:PSS | 3.68E+01 | [260] |
| | Sheet (Paper) | PEDOT:PSS | 1.55E+01 | [260] |
| | Textile (CNF) | | 1.28E+01 | [250] |
| | Textile (CNTs) | | 2.63E+02 | [249] |
| | Textile (Cotton) | | 1.29e+00 to 5.53e+00 | [242] |
| | Textile (Cotton) | Ni | 6.82E+01 | [248] |
| **Graphene + MXene** | Sheet (PET) | | 3.30e+01 to 8.00e+01 | [72-73] |
| | Sheet (Polyester) | | 6.98E+02 | |
| **MXene** | Sheet (PS) | | 9.50E+01 | [272] |
| | Sheet (Scotch Tape) | | 3.57E+02 | [268] |
| | Textile (CNTs) | CNTs | 2.03e+02 to 1.08e+03 | [184] |
| | Textile (Cellulose) | | 1.42E+03 | [187] |

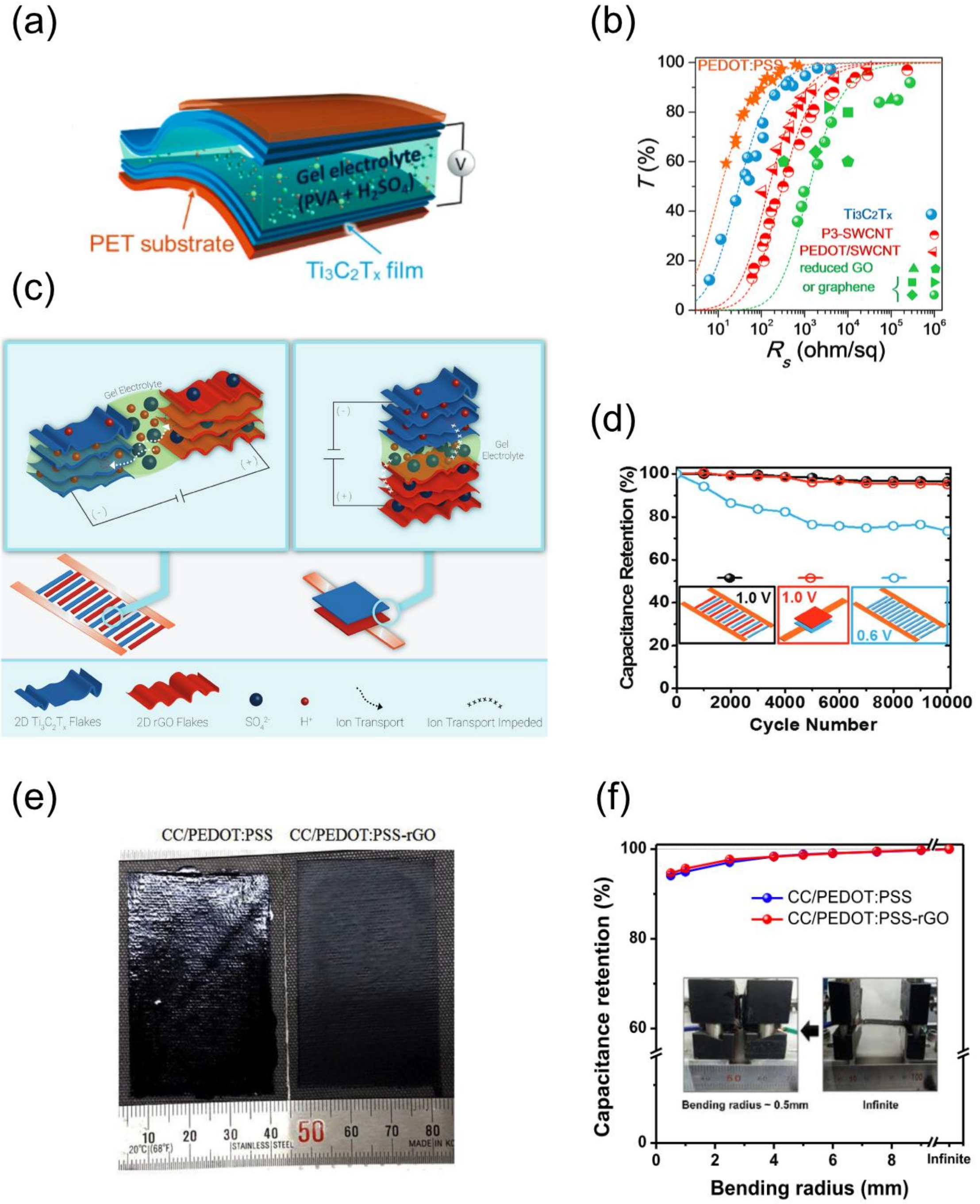

**Figure 7**: (a) Symmetric Flexible SC, (b) transmittance vs. sheet resistance for thin films. Adapted with permission.[55] Copyright 2017, Wiley. (c) Cofacial and coplanar asymmetric MSCs, (d) capacitance retention post cycling for asymmetric coplanar (black), asymmetric cofacial (red) and symmetric planar (blue) MSCs. Adapted with permission.[72] Copyright 2017, Wiley. (e) Carbon cloth coated SC electrodes, (f) capacitance retention with bending of SC. Adapted with permission.[93] Copyright 2018, Elsevier Ltd.

Zhang et al. [55] leveraged the thin film properties of $Ti_3C_2T_x$ to produce both symmetric and asymmetric (CNT anode), transparent and flexible layered SCs on PET substrates, as seen in **Figure 7**(a). To optimise transparency and SC performance, $Ti_3C_2T_x$ coatings of increasing thicknesses were evaluated for sheet resistance, light transmittance, and intrinsic capacitance. The findings, shown in Figure 7(b), demonstrated that the $Ti_3C_2T_x$ thin films were able to avoid the percolative effects seen at high transmittance (> 90%) in graphene, i.e., continue operating in a bulk-like conductivity regime. In practice, a coating was found to balance a high 91% transmittance with high areal capacitance 0.48 mF $cm^{-2}$.

Couley et al. [72] manufactured binder and current collector free asymmetric MSCs with a $Ti_3C_2T_x$ cathode and rGO anode in both coplanar and co-facial configurations, by spray coating PET sheets using masks for patterning as seen in Figure 7(c). They found increased power and energy densities over a symmetric $Ti_3C_2T_x$ configuration, and this performance gain is unlocked through the extended voltage window available to the rGO. Cycling tests, shown in Figure 7(d), evidenced improved electrochemical stability using the rGO anode, as the symmetric $Ti_3C_2T_x$ device saw capacitance drop to 75% after 10,000 cycles as opposed to 97% for the asymmetric set-ups. The coplanar configuration was better able to sustain performance at higher scan rates, and this is attributed to the architecture enabling in-plane diffusion of ions.

Kumar et al. [93] developed large area (up to 25 $cm^2$) cofacial textile-based supercapacitors by spray coating conductive carbon cloth with aqueous PEDOT:PSS and rGO dispersions, as shown in Figure 7(e) and subsequently treating with phosphoric acid. This latter step unlocked a larger operating window (0 to 2 V), which in turn enabled higher power and energy densities.

The addition of 20 wt.% rGO increased the areal capacitance from 1600 to 3000 mF $cm^{-2}$, as well as cycled capacitance retention from 93% to 100% after 2000 cycles. This device retained 94% capacitance at a bending radius of 0.5 mm, as shown in Figure 7(f), thus retaining some of the textile's flexibility in practice.

Flexible supercapacitors are an essential energy storage component for enabling novel wearable electronics applications. The practical application and promising capacitive properties of conductive 2D material dispersions, make them ideal electrode components. The advantages of these 2D material electrodes, such as high capacitance, good processability and large surface area, are balanced against their disadvantages, such as flake stacking, re-agglomeration and, in the case of MXenes, oxidation limiting the potential window. The use of asymmetric systems and hybrid electrodes has unlocked synergies for higher energy and power densities as well as improved capacitance retention. However, the complexity behind correctly identifying and characterising capacitive processes, in combination with the many combinations of device sizes, set-ups and configurations, make comparing performance a significant challenge and slowing progress. Further work should be done to standardise methodologies and reporting practices.

### *3.2.2 Thermoelectric Generators*

Countless devices and manufacturing procedures develop heat as a side effect to their function, e.g., air conditioners, microprocessors, combustion engines, industrial furnaces, and server rooms, to cite a few. The energy dissipated by this heat is wasted and is distant from being negligible. For example, the industrial heat waste in the US was estimated to be between 20 to 50% of the entire industrial energy consumed[273], and the total waste heat in the EU is quantified to be about 300 TWh/year[274], which, considering the average consume of a EU household to be around 3500 KWh/year would permit to power about 85 M houses. Therefore, utilizing heat waste is a huge opportunity to reduce energy consumption and increase sustainability of many sectors.

One of the most promising approach to use heat waste is exploiting the thermoelectric effect, which transforms a temperature difference in an electric voltage. To perform such task, thermoelectric (TE) materials are fundamental, and massive research effort has been conducted to optimize the existing solid-state TE materials and create new opportunities translating the TE design to flexible and conformable. Engineering a good TE material is not trivial because it should have a high electrical conductivity ($\sigma$) and Seebeck coefficient ($S$) and a low thermal conductivity ($k$) following the dimensionless TE figure of merit ($zT$)[275]:

$$zT = \frac{\sigma S^2 T}{k} \tag{4}$$

Where $T$ is the temperature. Another essential quantity to characterize a TE material is the power factor which is the product of $\sigma$ and $S^2$. TE materials can be p-type or n-type depending on the electrical charge carrier, exhibiting a positive and a negative Seebeck coefficient, respectively. The most common TE generator is constituted by p- and n-type materials joined at their end thermally in parallel and electrically in series.

Most performing and commonly used TE materials are rigid and bulky alloys of inorganic semiconducting tellurium, lead, and germanium that are p-/n-doped. These materials present issues linked to their sustainability, fabrication challenges, price, and availability.[30] Moreover, considering that more than 60 % of excess global heat is below 100°C, and the colder it is, the more difficult it is to extract valuable energy from it, one of the biggest technological challenges is pulling energy from the diversified spectrum of waste heat.[276] Considering this, an expansion of the TE material family is needed, possibly also to flexible and compliant materials that are more versatile to be interfaced with such diverse waste heat forms that diverge in terms of the frequency of occurrence, the temperature range, and the state (i.e., fluid and gaseous).[276-277] Layered 2D nanomaterials, which ensure the possibility to create flexible materials with many of the abovementioned properties, represent an interesting opportunity. Hence, in the last years, especially carbon-based nanomaterials such as graphenes have increased interest in the research community, especially in the case of coating of flexible substrate and textiles and hybridized with other carbon-based nanomaterials such as carbon nanotubes, as shown in **Table *10***. Mxenes for thermoelectric energy generation were predicted to have a good potential in light of the high electrical conductivity and Seebeck coefficient, but were not investigated thoroughly yet, especially in the form of coating.[278]

Cho C. et al.[31] dip-coated flexible PET substrates in a layer by layer fashion with PANi, GNPs, and CNTs from aqueous dispersion, obtaining a fully-organic TE material, as shown in **Figure 8**(a). GNPs and CNTs were stabilized using electrically conductive PEDOT:PSS. 1 micron thick layer comprising PANi/GNPs-PEDOT:PSS/PANi/CNTs-PEDOT:PSS repeating units displays a high electrical conductivity ($\sigma \approx 1.9 \times 10^5$ S $m^{-1}$) and Seebeck coefficient ($S \approx 120$ μV $K^{-1}$), as shown in Table 10, yielding a power factor of 2710 μW $m^{-1}$ $K^{-2}$. The authors of the papers claim that this water-based TE nanocomposite has power factors competitive with bismuth telluride at room temperature and has the benefit that could be applied as a coating to flexible surface (e.g., textiles). The authors also claim that this novel TE nanocomposites discloses opportunities in powering portable electronics with clothing that transforms wasted body heat to significant voltage.

The same authors developed an n-type organic TE material[90] that is a polymer nanocomposite assembled by dip-coating one after the other flexible PET with CNTs, stabilized with polyethyleneimine (PEI), and GO from aqueous solutions. The obtained material is flexible and can withstand bending and twisting (see Figure 8(b)). A thin film of ~610 nm thickness showed an electrical conductivity of 2730 S $m^{-1}$ and a Seebeck coefficient of −30 µV $K^{-1}$, resulting in a power factor of 2.5 µW $m^{-1}$ $K^{-2}$. A heat curing for 30 minutes at 150 °C reduces the GO and boosts the electrical conductivity and Seebeck coefficient to 46000 S $m^{-1}$ and −93 µV $K^{-1}$, respectively, providing a power factor of 400 µW $m^{-1}$ $K^{-2}$. The authors claim that this is among the highest power factors recorded for a fully organic negative Seebeck TE material. Furthermore, the authors demonstrate that depositing a polymer-clay thin film onto the TE materials provides air-stability over an extended period. They state that this n-type nanocomposite is promising for low-cost, scalable, flexible, environmentally benign, and efficient TE materials.

A previous work of ours[30] enabled the fabrication of p-type and n-type flexible thermoelectric textiles produced with low-cost and sustainable materials and scalable and green processes. Cotton was spray-coated with TE inks made with biopolyester and carbon nanomaterials. Different nanofillers permitted to achieve positive or negative Seebeck coefficient, with graphene nanoplatelets, showing a positive value of 16 µV $K^{-1}$. The best coating allows to perform repeated bending and washing cycles with an increase of their electrical resistance by five times after repeated bending cycles and only by 30% after washing. In-plane flexible thermoelectric generators coupling the best p- and n-type materials were fabricated (see Figure *8*(c)), resulting in an output voltage of ≈1.65 mV and maximum output power of ≈1.0 nW by connecting only two p/n thermocouples at a temperature difference of 70 °C. With their ease of fabrication, green manufacturing, and the biocomposites flexibility and sustainability, such proposed TE materials showed a path for a future generation of wearable TE devices, more harmless for the environment and more easily disposable.

**Table 10**: Resume on the composition of the conductive materials, employed techniques, substrate, and important features of the thermoelectric coating based on 2D nanomaterials. σ, Rs, t, and S correspond to the maximum electrical conductivity, the minimum sheet resistance, the thickness, and the Seebeck coefficient at room temperature, respectively.

| 2D material | Substrate | Other Conductive Materials | σ (S $m^{-1}$) | Rs (Ohm $sq^{-1}$) | t (μm) | S (μV $K^{-1}$) | Ref. |
|---|---|---|---|---|---|---|---|
| **GNP** | Textile (Cotton) | PEDOT:PSS | 5.44E+02 | 8.00E+01 | 2.30E+01 | 1.60E+01 | [30] |
| **GNP** | Sheet (PET) | CNTs, PEDOT:PSS, PANi | 1.79E+05 | 5.60E+00 | 9.98E-01 | 1.20E+02 | [31] |
| **rGO** | Sheet (PET) | CNTs | 4.59E+04 | 3.30E+01 | 4.47E-01 | -9.30E+01 | [90] |
| **GNP** | Sheet (PET) | CNTs,PANi | 1.08E+05 | 1.98E+01 | 4.68E-01 | 1.30E+02 | [117] |
| **GNP** | Sheet (PET) | CNTs | 3.00E+04 | 1.11E+02 | 3.00E-01 | -8.00E+01 | [116] |
| **GNP** | Textile (Cotton) | PANi | 2.80E+01 | - | - | 1.20E+01 | [279] |
| **GNP** | Sheet (Paper) | PEDOT:PSS, Ionic Liquid | 7.50E+03 | - | - | 3.10E+01 | [280] |

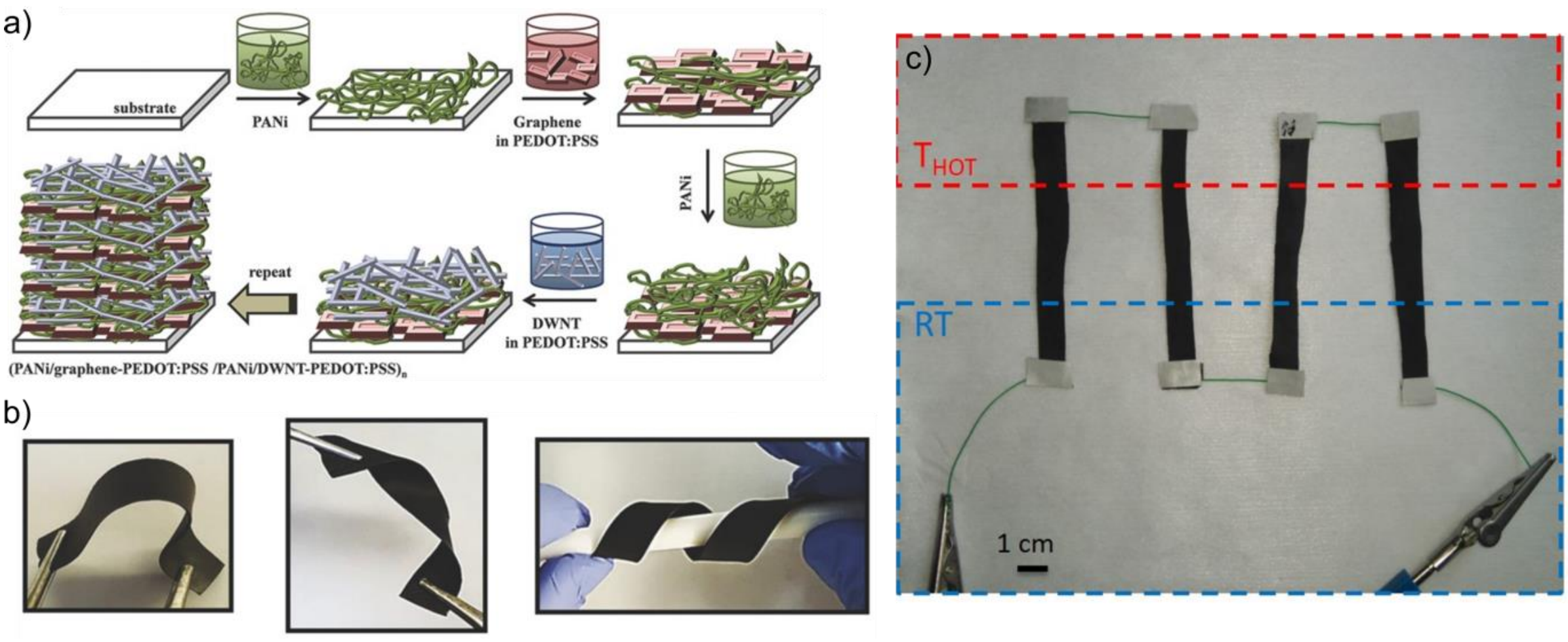

**Figure 8**: a) Schematic of the layer-by-layer deposition process followed to obtain a flexible p-type organic TE coating based on 2D nanomaterials on PET. Homogenous dispersions of PANi in water are deposited first. Afterward, graphene and CNTs in aqueous PEDOT:PSS solutions are dip-coated. This procedure is repeated several times to obtain a layer-by-layer

structure. Adapted with permission.[31]Copyright 2016, Wiley. b) Photos displaying the mechanical compliance of multilayer thin films of CNTs, stabilized with polyethyleneimine (PEI), and GO from aqueous solutions. Adapted with permission.[90]Copyright 2018, Wiley. c) Cotton-based in-plane TE device with 2 TE elements composed of p-type GNP/carbon nanofibers coatings and n-type CnTs coating (red box = hot plate; blue box = room temperature plate). Reused with permission.[30] Copyright 2019, Wiley.

## 3.3 Other Applications of Flexible and Conductive 2D Material-based Coatings

### *3.3.1 Electromagnetic Interference Shielding*

Miniaturization of electronics is boosting the performances of microprocessors, but, at the same time, it is increasing the density of electronics components. Too close components can interfere and modify/disrupt the functioning of devices.[281] Electromagnetic Interference (EMI) shielding is therefore necessary for electronic devices used daily. At the same time, it can be crucial for the safe functioning of large appliances, such as aircrafts in which lightning, solar flares, radar, or even communications can interfere and endanger passengers' security. Hence, airplanes need shielded enclosures, harnessers, and connectors.

EMI shielding is, therefore, an issue that can occur at different scales, and, as such, it could benefit from the versatility of coating techniques that are diversified in terms of the resolution of the deposition, of tuneable electro-mechanical properties, and of adhesions of the inks. Furthermore, an additional benefit could be given by applying flexible and lightweight coatings on curved and motile parts of the airplanes, reducing the weight sensibly and thus the fuel consumption.

When an electromagnetic wave travels and passes through a material, reflection, absorption, multiple internal reflection, and transmission occur. The total EMI shielding effectiveness (*SE*) depends on these quantities following[282]

$$SE = SE_R + SE_A = 10 \log \left(\frac{1}{T}\right) \quad (5)$$

Where $SE_R$ characterises the reflection contribution, $SE_A$ represents the absorption and internal reflections, and $T$ is the transmittance that denotes the fraction of incident electromagnetic wave transmitted. An ideal EMI shielding material should have a high EMI *SE* while being the thinnest possible. The threshold for practical commercial application is a shielding effectiveness ≥ 20 dB.[45]

The most widely used materials for EMI shielding are metals in the form of sheets, foams, or coatings.[283] Most exploited are copper, silver, steel, tin, brass, and nickel. Metals ensure high performances but have drawbacks mainly in terms of costs. Besides, they generally have high density, low resistance to oxidation/corrosion, and high processing expenses, and therefore are not adaptable to ensure lightweight and a capillary diffusion required in specific EMI shielding devices.[284] Furthermore, in the form of inks constituted of carrier material loaded with small metallic particulates typically of copper or nickel, these dispersions are challenging to be stable and generally require stabilizer and sintering post-processing after deposition.

In this context, coatings made with simple procedures on flexible and lightweight materials such as plastics, textiles, or foams, and constituted by conductive 2D nanomaterials such as graphenes and MXenes represent a significant opportunity to decrease the costs of inks for EMI shielding, reduce the weight of shielding components and enable the shield of flexible, curved, and deformable components with thin conductive materials.[284-286]

Valles C. et al. [114] spray-coated micrometric film of alternating electrically conductive negatively charged rGO and a positively charged polyelectrolyte (PEI) on PET in a scalable, thin, and layer by layer fashion. They demonstrated that the EMI shielding properties of such structures could be adjusted by changing the electrical characteristics of the graphene employed and the number of layers deposited. Excellent EMI shielding features were observed for these coatings through a mechanism of absorption. They obtained a maximum EMI *SE* of 29 dB for a 6 μm thick (PEI/RGO) multilayer structure corresponding to 10 coatings of the double component, with 19 vol % loading of rGO flakes with diameter of ∼ 3 μm. Normalizing by the thickness, the EMI *SE* is ≥ 4830 dB/mm. The authors claim that this performance exceeds those previously reported for thicker graphene papers and graphene/polymer composite films with higher GRM contents. They claim that this achievement represents an appropriate step to fabricating lightweight and thin, high-performance EMI shielding structures.

Wang Y. et al.[287] dip-coated cotton into thiol-modified rGO and waterborne polyurethane (WPU) solutions. The WPU molecules were changed with -ene group and acted as polymer matrix to covalently bind the conductive nanofiller and the textile with a thiol-ene click reaction. The authors claim that the interconnected network in the rGO-WPU/cotton sample imparted it with a satisfying value of sheet resistance reaching ~4 kΩ/sq and enhancing the mechanical properties. The EMI *SE* of the material reached 48 dB at only 2 wt% concentration of the modified-rGO, a higher value compared to the same sample made with simple rGO that reached 32.5 dB. The authors claim that the high EMI shielding feature of the samples is due to the better dispersion of the nanofillers modified with the thiol group that increased the interfaces for multiple reflections and scattering. The presented fabric can withstand multiple stresses (1000 cycle bending, 10 cycles of washing/friction), still preserving excellent EMI *SE* (>90% of the initial value).

Shen B. et al.[101] dip-coated polyurethane sponges with GO and afterwards reduced the nanoflakes chemically and thermally, as shown in **Figure 9**(a). The obtained composite foam was compressible and ultralightweight, showing a ~0.027–0.030 g/cm$^3$ density, and exhibited EMI shielding features dominated by absorption. The compression of the foams ensured the tuning of the shielding properties. Indeed, different compressive strains, i.e., 0, 25, 50, and 75%, decreased the EMI *SE* from 39.4 dB to 34.9, 26.2, and 23.4 dB, respectively, as shown in Figure 9(a). The authors justify this reduction with the shrinkage of the conductive graphene network under compression that decreases the foams void spaces and significantly weakens the scattering and multiple reflections of EM waves between the cell surfaces. This diminution occurs despite the slight increase of the electrical conductivity with compression. Notably, the SE of the foams did not change during 50 cycles of stress-release, indicating outstanding cycling stability. This finding indicated that the foams *SE* could be adjusted reversibly by compression or release, enabling adjustable EMI shielding. The authors claim that the simplicity of manufacturing could promote the large-scale production of such lightweight and deformable material for EMI shielding.

In a previous work of ours[32], deformable conductive coatings were manufactured by spray-coating a GNP–elastomer (thermoplastic polyurethane) suspension onto a nitrile rubber substrate, as displayed in Figure 9(b). The electrical resistance of the conductors displayed an increase of 40% after 20 folding-unfolding cycles because of crack formation. On the contrary, it decreased by around 5% after thousands of bending events due to favourable recombination of the conductive GNP in the elastomer matrix. The conformable coatings double their electrical resistance at 12% strain and are washable without modifying their electrical features. The production process can be modified by stretching the nitrile substrate before spraying. This procedure meant that the electrical resistance doubles at 25% strain. The as-produced coatings provide an EMI *SE* of ~20 dB in the 8–12 GHz electromagnetic band. The physical and electrical properties of the flexible conductors, deteriorate upon stretch-release cycling but are restored to the initial value upon heating the coating. For example the EMI *SE* is reduced by 25 % upon 9 cycles of stretch release, but restores the initial EMI shielding after a heating procedure, as shown in Figure 9(b). Such conformable conductive coatings with healable electrical features could permit the realization of truly deformable printed circuit boards, actuators, and tactile sensors.

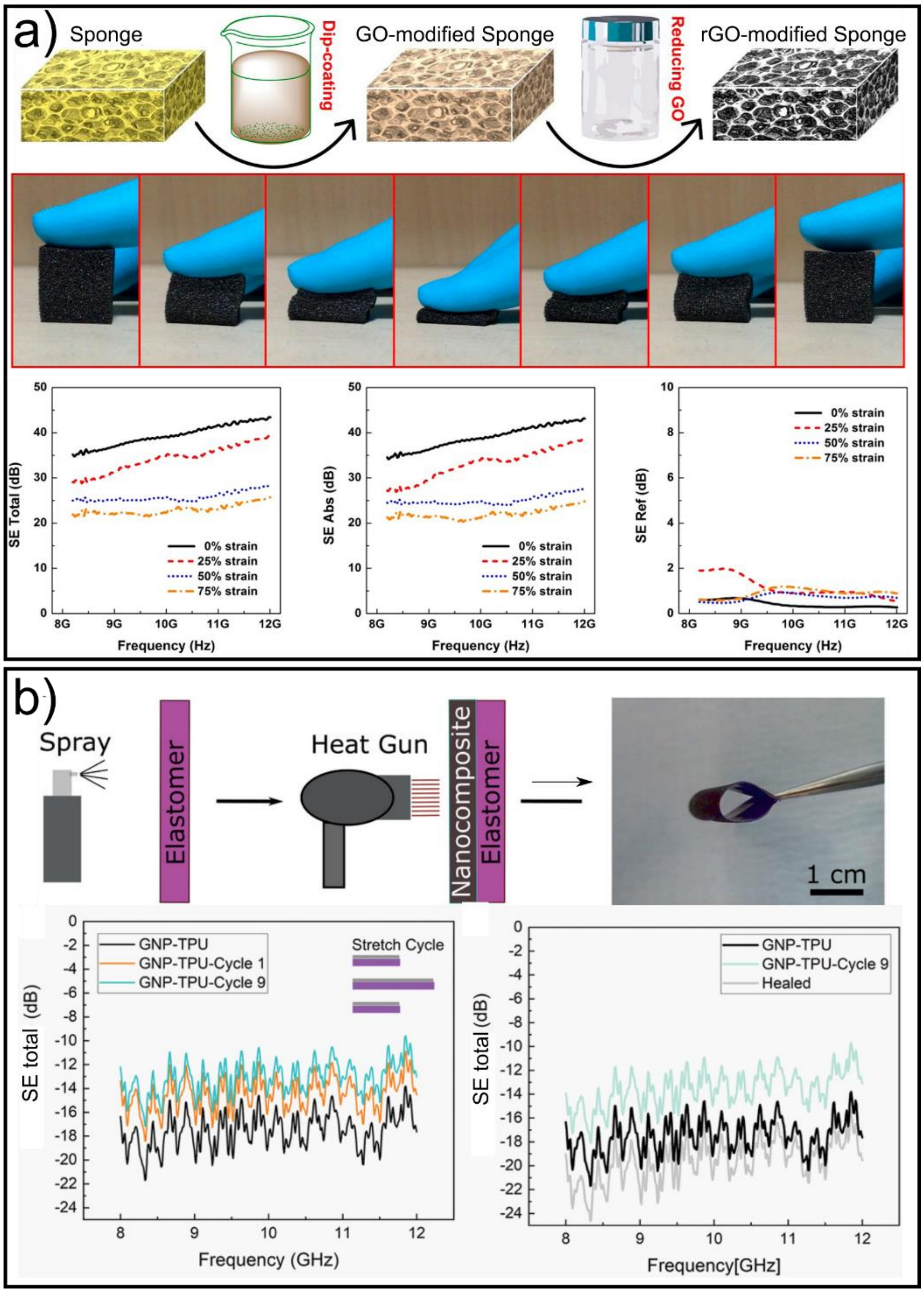

**Figure 9**: EMI shielding of graphene-related materials coated on elastic substrates. a) Top: Overall manufacturing method of the rGO-functionalized foams, comprising dip-coating GO

nanoflakes on the sponge structures and reducing hydrothermally with hydrazine vapor. Middle: Compress-release process of the rGO foam, exhibiting outstanding compressibility. Bottom from left to right: Total Shielding effectiveness (SE), absorption SE, and reflection SE of the polyurethane sponge under various compressions. Adapted with permission.[101] Copyright 2016, American Chemical Society. b) Top: a schematic of the preparation of the electrode and its flexibility. Thermoplastic Polyurethane (TPU) mixed with GNP constitutes the conductive inks sprayed onto nitrile rubber substrates followed by heating by a heat gun. Bottom Left: EMI SE of the GNP–TPU coating before and after 9 stretch–release cycles at 100% elongation. Bottom right: healing of the EMI SE with a heat gun process similar to the fabrication method for the GNP–TPU coatings. Adapted with permission.[32] Copyright 2020, Wiley.

Hu D. et al.[179] manufactured an electrically and thermally conductive MXene coating on cellulose via dip-coating. The as-obtained paper displays electrical conductivity of ~ 2750 S/m. After a PDMS coating, the sample can sustain 2000 bend-release cycles and decrease only of the 10% the starting EMI *SE* of 43 dB measured between 8 and 18 GHz. Besides, a planar thermal conductivity of ~3.9 W/(m·K) is achieved, being 540% higher than the material without MXenes. By cumulative dip-coating cycles, the EMI *SE* and thermal conductivity can be finely tuned. The authors claim that their material can be fabricated on a large scale and with green production methods. They add that considering the flexibility, multifunctionality, and remarkable EMI shielding performances, the nanocomposite paper can have applications in electronics and aerospace.

Uzun S. et al.[288] dip-coated cotton or linen in colloidal dispersion of $Ti_3C_2T_x$ MXene dyes dispersed in water, soaking the ink into the textiles and attaching it to the hydrophilic fabrics, as shown in Figure 10(a). Increasing dip-coating steps increased the conductive filler concentration and the electrical conductivity of the textiles. The samples with four dip-coating steps deposited a coating weight corresponding to ~ 15 wt% of the fabrics, reaching an EMI SE of ∼40 dB over the X-band range (8-12 GHz). After 24 coating, the EMI *SE* augmented to ∼80 dB. The authors showed that the MXene-based fabrics display better performances than commercial metal-based conductive ones. After storing under ambient conditions for two years, the EMI *SE* of MXene-coated fabrics only decreased by ∼8% and ∼13% for cotton and linen, respectively, as shown in Figure 10(a). Besides, the conductive fabrics were demonstrated to shield mobile phone communication. The authors claim that such technology enables the scalable and straightforward transferal of advantageous MXene features to textile-based goods, such as clothes, laptop cases, and wallets, to shield people from electromagnetic pollution or protect digital information. They affirm that their work proposes an attractive alternative to metal-based conductive textiles and presents helpful insights into producing environmentally durable wearable EMI shielding fabrics.

Li Y. et al.[74] fabricated an ultra-stretchable conductor depositing a crumple-textured coating made of $Ti_3C_2T_x$ nanoflakes and SWNTs on a pre-stretched latex rubber balloon, as shown in Figure 10(b). A GO layer was deposited before the MXene-SWNT layer to improve the adhesion of the conductive sheet to the latex. The resulting conductors can sustain up to an 800% areal strain and display strain-invariant resistance profiles through a 500-cycle fatigue experiment, as shown in Figure 10(b). A layer of the stretchable conductor (1 µm thick) demonstrated strain-insensitive EMI shielding properties of ~30 dB under broad areal strains (up to 800%), and the shielding feature was augmented to 47 and 52 dB by layering 5 and 10 films of the MXene conductors, respectively. The authors also demonstrated the manufacturing of a dipole antenna composed of stretchable MXene-based conductors. With a uniaxial stretch of 150%, the antenna resonant frequency was changed from 1.575 to 1.375 GHz, while the reflection |S11| stayed constant at −30 dB, corresponding to steadily reflected powers smaller than 0.1%. Combining the EMI absorbers and the stretchable antennas permitted the authors to demonstrate a wearable wireless device that produced mechanically stable wireless communication while attenuating EM absorption by humans.

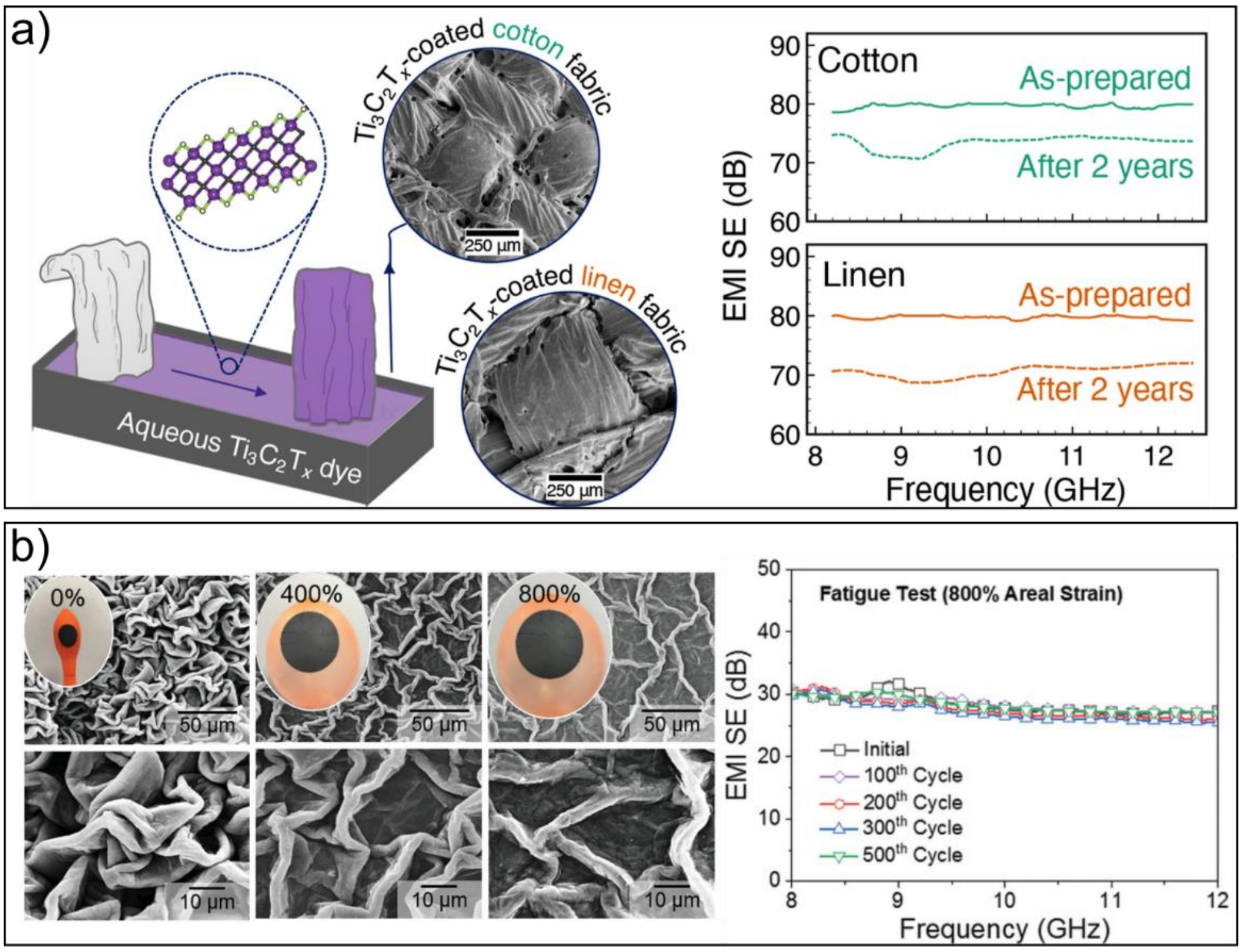

**Figure 10**: EMI shielding of MXenes coated on diverse substrates. a) Top: manufacturing procedure of the fabrics based conductive material and SEM images of the cotton or linen based material. The materials just slightly change the EMI SE after two years that were fabricated. Reused with permission.[288] Copyright 2021, Elsevier. b) Left: SEM pictures and photos of the SWNT-MXene coating on latex under different areal strains. Right: EMI SE of the stretchable SWCT-MXene-based material through 500 stretch-release cycles with areal strain up to 800%. Adapted with permission.[74] Copyright 2020, Wiley.

**Table 11**: EMI Shielding Applications in Surveyed Literature.

| 2D Material | Substrate | Other Conductive Materials | EMI SE (dB) | Ref. |
|---|---|---|---|---|
| **Graphene** | Foam (Latex, Natural Rubber) | | 2.40E+01 | [97] |
| | Foam (PU) | | 5.80E+01 | [101] |
| | Sheet () | Co NPs | 5.50E+01 | [105] |
| | Sheet (Cellulose) | | 4.00E+01 | [69] |
| | Sheet (Nitrile) | | 1.70E+01 | [32] |
| | Sheet (PET) | | 1.50e+01 to 2.90e+01 | [114] |
| | Sheet (PET) | Ag NWs | 4.29E+01 | [120] |
| | Sheet (Paper) | | 3.60E+01 | [86] |
| | Textile (Carbon Cloth) | | 3.79e+01 to 4.00e+01 | [71] |
| | Textile (Cellulose) | | 4.50E+01 | [137] |
| | Textile (Cotton) | | 2.00e+01 to 4.81e+01 | [141, 287, 289] |
| | Textile (Cotton) | Ag NPs | 2.74E+01 | [290] |
| | Textile (Cotton) | ZnO NPs | 5.47E+01 | [92] |
| | Textile (PET) | AgNO3 | 6.50E+01 | [163] |
| | Textile (Polyester) | | 2.60E+01 | [171] |
| | Textile (Wool) | PPy | 2.20E+01 | [291] |
| **Graphene + MXene** | Sheet (Latex) | CNTs | 3.00e+01 to 5.20e+01 | [74] |
| | Textile (Carbon Cloth) | | 3.77E+01 | [71] |
| **MXene** | Sheet (Filter Paper) | | 2.46e+01 to 4.30e+01 | [179] |
| | Sheet (PTFE) | | 5.07E+01 | [180] |
| | Textile (Carbon Cloth) | | 4.32E+01 | [185] |
| | Textile (Carbon Cloth) | CNTs | 4.67E+01 | [186] |
| | Textile (Cotton) | | 2.60e+01 to 8.00e+01 | [188-189, 288] [292] |
| | Textile (Linen) | | 4.00e+01 to 8.00e+01 | [288] |
| | Textile (PET) | PPy | 4.20e+01 to 9.00e+01 | [194] |
| | Textile (Silk) | Ag NWs | 4.20e+01 to 8.50e+01 | [195] |

From Table 11, textiles results by far as the prevalent adopted substrates (66%), followed by sheets (31%), while foams are only the 3% of the surveyed literature. The appeal to realize wearable conductive EMI shields reflects these numbers, together with the ease of functionalizing fabrics that absorb liquid inks.

### *3.3.2 Joule Heaters*

When a current flows through a media of finite electrical resistance under an applied voltage, resistive heating takes place following the first Joule law

$$P \propto I^2 R \propto \frac{V^2}{R} \qquad (6)$$

where $P$ is the power that corresponds to the thermal energy generated per unit of time. This phenomenon is called Joule heating and is often a side effect in applications such as air conditioners, microprocessors of laptops, smartphones, etc. In contrast, it is the basic principle exploited for the functioning of appliances such as hot plates, ovens, floor heating, toasters, and hair dryers, to name a few.[293] It is sometimes used for de-icing aerodynamic and structural surfaces, especially in the aeronautic and wind turbine sectors.[294-295]

Joule heating points of strength are the high release of heat in a small volume, the remarkable heating uniformity/non-uniformity and tuneable heating time depending on the design of the resistive path, the easiness of the power control (i.e., voltage-driven), and the high temperature achievable depending on the resistive material used that can reach even thousands of °C. The most commonly used materials for Joule heating are metal alloys such as nickel-chrome and iron-chromium-aluminum compounds (known as Nichrome and Kanthal, respectively), ceramics, and semiconductors such as molybdenum disilicide and silicon carbide. Such materials ensure heat resistance, high-temperature strength (i.e., absence of deformation during operation), an adequate electrical resistivity, and low thermal expansion coefficient. As such, they are ideal also for high-performance applications since they can sustain temperatures between 1000 and 1300 °C, needed in large industrial furnaces. Nevertheless, they are rigid and difficult to be incorporated as heaters in flexible and wearable devices, and in motile parts of robots.[296] Therefore, materials designed ad-hoc to comply with flexibility and conformability necessities are needed.

Electrical conductive coatings appear as a natural candidate to fill the gap between existing rigid heating materials and flexible ones.[293] Indeed the versatility of coating techniques in terms of surface (plane, curved, rough, and porous) and materials (rigid, flexible, and wearable) that can be coated, and the compatibility with the existing industrial processes, ensure a potentially short-term and straightforward expansion of heaters to the field of wearables and flexibles.[297] In this context, graphene- and MXene-based electrically conductive inks represent a great chance of realizing flexible Joule heating coating on a curved surface and achieve conformable and/or bendable materials that can be heated simply through an applied voltage. Indeed, nanostructured materials allow an exceptional spreading of the heating spots.[298] Hence, 2D nanomaterials are emerging as extraordinarily beneficial and favourable for Joule heaters expansion, as shown in Table 12.[294, 299]

Hao Y. et al.[26] developed a cotton-based heater spraying in sequence water-dispersed polyurethane (WPU), a GNP-tourmaline-WPU composite, and a final WPU protective layer, as shown in Figure 11(a). In the composite layer, tourmaline and GNP displayed a synergistic effect, with superior electro-thermal features than single-component fabrics, achieving a temperature of ~75 °C in 30 s under 10 V, while the pure GNP achieved ~ 45 °C at the same conditions, as shown in Figure 11(a). The composite fabric showed a high heating power density of $2 \times 10^3$ W/m$^2$. The flexible heater was significantly abrasion resistant, resulting in 21 % electrical conductivity reduction after a 2500-cycles abrasion test, far less than the 87 % of fabrics without a protective WPU layer. The single components were all necessary since: i) GNPs provided the electrical conductivity and supported high electro-thermal efficacy, ii) cotton ensured the flexibility and wearability of the material, iii) WPU acted as a binder that also contributed to the composite fabrics abrasive resistance, and iv) adding tourmaline nanoparticles enhanced prominently the maximum temperature reached at 10 V. The authors forecast application of the proposed Joule heater as medical electro-heating wearable devices and functional protective garments.

Karim N. et al.[300] displayed a process to make conductive graphene-coated glass fibre roving with fast de-icing features. The graphene ink is obtained through a large scale technique based on a microfluidic exfoliation. Such ink is coated using a dip-dry-cure technique that the authors claim to be potentially suitable for the industry. The graphene-coated glass roving can heat up rapidly to a required temperature. For example, it can heat up at 72 °C in 30 s at 10 V, it can reach 101 °C after 180 seconds, and 120°C after ~ 10 minutes. The authors integrated these graphene-coated glass rovings into a vacuum-infused epoxy–glass fabric composite and demonstrated their use for de-icing applications. They dipped it into a container full of ice and compared the melting of ice with another bucket containing only ice, applying 10 V. A rapid increase in the speed of the melting of ice in the bowl containing graphene-based composites by Joule heating was observed. Indeed, the temperature increases to 27 °C within 5 minutes, whereas the bucket temperature containing only ice remains constant even after 30 minutes.

Ahmed A. et al.[152] presented a mass and straightforward producible dip-coating technique to obtain an electro-conductive stretchable composite based on cotton with rGO and PEDOT:PSS as a conductive additive, as shown in Figure 11(b). The authors claim that the composite exhibited exceptional heating even after mechanical deformation. Indeed the heaters maintain their value of sheet resistance of ~ 154 Ω/sq after 100 folding cycles and decrease their heating performance of 4 °C (from 40 to 36 °C) at 60% applied strain, and under a voltage of 15 V. The heater provided rapid response and a consistent heat supply in the heating zone at low voltage. Indeed, under different voltages (ranging from 5 to 30 V), the maximum temperature was reached in 15 to 25 s with the highest temperature of 70 °C at 30 V, as shown in Figure 11(b). The materials exhibited durable heating/cooling performance under repeated cycle (10) and excellent stability to wash, increasing the initial value of sheet resistance by ~ 1% after 15 washing cycles.

The possibility to combine the mechanical properties of flexible substrates and multifunctionality of 2D nanofillers coatings, is enabling new multifunctional materials that are able to address several problems and tasks with a single configuration. For example, coatings containing MXenes with applications as Joule heaters and, at the same time, a combination of EMI shielding, strain sensing, hydrophobic, and transparency features are a growing trend in literature. [189, 193-194, 301]

Wang Q.-W. et al.[194] fabricated an electrically conductive flexible PET fibrous textile (~1000 S $m^{-1}$) dip-coating it with polypyrrole and MXene. A PDMS treatment was then performed to ensure hydrophobicity (water contact angle of ~126°) and prevent the degradation/oxidation of the MXene at high humidity while retaining the considerable air permeability of the textile. The authors claim that the coating ensures effective Joule heating performance together with high EMI shielding (~90 dB with a 1.3 mm thick sample). Polypyrrole improves the adhesion of the MXene nanoflakes and the PET fibers and increases the EMI shielding performance. The Joule heating of the multifunctional textile needs low voltage for an efficient control. Indeed, changing the input voltage to 2, 3, and 4 V permits reaching a saturated temperature of 40, 57, and 79 °C, respectively. Therefore, high tunability of the temperature in a range of voltages that are safe for operation is ensured. The authors show that stable cyclic heating performances are achieved during 50 cycles. Thus, they claim that such material is suitable for thermotherapy and personal heating garments. Moreover, in light of the high electrical conductivity, flexibility, and water-resistant features, flexible electronics, wearable smart clothes, and EMI shielding garments are highly promising.

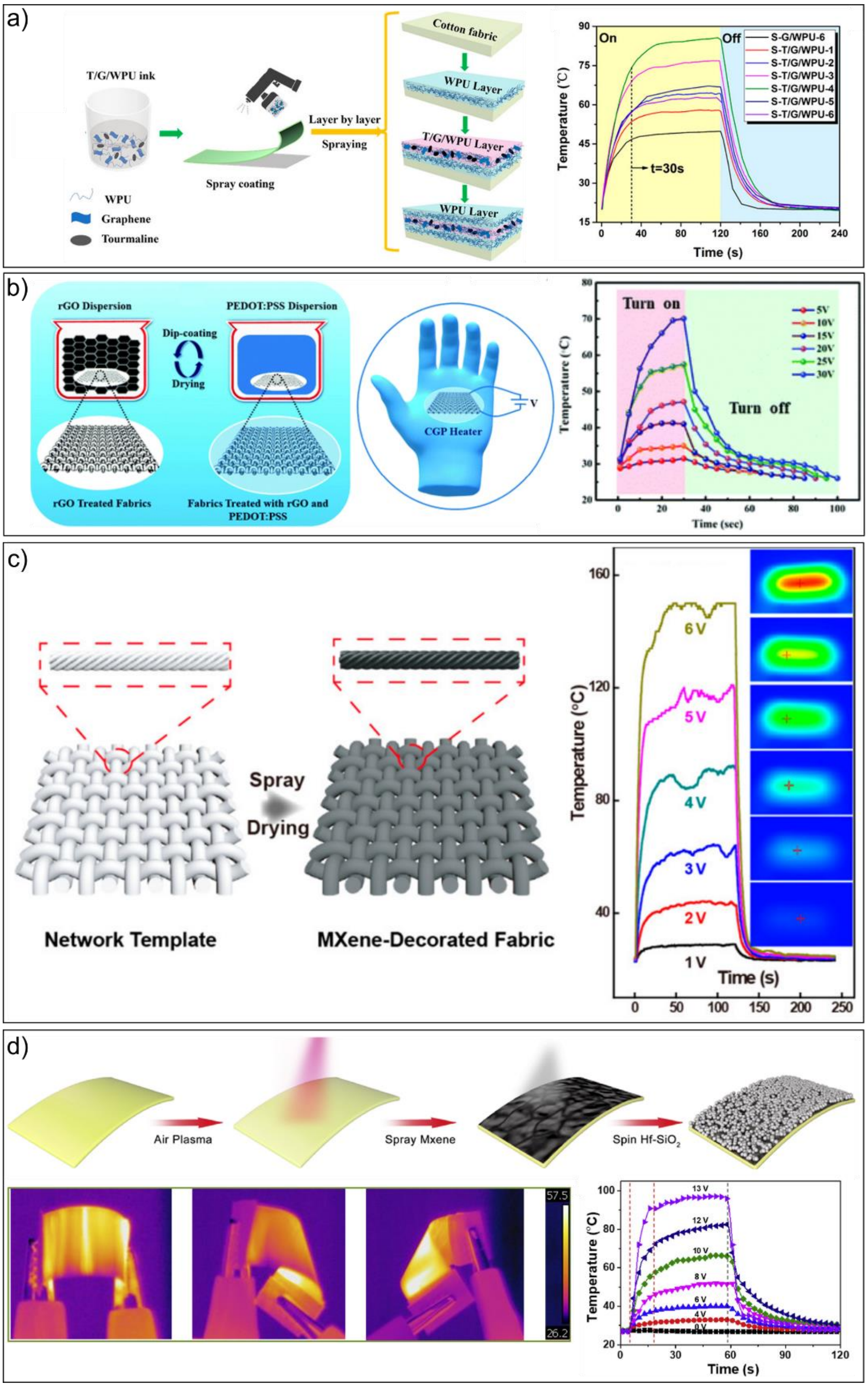

**Figure 11**: a) left: process to fabricate a sandwich structure with heating properties given by a GNP (G)-tourmaline (T) - water-dispersed polyurethane (WPU) composite. The substrate and the protective layer are realized with pure WPU. Right: temperature variation of samples with different T:G weight ratio under 10 V voltage; in black, the sample with only G. In green the best sample (weight ratio T:G=6:4). Adapted with permission.[26] Copyright 2018, American Chemical Society. b) Left: scheme to fabricate the conductive coating made of rGO and PEDOT:PSS. Right: Electro-thermal characterization of the as-obtained heater under different applied voltages (5 to 30 V). Adapted with permission.[152] Copyright 2020, Royal Society of Chemistry. c) Left: assembly of MXene-decorated cotton textile by spray-coating technique. Right: time-temperature curves of the sample with an applied voltage changing from 1 to 6 V. Reused with permission.[189] Copyright 2020, American Chemical Society. d) Top: diagram of the production of hydrophobic and transparent Joule heaters. Spin-coating hydrophobic fumed silica nanoparticles achieved hydrophobicity on top of a spray coated layer of Mxene. Bottom left: infrared images of the sample under mechanical deformation and with 8 voltage applied. Bottom right: change of the coating temperature in time under voltages ranging from 2 to 13 V. Adapted with permission.[301] Copyright 2021, Elsevier.

Zhang X. et al.[189] spray-coated $Ti_3C_2T_x$ nanosheets on cotton (Figure 11(c)), obtaining low sheet resistance (5 Ω $sq^{-1}$ at 6 wt.% nanoflakes loadings) and breathable fabrics. The material had remarkable Joule heating performance at low applied voltages. In particular, the temperature changed from 29 to 150 °C with voltage varying from 1 to 6 V (Figure 11(c)), which the authors claim exceed the change of most of the previously reported textile heaters coated with other conductive fillers. The Joule heating stability was also studied by subjecting the samples to 100 cycles at the voltage of 4 V, exhibiting remarkable durability. The flexible heaters were stitched onto diverse wearables, displaying to warm up very homogeneously to 60°C, as shown in Figure 11(c). The coating provided multifunctional properties to the cotton that exhibited outstanding EMI shielding features, up to 36 dB, and bend-induced strain sensing that monitored human motions. Therefore, the authors claim that their work detailed multifunctional and wearable textiles promising flexible voltage-driven heaters, EMI shielding clothes, and integrated strain-sensitive garments.

Zhou B. et al.[301] realized a flexible hydrophobic MXene-based transparent material with applications as a Joule heater and in EMI shielding. MXene inks were spray-coated on transparent polycarbonate substrates (Figure 11(d)). Afterward, spin-coating was employed to deposit hydrophobic fumed silica and obtain superhydrophobic (contact angle of 150.7°) and self-cleaning (sliding angle of 3°) surfaces that also prevented the oxidation of MXenes. The multilayer structure displays low sheet resistance (~35.1 Ω/sq) with performance diminishing only by 10% after 1000 bending cycles, a transmittance of 33%, and electromagnetic interference shielding effectiveness major of 20 dB. The material was exposed to outdoor conditions for 100 days without significant change of its properties. In particular, the coating showed rapid Joule heating at safe voltages. At voltages between 4 and 13 V, the steady temperature changed from 32 to 101 °C, as shown in Figure 11(d). The time needed to reach 101 °C was ~15 s and ~30 s to cool down. The temperature change was stable under several cycles. The authors claim that such a mixture of properties, comprising the outdoor stability for 100 days, bode well for a straightforward application of the material in many fields such as smart windows, flexible devices, and electrical heating systems.

**Table 12**: Joule Heater Coatings in Surveyed Literature.

| **2D Materials** | **Substrate** | **Other Conductive Materials** | **Temperature (C)** | **Voltage (V)** | **Ref.** |
|---|---|---|---|---|---|
| **Graphene** | Sheet (PDMS) | | 1.80E+02 | 8.00E+01 | [302] |
| | Sheet (PET) | Ag NWs | 2.80E+02 | 1.20E+01 | [121] |
| | Textile (Cotton) | | 7.50E+01 | 1.00E+01 | [26] |
| | Textile (Cotton) | | 1.63E+02 | 1.20E+01 | [303] |
| | Textile (Cotton) | CB | 1.03E+02 | 2.00E+01 | [208] |
| | Textile (Cotton) | MnO2 | 3.70E+01 | 1.50E+01 | [25] |
| | Textile (Cotton) | PEDOT:PSS | 7.00E+01 | 3.00E+01 | [152] |
| | Textile (Glass) | | 1.20E+02 | 1.00E+01 | [300] |
| | Textile (PET) | | 1.39E+02 | 1.40E+01 | [27] |
| **MXene** | Sheet (PTFE) | | 2.00E+00 | 3.96E+01 | [180] |
| | Sheet (Polycarbonate) | | 1.00E+02 | 1.30E+01 | [301] |
| | Textile (Cotton) | | 6.43E+01 | 4.00E+00 | [188] |
| | Textile (Cotton) | | 1.50E+02 | 6.00E+00 | [189] |
| | Textile (PET) | | 1.29E+02 | 1.50E+01 | [193] |
| | Textile (PET) | PPy | 7.90E+01 | 4.00E+00 | [194] |

As a general trend, textiles (77% of the surveyed literature) involving electrically conductive coatings are predominant with respect to sheets as a Joule heating substrate. This is due to the intrinsic wearable characteristic of such substrates that enable thermal therapy and intelligent garments. MXenes are so far rarer but exhibit performance advantages, as shown in Figure 12.

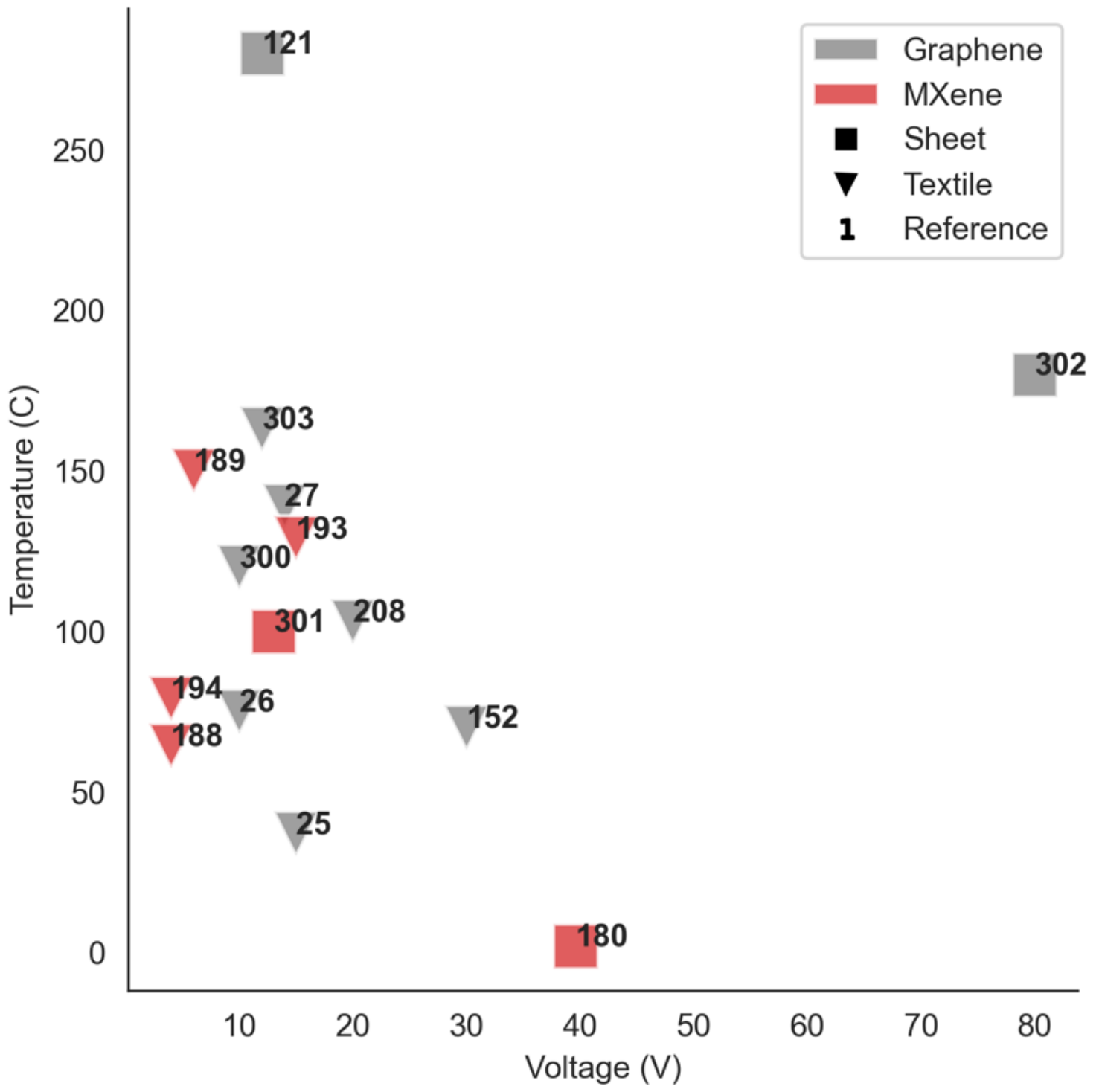

**Figure 12**: Plot of the Temperature versus voltage of the surveyed literature in Table 12. In the figure square (■) represents sheets and triangle (▼) textiles.

# 4 Conclusions and Future Perspectives

The more than decade-long momentum behind the continuous growth of 2D materials is due in part to the vast array of applications. As such, this review has sought to compare the progress of two of the most promising families of conductive coatings comprising 2D materials: graphenes and MXenes. A wide range of applications in the field of large-area flexible and stretchable electronics is summarized. This has enabled trends to be drawn both within and across diverse sectors, providing a comparative reference of past progress and insight into future possibilities.

We surveyed three key components; the formulation of the 2D material ink, the coating technique and substrate used, and the performance of the final device. In terms of ink formulations, water is generally preferred since it is easily sourced, non-toxic, and renewable. The next most popular solvents are ethanol, DMF, IPA, and NMP. GNP dispersibility in water is usually assisted with solvents, binders, or surfactants to ensure stable inks, whereas GO, rGO, and MXenes are more easily dispersible. Hybrid formulations combining 2D materials with other conductive fillers to enhance electrical properties or create synergies between conduction mechanisms are often exploited. Conductive polymers, nanometals, or other nanocarbons (i.e., carbon black and CNTs) are often used for such purposes.

We find that dip coating is preferred for flexible textiles and for deformable foams due to the self-soaking nature of these substrates, which permit an efficient coverage. Spraying is mostly selected for sheets because it enables a precise deposition through masking and prevents smearing and staining since the inks are almost dry when they impact the target surface. Dip coating is preferably performed with water-based conductive inks, while spray coating needs solvents that partially/fully evaporate before reaching the substrate to ensure sufficient substrate contact and prevent nonuniformity. It is promising that most large area coating techniques used for 2D material inks research are already readily scalable, and thus do not need a special integration effort for their assimilation in production processes. Their ease of use also facilitates the exploration of multiple variables e.g., different coating thicknesses and compositions, as such articles often compare multiple device configurations. This is advantageous as many performance comparisons can be within a single article. However, as is often the case, these lab-based devices are usually crude compared to industrially ready devices, e.g., lacking encapsulation and packaging. Meaning that scaling performance can be difficult, slowing technological uptake.

The rapid growth and intense interest in the field flexible electronics has so far outpaced industrial standardisation. As such, there is no clear approach for evaluating performance, e.g., most articles measure the intended application performance while the device is mechanically relaxed, and separately showcase the device's ability to flex and stretch. However, in practice a flexible and stretchable electronic device whether an energy storage component or, EMI shield or heater is likely to be required to operate under strain. This is set to change, as over the last three years the IPC (association connecting electronics industries) has begun to release a series of standards relating to eTextiles and SEMI (the global industry association that unites the electronics design and manufacturing supply chain) has established a technical committee to draft future standards.[304-305]

The combination of fillers with different dimensionality has been used to exploit the various micro-mechanical mechanisms that make and break percolating conductive pathways in flexible strain and pressure sensing devices. Achieving high, linear, and reliable sensitivities across a large working range, requires synergetic effects across the conductive fillers and substrate used. Textile substrates were most used for strain-sensors, as their component fibres provide a network for deformation, enabling higher sensitivities. Foam substrates were most used for pressure-sensors, as they can modulate the applied force through their porous structure.

The use of asymmetric and hybrid supercapacitor electrodes configured in cofacial, coplanar or fibrous set-ups has led to a diverse array of flexible SCs and MSCs that utilise graphenes and MXenes. The advantages of these 2D material electrodes, such as high capacitance, good processability and large surface area, are balanced against their disadvantages, such as flake stacking, re-agglomeration and, in the case of MXenes, oxidation limiting the potential window.

In thermoelectrics, graphene-based coatings are explored, while MXenes ones are not. GRMs are mixed with organic conductive polymers such as PEDOT:PSS and PANi to boost the performance of organic TE inks. A recent trend consists in functionalizing textiles to realize wearable thermoelectric materials. The high value of the graphene's thermal conductivity is hampering the generation of high power with TE generators based on such materials. The diminution of the high thermal conductivity of GRMs could rely on the functionalization of these 2D nanoparticles with additives with low thermal conductivity. However, the possibility of coating large surfaces and developing flexible and wearable devices is not reachable with state-of-the-art solid-state TE materials that are mainly based on bulk bismuth and tellurium alloys. Thus, graphene-based inks are of interest for the development of the field. MXenes have the potential to perform well in TE generations due to their high electrical conductivity and Seebeck coefficient.[278]

Regarding EMI shielding applications, textiles are the prevailing substrates, accompanied by sheets, while foams are only a few % of the examined reports. A recent trend for this application with 2D materials relies upon making stretchable electromagnetic shielding coatings, more frequent with sheets or foams. Flexibility, stretchability, and, in some cases, even healability[32] of the conductive EMI shielding coatings made with 2D materials establish valuable improvements compared to conventional metallic ones.

As a general trend, across most applications surveyed, the use of MXenes is gaining traction as they combine high electrical conductivity with key structural and processability advantages: They do not need the thermal or chemical reduction process required for GO, they readily disperse in aqueous substrates without the aid of additives unlike GNPs, all while sharing many of graphene's advantages, including good adhesion to substrates, high electrical conductivity and 2D morphology. In Joule heaters this translates to lower potential requirements to achieve the desired temperatures than that of graphenes, even rivalling that of metallic nanomaterials.[306] In EMI shielding their improved conductivity enables higher signal attenuation at the same thickness than that of graphenes.[284, 307-308] However, the tendency of MXenes to oxidise and degrade electrically, even within polymer matrices, may require further research effort to engineer more robust surface chemistry and larger lateral sizes.[309] It is worth nothing that $Ti_3C_2T_x$[57] is the only form of MXene used in all the surveyed literature, despite 29 other forms having been experimentally synthesised[33]. This fact leaves wide room for further explorations for this class of materials in all the applications surveyed.

There were fewer examples of coatings made up of a combination of graphene and MXene in the literature than initially expected. [71-74] Although, this makes sense considering their many similarities leave little potential for synergy at the cost of added system complexity. In practice, there is ample opportunity to share progress in experimental technique and theoretical understanding between 2D materials. As such, we encourage researchers to not limit their literature study to any one material family. Additionally, the composites plots presented for the different applications, show that despite the diverse array of materials used, device performance tends to cluster within 1 or 2 orders of magnitude. Further permutations of ink formulations, coating techniques and substrate types that do not significantly advance performance, should instead pay greater focus to developing theoretical understanding and challenging proof-of-concept devices with more robust industry-level testing.

## Acknowledgments

This work was supported by the Graphene NOWNANO CDT and Haydale Ltd. The authors acknowledge "Graphene Core 2" and "Graphene Core 3" GA: 785219 and GA:881603, respectively, which are implemented under the EU-Horizon 2020 Research & Innovation Actions (RIA) and are financially supported by EC-financed parts of the Graphene Flagship. P. C. and M. C. acknowledge the partial support of the European Research Council (ERC) under the European Union's Horizon 2020 research and innovation programme "ELFO", Grant Agreement No. 864299. P.C. acknowledges funding from the Marie Skłodowska-Curie actions (project name: BioConTact, grant agreement no. 101022279) under the European Union's Horizon 2020 research and innovation programme. I.A.K. thanks the Royal Academy of Engineering (RCSRF1819\7\3) and Morgan Advanced Materials for funding this Chair. Table 1 first and second icon made by Freepik from www.flaticon.com. Table 1 third icon made by monkik from www.flaticon.com.

## References

[1] W. Wong, A. Salleo, *Flexible Electronics: Materials and Applications*, Springer US, **2009**.
[2] M. S. Ergoktas, G. Bakan, P. Steiner, C. Bartlam, Y. Malevich, E. Ozden-Yenigun, G. He, N. Karim, P. Cataldi, M. A. Bissett, I. A. Kinloch, K. S. Novoselov, C. Kocabas, *Nano Lett* **2020**, 20, 5346.
[3] M. Kaya, in *Electronic Waste and Printed Circuit Board Recycling Technologies*, DOI: 10.1007/978-3-030-26593-9_2 **2019**, Ch. Chapter 2, p. 33.
[4] D. F. Fernandes, C. Majidi, M. Tavakoli, *J Mater Chem C* **2019**, 7, 14035.
[5] D. Grosso, *Journal of Materials Chemistry* **2011**, 21, 17033.
[6] A. Nathan, A. Ahnood, M. T. Cole, S. Lee, Y. Suzuki, P. Hiralal, F. Bonaccorso, T. Hasan, L. Garcia-Gancedo, A. Dyadyusha, S. Haque, P. Andrew, S. Hofmann, J.

Moultrie, D. Chu, A. J. Flewitt, A. C. Ferrari, M. J. Kelly, J. Robertson, G. A. J. Amaratunga, W. I. Milne, *Proceedings of the IEEE* **2012**, 100, 1486.
[7] B. Wang, A. Facchetti, *Adv Mater* **2019**, 31, 1901408.
[8] Q. Huang, Y. Zhu, *Adv Mater Technologies* **2019**, 4, 1800546.
[9] V. K. R. R, V. A. K, K. P. S, S. P. Singh, *Rsc Adv* **2015**, 5, 77760.
[10] M. Zenou, L. Grainger, in *Additive Manufacturing* (Eds: J. Zhang, Y.-G. Jung), Butterworth-Heinemann **2018**, p. 53.
[11] J. Kim, R. Kumar, A. J. Bandodkar, J. Wang, **2017**, 15.
[12] F. Torrisi, J. N. Coleman, *Nat Nanotechnol* **2014**, 9, 738.
[13] G. Hu, J. Kang, L. W. T. Ng, X. Zhu, R. C. T. Howe, C. G. Jones, M. C. Hersam, T. Hasan, *Chem Soc Rev* **2018**, 47, 3265.
[14] Conductive Pastes, https://www.jesolutionsconsultancyltd.com/conductive-pastes, accessed: 29th November, 2021.
[15] L. Nayak, S. Mohanty, S. K. Nayak, A. Ramadoss, *J Mater Chem C* **2019**, 7, 8771.
[16] E. Jabari, F. Ahmed, F. Liravi, E. B. Secor, L. Lin, E. Toyserkani, *2d Mater* **2019**, 6, 042004.
[17] V. Beedasy, P. J. Smith, *Materials (Basel)* **2020**, 13.
[18] C. Cano-Raya, Z. Z. Denchev, S. F. Cruz, J. C. Viana, *Appl Mater Today* **2019**, 15, 416.
[19] H.-Y. Jun, S.-J. Kim, C.-H. Choi, *Nanomaterials-basel* **2021**, 11, 3441.
[20] Y. Z. Zhang, Y. Wang, Q. Jiang, J. K. El-Demellawi, H. Kim, H. N. Alshareef, *Adv Mater* **2020**, 32, e1908486.
[21] S. Abdolhosseinzadeh, X. Jiang, H. Zhang, J. Qiu, C. Zhang, *Mater Today* **2021**, 48, 214.
[22] P. Cataldi, S. Dussoni, L. Ceseracciu, M. Maggiali, L. Natale, G. Metta, A. Athanassiou, I. S. Bayer, *Adv Sci (Weinh)* **2018**, 5, 1700587.
[23] Q. Jiang, N. Kurra, K. Maleski, Y. Lei, H. Liang, Y. Zhang, Y. Gogotsi, H. N. Alshareef, *Adv Energy Mater* **2019**, 9, 1901061.
[24] G. Wang, V. Babaahmadi, N. He, Y. Liu, Q. Pan, M. Montazer, W. Gao, *J Power Sources* **2017**, 367, 34.
[25] M. Tian, M. Du, L. Qu, K. Zhang, H. Li, S. Zhu, D. Liu, *J Power Sources* **2016**, 326, 428.
[26] Y. Hao, M. Tian, H. Zhao, L. Qu, S. Zhu, X. Zhang, S. Chen, K. Wang, J. Ran, *Ind Eng Chem Res* **2018**, 57, 13437.
[27] D. Wang, D. Li, M. Zhao, Y. Xu, Q. Wei, *Appl Surf Sci* **2018**, 454, 218.
[28] Y. Huang, X. He, L. Gao, Y. Wang, C. Liu, P. Liu, *Journal of Materials Science: Materials in Electronics* **2017**, 28, 9495.
[29] P. Wang, B. Sun, Y. Liang, H. Han, X. Fan, W. Wang, Z. Yang, *J Mater Chem A* **2018**, 6, 10404.
[30] P. Cataldi, M. Cassinelli, J. A. Heredia-Guerrero, S. Guzman-Puyol, S. Naderizadeh, A. Athanassiou, M. Caironi, *Adv Funct Mater* **2019**, 30, 1907301.
[31] C. Cho, K. L. Wallace, P. Tzeng, J.-H. Hsu, C. Yu, J. C. Grunlan, *Adv Energy Mater* **2016**, 6, 1502168.
[32] P. Cataldi, D. G. Papageorgiou, G. Pinter, A. V. Kretinin, W. W. Sampson, R. J. Young, M. Bissett, I. A. Kinloch, *Advanced Electronic Materials* **2020**, 6, 2000429.
[33] Y. Gogotsi, B. Anasori, *Acs Nano* **2019**, 13, 8491.
[34] E. Shim, in *Smart Textile Coatings and Laminates (Second Edition)*, DOI: https://doi.org/10.1016/B978-0-08-102428-7.00002-X (Ed: W. C. Smith), Woodhead Publishing **2019**, p. 11.

[35] J. J. Licari, in *Coating Materials for Electronic Applications*, DOI: https://doi.org/10.1016/B978-081551492-3.50005-2 (Ed: J. J. Licari), William Andrew Publishing, Norwich, NY **2003**, p. 201.
[36] T. Carey, C. Jones, F. Le Moal, D. Deganello, F. Torrisi, *ACS Appl Mater Interfaces* **2018**, 10, 19948.
[37] R. Cherrington, J. Liang, in *Design and Manufacture of Plastic Components for Multifunctionality*, DOI: https://doi.org/10.1016/B978-0-323-34061-8.00002-8 (Eds: V. Goodship, B. Middleton, R. Cherrington), William Andrew Publishing, Oxford **2016**, p. 19.
[38] S. Stankovich, D. A. Dikin, G. H. Dommett, K. M. Kohlhaas, E. J. Zimney, E. A. Stach, R. D. Piner, S. T. Nguyen, R. S. Ruoff, *Nature* **2006**, 442, 282.
[39] A. Kausar, *Journal of Composites Science* **2021**, 5, 292.
[40] A. Zamhuri, G. P. Lim, N. L. Ma, K. S. Tee, C. F. Soon, *BioMedical Engineering OnLine* **2021**, 20, 33.
[41] K. S. Novoselov, *Science* **2004**, 306, 666.
[42] I. A. Kinloch, J. Suhr, J. Lou, R. J. Young, P. M. Ajayan, *Science* **2018**, 362, 547.
[43] H. Döscher, T. Schmaltz, C. Neef, A. Thielmann, T. Reiss, *2d Mater* **2021**, 8, 022005.
[44] A. Kovtun, E. Treossi, N. Mirotta, A. Scidà, A. Liscio, M. Christian, F. Valorosi, A. Boschi, R. J. Young, C. Galiotis, I. A. Kinloch, V. Morandi, V. Palermo, *2d Mater* **2019**, 6.
[45] P. Cataldi, P. Steiner, T. Raine, K. Lin, C. Kocabas, R. J. Young, M. Bissett, I. A. Kinloch, D. G. Papageorgiou, *ACS Applied Polymer Materials* **2020**, 2, 3525.
[46] P. Cataldi, A. Athanassiou, I. Bayer, *Applied Sciences* **2018**, 8, 1438.
[47] D. Verma, K. L. Goh, in *Functionalized Graphene Nanocomposites and their Derivatives*, DOI: https://doi.org/10.1016/B978-0-12-814548-7.00011-8 (Eds: M. Jawaid, R. Bouhfid, A. e. Kacem Qaiss), Elsevier **2019**, p. 219.
[48] S. Stankovich, D. A. Dikin, R. D. Piner, K. A. Kohlhaas, A. Kleinhammes, Y. Jia, Y. Wu, S. T. Nguyen, R. S. Ruoff, *Carbon* **2007**, 45, 1558.
[49] Y. Zhu, H. Ji, H.-M. Cheng, R. S. Ruoff, *National Science Review* **2018**, 5, 90.
[50] D. C. Marcano, D. V. Kosynkin, J. M. Berlin, A. Sinitskii, Z. Sun, A. Slesarev, L. B. Alemany, W. Lu, J. M. Tour, *Acs Nano* **2010**, 4, 4806.
[51] J. Cao, P. He, M. A. Mohammed, X. Zhao, R. J. Young, B. Derby, I. A. Kinloch, R. A. W. Dryfe, *Journal of the American Chemical Society* **2017**, 139, 17446.
[52] D. Beloin-Saint-Pierre, R. Hischier, *The International Journal of Life Cycle Assessment* **2021**, 26, 327.
[53] K. De Silva, H.-H. Huang, R. Joshi, M. Yoshimura, *Carbon* **2017**, 119, 190.
[54] M. Aunkor, I. Mahbubul, R. Saidur, H. Metselaar, *Rsc Advances* **2016**, 6, 27807.
[55] C. J. Zhang, B. Anasori, A. Seral-Ascaso, S. H. Park, N. McEvoy, A. Shmeliov, G. S. Duesberg, J. N. Coleman, Y. Gogotsi, V. Nicolosi, *Adv Mater* **2017**, 29, 1702678.
[56] B. Anasori, Y. Gogotsi, *2D Metal Carbides and Nitrides (MXenes) Structure, Properties and Applications*, Springer International Publishing, **2019**.
[57] M. Naguib, M. Kurtoglu, V. Presser, J. Lu, J. Niu, M. Heon, L. Hultman, Y. Gogotsi, M. W. Barsoum, *Adv Mater* **2011**, 23, 4248.
[58] M. W. Barsoum, *MAX Phases : Properties of Machinable Ternary Carbides and Nitrides*, John Wiley & Sons, Incorporated, Weinheim, GERMANY **2013**.
[59] L. Liu, M. Orbay, S. Luo, S. Duluard, H. Shao, J. Harmel, P. Rozier, P.-L. Taberna, P. Simon, *Acs Nano* **2021**, DOI: 10.1021/acsnano.1c08498.
[60] C. Wang, H. Shou, S. Chen, S. Wei, Y. Lin, P. Zhang, Z. Liu, K. Zhu, X. Guo, X. Wu, P. M. Ajayan, L. Song, *Adv Mater* **2021**, 33, 2101015.

[61] A. Jawaid, A. Hassan, G. Neher, D. Nepal, R. Pachter, W. J. Kennedy, S. Ramakrishnan, R. A. Vaia, *Acs Nano* **2021**, 15, 2771.
[62] W. Sun, S. A. Shah, Y. Chen, Z. Tan, H. Gao, T. Habib, M. Radovic, M. J. Green, *J Mater Chem A* **2017**, 5, 21663.
[63] C. E. Shuck, A. Sarycheva, M. Anayee, A. Levitt, Y. Zhu, S. Uzun, V. Balitskiy, V. Zahorodna, O. Gogotsi, Y. Gogotsi, *Adv Eng Mater* **2020**, 22.
[64] O. Mashtalir, K. M. Cook, V. N. Mochalin, M. Crowe, M. W. Barsoum, Y. Gogotsi, *J Mater Chem A* **2014**, 2, 14334.
[65] C. J. Zhang, S. Pinilla, N. McEvoy, C. P. Cullen, B. Anasori, E. Long, S.-H. Park, A. Seral-Ascaso, A. Shmeliov, D. Krishnan, C. Morant, X. Liu, G. S. Duesberg, Y. Gogotsi, V. Nicolosi, *Chem Mater* **2017**, 29, 4848.
[66] S. Yang, P. Zhang, F. Wang, A. G. Ricciardulli, M. R. Lohe, P. W. Blom, X. Feng, *Angewandte Chemie* **2018**, 130, 15717.
[67] S. Abdolhosseinzadeh, R. Schneider, A. Verma, J. Heier, F. Nüesch, C. Zhang, *Advanced Materials* **2020**, 32, 2000716.
[68] S. Abdolhosseinzadeh, J. Heier, C. Zhang, *Journal of Physics: Energy* **2020**, 2, 031004.
[69] P. Cataldi, F. Bonaccorso, A. Esau del Rio Castillo, V. Pellegrini, Z. Jiang, L. Liu, N. Boccardo, M. Canepa, R. Cingolani, A. Athanassiou, I. S. Bayer, *Advanced Electronic Materials* **2016**, 2, 1600245.
[70] P. Cataldi, O. Condurache, D. Spirito, R. Krahne, I. S. Bayer, A. Athanassiou, G. Perotto, *Acs Sustain Chem Eng* **2019**, 7, 12544.
[71] K. Raagulan, R. Braveenth, H. J. Jang, Y. Seon Lee, C. M. Yang, B. Mi Kim, J. J. Moon, K. Y. Chai, *Materials (Basel)* **2018**, 11, 1803.
[72] C. Couly, M. Alhabeb, K. L. Van Aken, N. Kurra, L. Gomes, A. M. Navarro-Suárez, B. Anasori, H. N. Alshareef, Y. Gogotsi, *Advanced Electronic Materials* **2018**, 4, 1700339.
[73] H. Li, Y. Hou, F. Wang, M. R. Lohe, X. Zhuang, L. Niu, X. Feng, *Adv Energy Mater* **2017**, 7, 1601847.
[74] Y. Li, X. Tian, S. P. Gao, L. Jing, K. Li, H. Yang, F. Fu, J. Y. Lee, Y. X. Guo, J. S. Ho, P. Y. Chen, *Adv Funct Mater* **2019**, 30, 1907451.
[75] B. Fadeel, C. Bussy, S. Merino, E. Vázquez, E. Flahaut, F. Mouchet, L. Evariste, L. Gauthier, A. J. Koivisto, U. Vogel, *ACS nano* **2018**, 12, 10582.
[76] G. P. Kotchey, B. L. Allen, H. Vedala, N. Yanamala, A. A. Kapralov, Y. Y. Tyurina, J. Klein-Seetharaman, V. E. Kagan, A. Star, *ACS nano* **2011**, 5, 2098.
[77] R. Kurapati, F. Bonachera, J. Russier, A. R. Sureshbabu, C. Ménard-Moyon, K. Kostarelos, A. Bianco, *2d Mater* **2017**, 5, 015020.
[78] R. Kurapati, J. Russier, M. A. Squillaci, E. Treossi, C. Ménard-Moyon, A. E. Del Rio-Castillo, E. Vazquez, P. Samorì, V. Palermo, A. Bianco, *Small* **2015**, 11, 3985.
[79] G. Lalwani, W. Xing, B. Sitharaman, *Journal of Materials Chemistry B* **2014**, 2, 6354.
[80] J. Chen, Q. Huang, H. Huang, L. Mao, M. Liu, X. Zhang, Y. Wei, *Nanoscale* **2020**, 12, 3574.
[81] I. Miccoli, F. Edler, H. Pfnur, C. Tegenkamp, *J Phys Condens Matter* **2015**, 27, 223201.
[82] S. Yilmaz, *Journal of Semiconductors* **2015**, 36, 082001.
[83] H. Topsoe, in *Bulletin*, Vol. 472-13, Haldor Topsoe Semiconductor Division, 1968.
[84] P. Cataldi, I. S. Bayer, F. Bonaccorso, V. Pellegrini, A. Athanassiou, R. Cingolani, *Advanced Electronic Materials* **2015**, 1, 1500224.
[85] P. Cataldi, L. Ceseracciu, A. Athanassiou, I. S. Bayer, *ACS Appl Mater Interfaces* **2017**, 9, 13825.
[86] J. E. Mates, I. S. Bayer, M. Salerno, P. J. Carroll, Z. Jiang, L. Liu, C. M. Megaridis, *Carbon* **2015**, 87, 163.

[87] X. Yang, X.-M. Li, Q.-Q. Kong, Z. Liu, J.-P. Chen, H. Jia, Y.-Z. Liu, L.-J. Xie, C.-M. Chen, *Sci China Mater* **2019**, 63, 392.
[88] M. Zahid, E. L. Papadopoulou, A. Athanassiou, I. S. Bayer, *Mater Design* **2017**, 135, 213.
[89] H. U. Lee, S. W. Kim, *J Mater Chem A* **2017**, 5, 13581.
[90] C. Cho, N. Bittner, W. Choi, J. H. Hsu, C. Yu, J. C. Grunlan, *Advanced Electronic Materials* **2018**, 5, 1800465.
[91] E. J. Ward, J. Lacey, C. Crua, M. K. Dymond, K. Maleski, K. Hantanasirisakul, Y. Gogotsi, S. Sandeman, *Adv Funct Mater* **2020**, DOI: 10.1002/adfm.2020008412000841.
[92] S. Gupta, C. Chang, A. K. Anbalagan, C.-H. Lee, N.-H. Tai, *Compos Sci Technol* **2020**, 188, 107994.
[93] N. Kumar, R. T. Ginting, J.-W. Kang, *Electrochim Acta* **2018**, 270, 37.
[94] K. AlHassoon, M. Han, Y. Malallah, V. Ananthakrishnan, R. Rakhmanov, W. Reil, Y. Gogotsi, A. S. Daryoush, *Appl Phys Lett* **2020**, 116, 184101.
[95] A. Sandwell, Z. Kockerbeck, C. I. Park, M. Hassani, R. Hugo, S. S. Park, *Flexible Print Electron* **2020**, 5, 015005.
[96] T. Habib, X. Zhao, S. A. Shah, Y. Chen, W. Sun, H. An, J. L. Lutkenhaus, M. Radovic, M. J. Green, *npj 2D Materials and Applications* **2019**, 3.
[97] Y. Sun, L. Ma, Y. Song, A. D. Phule, L. Li, Z. X. Zhang, *European Polymer Journal* **2021**, 147.
[98] A. Rinaldi, A. Tamburrano, M. Fortunato, M. S. Sarto, *Sensors (Basel)* **2016**, 16, 2148.
[99] J. Yang, Y. Ye, X. Li, X. Lü, R. Chen, *Compos Sci Technol* **2018**, 164, 187.
[100] C. S. Boland, U. Khan, M. Binions, S. Barwich, J. B. Boland, D. Weaire, J. N. Coleman, *Nanoscale* **2018**, 10, 5366.
[101] B. Shen, Y. Li, W. Zhai, W. Zheng, *ACS Appl Mater Interfaces* **2016**, 8, 8050.
[102] Z. Ma, A. Wei, J. Ma, L. Shao, H. Jiang, D. Dong, Z. Ji, Q. Wang, S. Kang, *Nanoscale* **2018**, 10, 7116.
[103] A. Tewari, S. Gandla, S. Bohm, C. R. McNeill, D. Gupta, *ACS Appl Mater Interfaces* **2018**, 10, 5185.
[104] X. Dong, Y. Wei, S. Chen, Y. Lin, L. Liu, J. Li, *Compos Sci Technol* **2018**, 155, 108.
[105] Z. Xu, M. Liang, X. He, Q. Long, J. Yu, K. Xie, L. Liao, *Compos Part Appl Sci Manuf* **2019**, 119, 111.
[106] J. Xu, J. Chen, M. Zhang, J.-D. Hong, G. Shi, *Advanced Electronic Materials* **2016**, 2, 1600022.
[107] S. Luo, Y. Wang, G. Wang, F. Liu, Y. Zhai, Y. Luo, *Carbon* **2018**, 139, 437.
[108] G. Jia, J. Plentz, M. Presselt, J. Dellith, A. Dellith, S. Patze, F. J. Tölle, R. Mülhaupt, G. Andrä, F. Falk, B. Dietzek, *Adv Mater Interfaces* **2017**, 4, 1700758.
[109] Q. Chen, F. Zabihi, M. Eslamian, *Synthetic Met* **2016**, 222, 309.
[110] D. S. Saidina, S. A. Zubir, S. Fontana, C. Hérold, M. Mariatti, *Journal of Electronic Materials* **2019**, 48, 5757.
[111] H. Zhao, M. Yang, D. He, Y. Liu, J. Bai, J. Bai, *Mater Res Express* **2019**, 6, 0850g1.
[112] Y. J. Yun, J. Ju, J. H. Lee, S.-H. Moon, S.-J. Park, Y. H. Kim, W. G. Hong, D. H. Ha, H. Jang, G. H. Lee, H.-M. Chung, J. Choi, S. W. Nam, S.-H. Lee, Y. Jun, *Adv Funct Mater* **2017**, 27, 1701513.
[113] T. Tomašević-Ilić, J. Pešić, I. Milošević, J. Vujin, A. Matković, M. Spasenović, R. Gajić, *Opt Quant Electron* **2016**, 48, 319.
[114] C. Vallés, X. Zhang, J. Cao, F. Lin, R. J. Young, A. Lombardo, A. C. Ferrari, L. Burk, R. Mülhaupt, I. A. Kinloch, *Acs Appl Nano Mater* **2019**, 2, 5272.
[115] C.-H. Huang, Y.-Y. Wang, T.-H. Lu, Y.-C. Li, *Polymers-basel* **2017**, 9, 28.

[116] C. Cho, M. Culebras, K. L. Wallace, Y. Song, K. Holder, J.-H. Hsu, C. Yu, J. C. Grunlan, *Nano Energy* **2016**, 28, 426.
[117] C. Cho, B. Stevens, J. H. Hsu, R. Bureau, D. A. Hagen, O. Regev, C. Yu, J. C. Grunlan, *Adv Mater* **2015**, 27, 2996.
[118] A. Aliprandi, T. Moreira, C. Anichini, M. A. Stoeckel, M. Eredia, U. Sassi, M. Bruna, C. Pinheiro, C. A. T. Laia, S. Bonacchi, P. Samori, *Adv Mater* **2017**, 29.
[119] W.-H. Khoh, B.-H. Wee, J.-D. Hong, *Colloids Surfaces Physicochem Eng Aspects* **2019**, 581, 123815.
[120] C.-C. Huang, S. Gupta, C.-Y. Lo, N.-H. Tai, *Mater Lett* **2019**, 253, 152.
[121] B. Sharma, J.-S. Kim, A. Sharma, *Mater Res Express* **2019**, 6, 066312.
[122] L. Liu, Z. Shen, X. Zhang, H. Ma, *J Colloid Interface Sci* **2021**, 582, 12.
[123] T. Wang, L.-C. Jing, Q. Zhu, A. Sagadevan Ethiraj, Y. Tian, H. Zhao, X.-T. Yuan, J.-G. Wen, L.-K. Li, H.-Z. Geng, *Appl Surf Sci* **2020**, 500, 143997.
[124] X. Fang, Z. Fan, Y. Gu, J. Shi, M. Chen, X. Chen, S. Qiu, F. Zabihi, M. Eslamian, Q. Chen, *J Shanghai Jiaotong Univ Sci* **2018**, 23, 106.
[125] Q. Chang, L. Li, L. Sai, W. Shi, L. Huang, *Advanced Electronic Materials* **2018**, 4, 1800059.
[126] A. Chhetry, M. Sharifuzzaman, H. Yoon, S. Sharma, X. Xuan, J. Y. Park, *Acs Appl Mater Inter* **2019**, 11, 22531.
[127] C. Huang, L. Kang, J. Zhang, J. Li, S. Wan, N. Zhang, H. Xu, C. Wang, Y. Yu, C. Luo, X. Wu, *Acs Appl Energy Mater* **2018**, 1, 7182.
[128] V. Kumar, S. Forsberg, A.-C. Engström, M. Nurmi, B. Andres, C. Dahlström, M. Toivakka, *Flexible Print Electron* **2017**, 2, 035002.
[129] X. Wu, P. Steiner, T. Raine, G. Pinter, A. Kretinin, C. Kocabas, M. Bissett, P. Cataldi, *Advanced Electronic Materials* **2020**, 6, 2000232.
[130] J.-K. Chih, A. Jamaluddin, F. Chen, J.-K. Chang, C.-Y. Su, *J Mater Chem A* **2019**, 7, 12779.
[131] S. Lv, F. Fu, S. Wang, J. Huang, L. Hu, *Electron Mater Lett* **2015**, 11, 633.
[132] M. Akter Shathi, C. Minzhi, N. A. Khoso, H. Deb, A. Ahmed, W. Sai Sai, *Synthetic Met* **2020**, 263, 116329.
[133] J. Zhang, Y. Cao, M. Qiao, L. Ai, K. Sun, Q. Mi, S. Zang, Y. Zuo, X. Yuan, Q. Wang, *Sensors and Actuators A: Physical* **2018**, 274, 132.
[134] S. Zang, Q. Wang, Q. Mi, J. Zhang, X. Ren, *Sensors and Actuators A: Physical* **2017**, 267, 532.
[135] H. Kim, S. Lee, *Fiber Polym* **2018**, 19, 2351.
[136] Y. Wang, S. Tang, S. Vongehr, J. A. Syed, X. Wang, X. Meng, *Sci Rep* **2016**, 6, 12883.
[137] P. Cataldi, J. A. Heredia-Guerrero, S. Guzman-Puyol, L. Ceseracciu, L. La Notte, A. Reale, J. Ren, Y. Zhang, L. Liu, M. Miscuglio, P. Savi, S. Piazza, M. Duocastella, G. Perotto, A. Athanassiou, I. S. Bayer, *Adv Sustain Syst* **2018**, 2, 1800069.
[138] S. Afroj, S. Tan, A. M. Abdelkader, K. S. Novoselov, N. Karim, *Adv Funct Mater* **2020**, 30, 2000293.
[139] X. Hu, M. Tian, L. Qu, S. Zhu, G. Han, *Carbon* **2015**, 95, 625.
[140] Y. Zheng, Y. Li, Y. Zhou, K. Dai, G. Zheng, B. Zhang, C. Liu, C. Shen, *ACS Appl Mater Interfaces* **2020**, 12, 1474.
[141] M. Tian, M. Du, L. Qu, S. Chen, S. Zhu, G. Han, *Rsc Adv* **2017**, 7, 42641.
[142] A. M. Abdelkader, N. Karim, C. Vallés, S. Afroj, K. S. Novoselov, S. G. Yeates, *2d Mater* **2017**, 4, 035016.
[143] N. Karim, S. Afroj, S. Tan, P. He, A. Fernando, C. Carr, K. S. Novoselov, *Acs Nano* **2017**, 11, 12266.
[144] P. Zhang, H. Zhang, C. Yan, Z. Zheng, Y. Yu, *Mater Res Express* **2017**, 4, 075602.

[145] G. Cai, Z. Xu, M. Yang, B. Tang, X. Wang, *Appl Surf Sci* **2017**, 393, 441.
[146] A. Chatterjee, M. Nivas Kumar, S. Maity, *J Text Inst* **2017**, 108, 1910.
[147] S. He, B. Xin, Z. Chen, Y. Liu, *Text Res J* **2018**, 89, 1038.
[148] L.-L. Xu, M.-X. Guo, S. Liu, S.-W. Bian, *Rsc Adv* **2015**, 5, 25244.
[149] J. Xu, D. Wang, Y. Yuan, W. Wei, L. Duan, L. Wang, H. Bao, W. Xu, *Org Electron* **2015**, 24, 153.
[150] Y. Liang, W. Weng, J. Yang, L. Liu, Y. Zhang, L. Yang, X. Luo, Y. Cheng, M. Zhu, *RSC Adv.* **2017**, 7, 48934.
[151] M. Tian, X. Hu, L. Qu, S. Zhu, Y. Sun, G. Han, *Carbon* **2016**, 96, 1166.
[152] A. Ahmed, M. A. Jalil, M. M. Hossain, M. Moniruzzaman, B. Adak, M. T. Islam, M. S. Parvez, S. Mukhopadhyay, *J Mater Chem C* **2020**, 8, 16204.
[153] R. Moriche, A. Jiménez-Suárez, M. Sánchez, S. G. Prolongo, A. Ureña, *Compos Sci Technol* **2017**, 146, 59.
[154] V. B. Mohan, D. Bhattacharyya, *Int J Smart Nano Mater* **2020**, 11, 78.
[155] V. B. Mohan, K. Jayaraman, D. Bhattacharyya, *Adv Polym Tech* **2018**, 37, 3438.
[156] Y. F. Fu, Y. Q. Li, Y. F. Liu, P. Huang, N. Hu, S. Y. Fu, *ACS Appl Mater Interfaces* **2018**, 10, 35503.
[157] E. Barjasteh, C. Sutanto, D. Nepal, *Langmuir* **2019**, 35, 2261.
[158] M. Tian, R. Zhao, L. Qu, Z. Chen, S. Chen, S. Zhu, W. Song, X. Zhang, Y. Sun, R. Fu, *Macromol Mater Eng* **2019**, 304, 1900244.
[159] M. K. Yapici, T. E. Alkhidir, *Sensors (Basel)* **2017**, 17, 875.
[160] M. K. Yapici, T. Alkhidir, Y. A. Samad, K. Liao, *Sensors and Actuators B: Chemical* **2015**, 221, 1469.
[161] G. Cai, M. Yang, Z. Xu, J. Liu, B. Tang, X. Wang, *Chem Eng J* **2017**, 325, 396.
[162] C. Zhao, K. Shu, C. Wang, S. Gambhir, G. G. Wallace, *Electrochim Acta* **2015**, 172, 12.
[163] C. Wang, C. Xiang, L. Tan, J. Lan, L. Peng, S. Jiang, R. Guo, *Rsc Adv* **2017**, 7, 40452.
[164] S. J. Woltornist, F. A. Alamer, A. McDannald, M. Jain, G. A. Sotzing, D. H. Adamson, *Carbon* **2015**, 81, 38.
[165] F. Sun, M. Tian, X. Sun, T. Xu, X. Liu, S. Zhu, X. Zhang, L. Qu, *Nano Lett* **2019**, 19, 6592.
[166] X. Li, H. Hu, T. Hua, B. Xu, S. Jiang, *Nano Res* **2018**, 11, 5799.
[167] S.-T. Hsiao, C.-C. M. Ma, H.-W. Tien, W.-H. Liao, Y.-S. Wang, S.-M. Li, W.-P. Chuang, *Compos Sci Technol* **2015**, 118, 171.
[168] Q. Mi, Q. Wang, S. Zang, G. Mao, J. Zhang, X. Ren, *Smart Mater Struct* **2018**, 27, 015014.
[169] G. Manasoglu, R. Celen, M. Kanik, Y. Ulcay, *J Appl Polym Sci* **2019**, 136, 48024.
[170] Y. Lu, M. Tian, X. Sun, N. Pan, F. Chen, S. Zhu, X. Zhang, S. Chen, *Compos Part Appl Sci Manuf* **2019**, 117, 202.
[171] B. Shen, Y. Li, D. Yi, W. Zhai, X. Wei, W. Zheng, *Carbon* **2017**, 113, 55.
[172] L. Zulan, L. Zhi, C. Lan, C. Sihao, W. Dayang, D. Fangyin, *Advanced Electronic Materials* **2019**, 5, 1800648.
[173] Z. Lu, C. Mao, H. Zhang, *J Mater Chem C* **2015**, 3, 4265.
[174] J. Cao, C. Wang, *Appl Surf Sci* **2017**, 405, 380.
[175] Y. Ji, Y. Li, G. Chen, T. Xing, *Mater Design* **2017**, 133, 528.
[176] H. Montazerian, A. Rashidi, A. Dalili, H. Najjaran, A. S. Milani, M. Hoorfar, *Small* **2019**, 15, 1804991.
[177] H. Liu, Q. Li, Y. Bu, N. Zhang, C. Wang, C. Pan, L. Mi, Z. Guo, C. Liu, C. Shen, *Nano Energy* **2019**, 66, 104143.
[178] L. Liao, D. Jiang, K. Zheng, M. Zhang, J. Liu, *Adv Funct Mater* **2021**, 31.

[179] D. Hu, X. Huang, S. Li, P. Jiang, *Compos Sci Technol* **2020**, 188, 107995.
[180] Q. Gao, Y. Pan, G. Zheng, C. Liu, C. Shen, X. Liu, *Advanced Composites and Hybrid Materials* **2021**, 4, 274.
[181] H. Hu, *J Mater Chem A* **2017**, 11.
[182] E. Quain, T. S. Mathis, N. Kurra, K. Maleski, K. L. Van Aken, M. Alhabeb, H. N. Alshareef, Y. Gogotsi, *Adv Mater Technologies* **2019**, 4, 1800256.
[183] N. Kurra, B. Ahmed, Y. Gogotsi, H. N. Alshareef, *Adv Energy Mater* **2016**, 6, 1601372.
[184] Z. Wang, S. Qin, S. Seyedin, J. Zhang, J. Wang, A. Levitt, N. Li, C. Haines, R. Ovalle-Robles, W. Lei, Y. Gogotsi, R. H. Baughman, J. M. Razal, *Small* **2018**, 14, e1802225.
[185] K. Raagulan, R. Braveenth, H. J. Jang, Y. S. Lee, C. M. Yang, B. M. Kim, J. J. Moon, K. Y. Chai, *B Korean Chem Soc* **2018**, 39, 1412.
[186] K. Raagulan, R. Braveenth, L. R. Lee, J. Lee, B. M. Kim, J. J. Moon, S. B. Lee, K. Y. Chai, *Nanomaterials-basel* **2019**, 9, 519.
[187] S. Uzun, S. Seyedin, A. L. Stoltzfus, A. S. Levitt, M. Alhabeb, M. Anayee, C. J. Strobel, J. M. Razal, G. Dion, Y. Gogotsi, *Adv Funct Mater* **2019**, 29, 1905015.
[188] W. Cheng, Y. Zhang, W. Tian, J. Liu, J. Lu, B. Wang, W. Xing, Y. Hu, *Ind Eng Chem Res* **2020**, 59, 14025.
[189] X. Zhang, X. Wang, Z. Lei, L. Wang, M. Tian, S. Zhu, H. Xiao, X. Tang, L. Qu, *ACS Appl Mater Interfaces* **2020**, 12, 14459.
[190] B. Wang, X. Lai, H. Li, C. Jiang, J. Gao, X. Zeng, *ACS Appl Mater Interfaces* **2021**, 13, 23020.
[191] C. Deng, S. Zhao, E. Su, Y. Li, F. Wu, *Adv Mater Technologies* **2021**, 6.
[192] H. An, T. Habib, S. Shah, H. Gao, M. Radovic, M. J. Green, J. L. Lutkenhaus, *Sci Adv* **2018**, 4, eaaq0118.
[193] T. H. Park, S. Yu, M. Koo, H. Kim, E. H. Kim, J. E. Park, B. Ok, B. Kim, S. H. Noh, C. Park, E. Kim, C. M. Koo, C. Park, *Acs Nano* **2019**, 13, 6835.
[194] Q.-W. Wang, H.-B. Zhang, J. Liu, S. Zhao, X. Xie, L. Liu, R. Yang, N. Koratkar, Z.-Z. Yu, *Adv Funct Mater* **2019**, 29, 1806819.
[195] L. X. Liu, W. Chen, H. B. Zhang, Q. W. Wang, F. Guan, Z. Z. Yu, *Adv Funct Mater* **2019**, 29, 1905197.
[196] S. Huang, Y. Liu, Y. Zhao, Z. Ren, C. F. Guo, *Adv Funct Mater* **2019**, 29, 1805924.
[197] M. Amjadi, K.-U. Kyung, I. Park, M. Sitti, *Adv Funct Mater* **2016**, 26, 1678.
[198] A. Qiu, P. Li, Z. Yang, Y. Yao, I. Lee, J. Ma, *Adv Funct Mater* **2019**, 29, 1806306.
[199] H. Souri, H. Banerjee, A. Jusufi, N. Radacsi, A. A. Stokes, I. Park, M. Sitti, M. Amjadi, *Advanced Intelligent Systems* **2020**, 2, 2000039.
[200] X. Shi, S. Liu, Y. Sun, J. Liang, Y. Chen, *Adv Funct Mater* **2018**, 28, 1800850.
[201] Y. Cai, J. Shen, G. Ge, Y. Zhang, W. Jin, W. Huang, J. Shao, J. Yang, X. Dong, *Acs Nano* **2018**, 12, 56.
[202] H. Souri, D. Bhattacharyya, *J Mater Chem C* **2018**, 6, 10524.
[203] S. Chun, S. B. Cho, W. Son, Y. Kim, H. Jung, Y.-J. Kim, C. Choi, *Nanotechnology* **2020**, 31, 085303.
[204] H. Lee, M. J. Kim, J. H. Kim, J.-Y. Lee, E. Ji, A. Capasso, H.-J. Choi, W. Shim, G.-H. Lee, *Mater Res Express* **2020**, 7, 045603.
[205] T. Sakorikar, M. K. Kavitha, P. Vayalamkuzhi, M. Jaiswal, *Sci Rep-uk* **2017**, 7, 2598.
[206] S. Lu, J. Ma, K. Ma, X. Wang, S. Wang, X. Yang, H. Tang, *Appl Phys* **2019**, 125, 471.
[207] B. Yin, Y. Wen, T. Hong, Z. Xie, G. Yuan, Q. Ji, H. Jia, *ACS Appl Mater Interfaces* **2017**, 9, 32054.
[208] H. Souri, D. Bhattacharyya, *ACS Appl Mater Interfaces* **2018**, 10, 20845.
[209] W. Yuan, Q. Zhou, Y. Li, G. Shi, *Nanoscale* **2015**, 7, 16361.
[210] Y. Huang, L. Gao, Y. Zhao, X. Guo, C. Liu, P. Liu, *J Appl Polym Sci* **2017**, 134, 45340.

[211] W. Yuan, J. Yang, K. Yang, H. Peng, F. Yin, *ACS Appl Mater Interfaces* **2018**, 10, 19906.
[212] D. Du, P. Li, J. Ouyang, *J Mater Chem C* **2016**, 4, 3224.
[213] Y. Huang, Y. Zhao, Y. Wang, X. Guo, Y. Zhang, P. Liu, C. Liu, Y. Zhang, *Smart Mater Struct* **2018**, 27, 035013.
[214] D. Kang, Y.-E. Shin, H. J. Jo, H. Ko, H. S. Shin, *Part Amp Part Syst Charact* **2017**, 34, 1600382.
[215] M. Bai, Y. Zhai, F. Liu, Y. Wang, S. Luo, *Sci Rep* **2019**, 9, 18644.
[216] Y. Zhao, Y. Huang, W. Hu, X. Guo, Y. Wang, P. Liu, C. Liu, Y. Zhang, *Smart Mater Struct* **2019**, 28, 035004.
[217] X. P. Li, Y. Li, X. Li, D. Song, P. Min, C. Hu, H. B. Zhang, N. Koratkar, Z. Z. Yu, *J Colloid Interface Sci* **2019**, 542, 54.
[218] H. Wang, R. Zhou, D. Li, L. Zhang, G. Ren, L. Wang, J. Liu, D. Wang, Z. Tang, G. Lu, G. Sun, H. D. Yu, W. Huang, *Acs Nano* **2021**, 15, 9690.
[219] Y. Ma, N. Liu, L. Li, X. Hu, Z. Zou, J. Wang, S. Luo, Y. Gao, *Nat Commun* **2017**, 8, 1207.
[220] T. Li, L. Chen, X. Yang, X. Chen, Z. Zhang, T. Zhao, X. Li, J. Zhang, *J Mater Chem C* **2019**, 7, 1022.
[221] G. Ge, Y. Cai, Q. Dong, Y. Zhang, J. Shao, W. Huang, X. Dong, *Nanoscale* **2018**, 10, 10033.
[222] Y. Guo, M. Zhong, Z. Fang, P. Wan, G. Yu, *Nano Lett* **2019**, 19, 1143.
[223] Y. Zang, F. Zhang, C.-a. Di, D. Zhu, *Mater Horizons* **2015**, 2, 140.
[224] S. Chun, H. Jung, Y. Choi, G. Bae, J. P. Kil, W. Park, *Carbon* **2015**, 94, 982.
[225] D.-J. Yao, Z. Tang, L. Zhang, Z.-G. Liu, Q.-J. Sun, S.-C. Hu, Q.-X. Liu, X.-G. Tang, J. Ouyang, *J Mater Chem C* **2021**, 9, 12642.
[226] Y. Zheng, R. Yin, Y. Zhao, H. Liu, D. Zhang, X. Shi, B. Zhang, C. Liu, C. Shen, *Chem Eng J* **2021**, 420.
[227] P. M. Biesheuvel, S. Porada, J. E. Dykstra, arXiv:1809.02930 [physics.chem-ph] 2021.
[228] B. E. Conway, *Electrochemical Supercapacitors*, Springer, Boston, MA **1999**.
[229] P. Simon, Y. Gogotsi, B. Dunn, *Science* **2014**, 343, 1210.
[230] M. R. Lukatskaya, S. Kota, Z. Lin, M.-Q. Zhao, N. Shpigel, M. D. Levi, J. Halim, P.-L. Taberna, M. W. Barsoum, P. Simon, Y. Gogotsi, *Nature Energy* **2017**, 2, 17105.
[231] Y. Liu, X. Peng, *Applied Materials Today* **2017**, 8, 104.
[232] J. Xia, F. Chen, J. Li, N. Tao, *Nat Nanotechnol* **2009**, 4, 505.
[233] F. Beguin, E. Frackowiak, *Supercapacitors: Materials, Systems, and Applications*, **2013**.
[234] C. Costentin, J.-M. Savéant, *Chemical Science* **2019**, 10, 5656.
[235] C. Zhan, M. Naguib, M. Lukatskaya, P. R. C. Kent, Y. Gogotsi, D.-e. Jiang, *The Journal of Physical Chemistry Letters* **2018**, 9, 1223.
[236] M. R. Lukatskaya, S.-M. Bak, X. Yu, X.-Q. Yang, M. W. Barsoum, Y. Gogotsi, *Adv Energy Mater* **2015**, 5, 1500589.
[237] M. Hu, R. Cheng, Z. Li, T. Hu, H. Zhang, C. Shi, J. Yang, C. Cui, C. Zhang, H. Wang, B. Fan, X. Wang, Q.-H. Yang, *Nanoscale* **2020**, 12, 763.
[238] U. Gulzar, S. Goriparti, E. Miele, T. Li, G. Maidecchi, A. Toma, F. De Angelis, C. Capiglia, R. P. Zaccaria, *J Mater Chem A* **2016**, 4, 16771.
[239] P. Forouzandeh, S. C. Pillai, *Mater Today Proc* **2021**, 41, 498.
[240] T. S. Mathis, N. Kurra, X. Wang, D. Pinto, P. Simon, Y. Gogotsi, *Adv Energy Mater* **2019**, 9, 1902007.
[241] T. Christen, M. W. Carlen, *J Power Sources* **2000**, 91, 210.
[242] Q. Zhou, X. Ye, Z. Wan, C. Jia, *J Power Sources* **2015**, 296, 186.

[243] B. B. Etana, S. Ramakrishnan, M. Dhakshnamoorthy, S. Saravanan, P. C Ramamurthy, T. A. Demissie, *Mater Res Express* **2020**, 6, 125708.
[244] M. Li, R. Jijie, A. Barras, P. Roussel, S. Szunerits, R. Boukherroub, *Electrochim Acta* **2019**, 302, 1.
[245] M. Khalid, A. M. B. Honorato, *J Energy Chem* **2018**, 27, 866.
[246] C. Cheng, J. Xu, W. Gao, S. Jiang, R. Guo, *Electrochim Acta* **2019**, 318, 23.
[247] J. Kim, J. Yin, X. Xuan, J. Y. Park, *Micro Nano Syst Lett* **2019**, 7, 4.
[248] L. Liu, Y. Yu, C. Yan, K. Li, Z. Zheng, *Nat Commun* **2015**, 6, 7260.
[249] Q. Zhou, X. Chen, F. Su, X. Lyu, M. Miao, *Ind Eng Chem Res* **2020**, 59, 5752.
[250] J.-h. Liu, X.-y. Xu, J. Yu, J.-l. Hong, C. Liu, X. Ouyang, S. Lei, X. Meng, J.-N. Tang, D.-Z. Chen, *Electrochim Acta* **2019**, 314, 9.
[251] C. Jin, H.-T. Wang, Y.-N. Liu, X.-H. Kang, P. Liu, J.-N. Zhang, L.-N. Jin, S.-W. Bian, Q. Zhu, *Electrochim Acta* **2018**, 270, 205.
[252] I. K. Moon, B. Ki, J. Oh, *Chem Eng J* **2020**, 392, 123794.
[253] B. C. Kim, H. T. Jeong, C. J. Raj, Y.-R. Kim, B.-B. Cho, K. H. Yu, *Synthetic Met* **2015**, 207, 116.
[254] M. Z. Esfahani, M. Khosravi, *J Power Sources* **2020**, 462, 228166.
[255] Y. Xiao, L. Huang, Q. Zhang, S. Xu, Q. Chen, W. Shi, *Appl Phys Lett* **2015**, 107, 013906.
[256] Q. Zhang, L. Huang, Q. Chang, W. Shi, L. Shen, Q. Chen, *Nanotechnology* **2016**, 27, 105401.
[257] N. Wang, G. Han, Y. Xiao, Y. Li, H. Song, Y. Zhang, *Electrochim Acta* **2018**, 270, 490.
[258] F. Tehrani, M. Beltrán-Gastélum, K. Sheth, A. Karajic, L. Yin, R. Kumar, F. Soto, J. Kim, J. Wang, S. Barton, M. Mueller, J. Wang, *Adv Mater Technologies* **2019**, 4, 1900162.
[259] K. Li, X. Liu, S. Chen, W. Pan, J. Zhang, *J Energy Chem* **2019**, 32, 166.
[260] X. Shi, Z. S. Wu, J. Qin, S. Zheng, S. Wang, F. Zhou, C. Sun, X. Bao, *Adv Mater* **2017**, 29, 1703034.
[261] Z. Chen, W. Liao, X. Ni, *Chem Eng J* **2017**, 327, 1198.
[262] Z. Li, M. Tian, X. Sun, H. Zhao, S. Zhu, X. Zhang, *J Alloy Compd* **2019**, 782, 986.
[263] N. Wang, J. Liu, Y. Zhao, M. Hu, R. Qin, G. Shan, *Chemnanomat* **2019**, 5, 658.
[264] C. J. Zhang, M. P. Kremer, A. Seral-Ascaso, S.-H. Park, N. McEvoy, B. Anasori, Y. Gogotsi, V. Nicolosi, *Adv Funct Mater* **2018**, 28, 1705506.
[265] S. Abdolhosseinzadeh, J. Heier, C. Zhang, *ChemElectroChem* **2021**, 8, 1911.
[266] H. Huang, X. Chu, H. Su, H. Zhang, Y. Xie, W. Deng, N. Chen, F. Liu, H. Zhang, B. Gu, W. Deng, W. Yang, *J Power Sources* **2019**, 415, 1.
[267] S. Xu, Y. Dall'Agnese, G. Wei, C. Zhang, Y. Gogotsi, W. Han, *Nano Energy* **2018**, 50, 479.
[268] Y.-Y. Peng, B. Akuzum, N. Kurra, M.-Q. Zhao, M. Alhabeb, B. Anasori, E. C. Kumbur, H. N. Alshareef, M.-D. Ger, Y. Gogotsi, *Energ Environ Sci* **2016**, 9, 2847.
[269] M. Hu, T. Hu, R. Cheng, J. Yang, C. Cui, C. Zhang, X. Wang, *J Energy Chem* **2018**, 27, 161.
[270] L. Wang, D. Shao, J. Guo, S. Zhang, Y. Lu, *Energy Technol-ger* **2020**, 8, 1901003.
[271] A. Levitt, D. Hegh, P. Phillips, S. Uzun, M. Anayee, J. M. Razal, Y. Gogotsi, G. Dion, *Mater Today* **2020**, 34, 17.
[272] T. H. Chang, T. Zhang, H. Yang, K. Li, Y. Tian, J. Y. Lee, P. Y. Chen, *Acs Nano* **2018**, 12, 8048.
[273] I. Johnson, W. T. Choate, A. Davidson, BCS, Inc., Laurel, MD (United States), 2008.
[274] M. Papapetrou, G. Kosmadakis, A. Cipollina, U. La Commare, G. Micale, *Applied Thermal Engineering* **2018**, 138, 207.

[275] A. F. Ioffe, L. S. Stil'bans, E. K. Iordanishvili, T. S. Stavitskaya, A. Gelbtuch, G. Vineyard, *Physics Today* **1959**, 12, 42.
[276] C. Forman, I. K. Muritala, R. Pardemann, B. Meyer, *Renewable and Sustainable Energy Reviews* **2016**, 57, 1568.
[277] L. Zhang, X.-L. Shi, Y.-L. Yang, Z.-G. Chen, *Mater Today* **2021**.
[278] Y. Wang, T. Guo, Z. Tian, K. Bibi, Y. Z. Zhang, H. N. Alshareef, *Advanced Materials* **2022**, 2108560.
[279] R. Amirabad, A. R. Saadatabadi, M. H. Siadati, *Materials for Renewable and Sustainable Energy* **2020**, 9, 1.
[280] S. Mardi, P. Cataldi, A. Athanassiou, A. Reale, *Applied Physics Letters* **2022**, 120, 033102.
[281] L.-C. Jia, C.-G. Zhou, W.-J. Sun, L. Xu, D.-X. Yan, Z.-M. Li, *Chem Eng J* **2020**, 384, 123368.
[282] Z. Li, Z. Wang, W. Lu, B. Hou, *Metals* **2018**, 8, 652.
[283] D. Chung, *Materials Chemistry and Physics* **2020**, 123587.
[284] F. M. Oliveira, R. Gusmao, *Acs Appl Electron Mater* **2020**, 2, 3048.
[285] X. Wu, B. Han, H.-B. Zhang, X. Xie, T. Tu, Y. Zhang, Y. Dai, R. Yang, Z.-Z. Yu, *Chem Eng J* **2020**, 381, 122622.
[286] Y. Wang, H.-K. Peng, T.-T. Li, B.-C. Shiu, H.-T. Ren, X. Zhang, C.-W. Lou, J.-H. Lin, *Chem Eng J* **2021**, 412, 128681.
[287] Y. Wang, W. Wang, R. Xu, M. Zhu, D. Yu, *Chem Eng J* **2019**, 360, 817.
[288] S. Uzun, M. Han, C. J. Strobel, K. Hantanasirisakul, A. Goad, G. Dion, Y. Gogotsi, *Carbon* **2021**, 174, 382.
[289] Z. Nooralian, M. Parvinzadeh Gashti, I. Ebrahimi, *Rsc Adv* **2016**, 6, 23288.
[290] S. Ghosh, S. Ganguly, P. Das, T. K. Das, M. Bose, N. K. Singha, A. K. Das, N. C. Das, *Fiber Polym* **2019**, 20, 1161.
[291] L. Zou, S. Zhang, X. Li, C. Lan, Y. Qiu, Y. Ma, *Adv Mater Interfaces* **2016**, 3, 1500476.
[292] L. Geng, P. Zhu, Y. Wei, R. Guo, C. Xiang, C. Cui, Y. Li, *Cellulose* **2019**, 26, 2833.
[293] J.-G. Lee, J.-H. Lee, S. An, D.-Y. Kim, T.-G. Kim, S. S. Al-Deyab, A. L. Yarin, S. S. Yoon, *J Mater Chem A* **2017**, 5, 6677.
[294] Q. Zhang, Y. Yu, K. Yang, B. Zhang, K. Zhao, G. Xiong, X. Zhang, *Carbon* **2017**, 124, 296.
[295] O. Redondo, S. Prolongo, M. Campo, C. Sbarufatti, M. Giglio, *Compos Sci Technol* **2018**, 164, 65.
[296] B. Zhou, X. Han, L. Li, Y. Feng, T. Fang, G. Zheng, B. Wang, K. Dai, C. Liu, C. Shen, *Compos Sci Technol* **2019**, 183, 107796.
[297] J. Park, *Polymers-basel* **2020**, 12, 189.
[298] P. Yang, T. Xia, S. Ghosh, J. Wang, S. D. Rawson, P. J. Withers, I. Kinloch, S. Barg, *2d Mater*.
[299] T. Xia, D. Zeng, Z. Li, R. J. Young, C. Vallés, I. A. Kinloch, *Compos Sci Technol* **2018**, 164, 304.
[300] N. Karim, M. Zhang, S. Afroj, V. Koncherry, P. Potluri, K. S. Novoselov, *Rsc Adv* **2018**, 8, 16815.
[301] B. Zhou, Z. Li, Y. Li, X. Liu, J. Ma, Y. Feng, D. Zhang, C. He, C. Liu, C. Shen, *Compos Sci Technol* **2021**, 201, 108531.
[302] Q. Bu, Y. Zhan, F. He, M. Lavorgna, H. Xia, *J Appl Polym Sci* **2016**, 133, n/a.
[303] M. Tian, Y. Hao, L. Qu, S. Zhu, X. Zhang, S. Chen, *Mater Lett* **2019**, 234, 101.
[304] IPC, in *IPC-8921*, IPC, 2019.
[305] SEMI, SEMI, 2020.
[306] A. S. Farooq, P. Zhang, *Compos Part Appl Sci Manuf* **2020**, 106249.

[307] A. Iqbal, P. Sambyal, C. M. Koo, *Adv Funct Mater* **2020**, 30, 2000883.
[308] F. Shahzad, M. Alhabeb, C. B. Hatter, B. Anasori, S. M. Hong, C. M. Koo, Y. Gogotsi, *Science* **2016**, 353, 1137.
[309] A. Iqbal, J. Hong, T. Y. Ko, C. M. Koo, *Nano Converg* **2021**, 8, 9.